\documentclass[12pt]{article}
\usepackage{fullpage,amssymb,amsmath}
\usepackage[dvips]{epsfig}

\def\##1{{\bf #1}}
\def\=#1{\underline{\underline{#1}}}

\def\+#1{\underline{\bf #1}}
\def\*#1{\underline{\underline{\bf #1}}}

\def\r#1{(\ref{#1})}
\def\l#1{\label{#1}}
\def\c#1{\cite{#1}}

\def\le{\left(}
\def\ri{\right)}
\def\les{\left[}
\def\ris{\right]}
\def\lec{\left\{}
\def\ric{\right\}}

\def\.{\mbox{ \tiny{$^\bullet$} }}

\def\epso{\epsilon_{\scriptscriptstyle 0}}

\def\muo{\mu_{\scriptscriptstyle 0}}

\def\ko{k_{\scriptscriptstyle 0}}
\def\co{c_{\scriptscriptstyle 0}}

\def\eps{\epsilon}

\begin{document}

\Large
\begin{center}
{\bf Positive--, Negative--, and Orthogonal--Phase--Velocity
Propagation of Electromagnetic Plane Waves in a Simply Moving
Medium}

\vspace{10mm} \large

Tom G. Mackay\footnote{Corresponding Author. Fax: + 44 131
650 6553; e--mail: T.Mackay@ed.ac.uk.}\\
{\em School of Mathematics, University of Edinburgh, Edinburgh EH9
3JZ, UK}\\ \vspace{4mm}
 Akhlesh  Lakhtakia\\
 {\em CATMAS~---~Computational \& Theoretical
Materials Sciences Group\\ Department of Engineering Science and
Mechanics\\ Pennsylvania State University, University Park, PA
16802--6812, USA}\\ \vspace{4mm}
Sandi Setiawan\\
{\em School of Mathematics,
University of Edinburgh, Edinburgh EH9 3JZ, UK}\\

\end{center}

\vspace{4mm}

\normalsize

\begin{abstract}
Planewave propagation in  a simply moving, dielectric--magnetic
medium that is isotropic in the co--moving reference frame, is
classified into three different categories: positive--,
negative--, and orthogonal--phase--velocity (PPV, NPV, and OPV).
Calculations from the perspective of an observer located in a
non--co--moving reference frame show that, whether the nature of
planewave propagation is PPV or NPV (or OPV in the case of
nondissipative mediums)  depends strongly upon the magnitude and
direction of that observer's velocity relative to the medium. PPV
propagation is characterized by a positive real wavenumber,  NPV
propagation by a negative real wavenumber. OPV propagation only
occurs for nondissipative mediums, but weakly dissipative mediums
can support nearly OPV propagation.
\end{abstract}


\section{Introduction}

Although the complexity of the response of materials to electromagnetic
radiation has long been recognized, the subdiscipline of
\emph{complex--mediums electromagnetics\/} (CME) came into
prominence only during the 1990s~---~as Weiglhofer recently demonstrated
in an extended review \c{Weigl}. Besides nonlinearity, general complex mediums
are not merely anisotropic, but are also magnetoelectric. Plasmas,
ferrites, isotropic chiral materials, structurally chiral materials, bianisotropic
materials~---~all are excellent examples of linear complex mediums. Some
of these materials occur in nature, others are artificially made. Add nonlinearity
to the mix, and a  bewilderingly huge palette of electromagnetic response
properties emerges \c{GSbook,LMbook}.

While that stream of CME continues
to flow unabated, a second stream of CME sprang in 2000. This new stream
initially contained quite simple mediums: isotropic and dielectric--magnetic,
but with the phase velocity of plane waves
therein being in opposition to the time--averaged Poynting
vector \c{LMW03}. There has been some intermingling of the two streams \c{Pen,LM05},
but much of today's focus in the second stream is still on
isotropic and dielectric--magnetic mediums.

One current in that stream is the visualization of the
negative--phase--velocity (NPV) scenario by inertial observers
that are moving at a fixed velocity with respect to a certain
medium (or vice versa!). This is a sensible issue, as human
vehicles continue to penetrate the universe beyond our planetary
atmosphere \c{MLcurr}. In a predecessor paper \c{NPV_STR}, we theoretically
demonstrated that isotropic  dielectric--magnetic mediums which do
not support NPV propagation when viewed by a co--moving observer, \emph{can} support NPV
propagation when they are viewed in a reference frame which is
translating at a fixed velocity of sufficiently high magnitude. Even more
recently, we deduced the possiblity of orthogonal--phase--velocity
(OPV) propagation~---~a phenomenon characterized by the
orthogonality of the phase velocity and the time--averaged
Poynting vector, when viewed by a non--co--moving inertial
observer \c{OPV_STR}.

In this paper, we report a comprehensive study of NPV, OPV,
and the conventional positive--phase--velocity (PPV) propagation of
plane waves in a simply moving, homogeneous, dielectric--magnetic
medium that is isotropic with respect to a co--moving observer.
The plan of this paper is as follows: Section \ref{analysis} is devoted
to the Minkowski constitutive relations and planewave propagation. Section \ref{num} contains numerical results for NPV, PPV, and OPV propagation
in both dissipative and nondissipative mediums. Conclusions are presented in Section \ref{con}.

A note on notation:
  ${\rm Re}\lec Q \ric$ and ${\rm
Im}\lec Q \ric$ represent the real and imaginary parts,
respectively, of a complex--valued $Q$. The complex conjugate is
written as $Q^*$. Vectors are identified by bold typeface and
3$\times$3 dyadics are double underlined; $\hat{\#v}$ is a unit
vector co--directional with $\#v$; the unit dyadic is $\=I$;  and
$\#r$  denotes the spatial coordinate vector.
 The
permittivity and permeability of free space (i.e., vacuum) are
 $\epso$ and $\muo$, respectively;
 $\co=(\epso\muo)^{-1/2}$ is the speed of light in free space;
 $\omega$ is the angular frequency; and $\ko = \omega/\co$.
 An $\exp(-i\omega t)$ time--dependence is implicit.

\section{Theory}\label{analysis}

\subsection{ Minkowski constitutive relations}

Let us consider a homogeneous,  isotropic,
dielectric--magnetic medium, characterized by its relative
permittivity $\eps_r$ and relative permeability $\mu_r$,
 in an inertial frame of reference $\Sigma'$. That is,  the
 frequency--domain constitutive
 relations
\begin{equation}
\left.
\begin{array}{l}
\#D' =   \epso \eps_r\#E'  \\ \vspace{-3mm} \\
\#B' =  \muo \mu_r\#B'
\end{array}
\right\}, \l{Con_rel_p}
\end{equation}
describe the medium from the perspective of an observer located in
$\Sigma'$.

The inertial reference frame $\Sigma'$ moves at
constant velocity $\#v=v\hat{\#v}$ relative to another inertial
reference frame
 $\Sigma$.
  From the perspective of an observer located in $\Sigma$, the medium is
 described by
the Minkowski constitutive relations  \c{Chen,NPV_STR}
\begin{equation}
\left.
\begin{array}{l}
\displaystyle{ \#D = \epso \eps_r\, \=\alpha\.\#E + \frac{m
\hat{\#v}\times\#H}{\co} }
\\ \vspace{-3mm} \\
\#B = \displaystyle{ -\, \frac{m \hat{\#v}\times\#E}{\co} + \muo
\mu_r\, \=\alpha\.\#H\ }
\end{array}
\label{cr} \right\},
\end{equation}
where
\begin{eqnarray}
\=\alpha &=& \alpha\,\=I + (1-\alpha)\,\hat{\#v}\hat{\#v} \,,\\
\alpha & = & \frac{1-\beta^2}{1- \eps_r\mu_r\beta^2} \,,\\
m &=& \beta\,\frac{ \eps_r\mu_r-1} {1- \eps_r\mu_r\beta^2}\,.
\end{eqnarray}
The transformation from \r{Con_rel_p} to \r{cr} is achieved via
\begin{eqnarray}
\#E' &=& \le \#E \.\hat{\#v} \ri \hat{\#v} +
 \frac{1}{\sqrt{1 - \beta^2}} \,
\les \le \=I - \hat{\#v}\hat{\#v} \ri \. \#E + \#v \times
\#B \ris, \l{Ep}  \\
\#B' &=& \le \#B \.\hat{\#v} \ri \hat{\#v} +
 \frac{1}{\sqrt{1 - \beta^2}} \,
\les \le \=I - \hat{\#v}\hat{\#v} \ri \. \#B - \frac{ \#v \times
\#E}{\co^2} \ris,  \\
\#H' &=& \le \#H \.\hat{\#v} \ri \hat{\#v} +
 \frac{1}{\sqrt{1 - \beta^2}} \,
\les \le \=I - \hat{\#v}\hat{\#v} \ri \. \#H - \#v \times
\#D \ris,  \\
\#D' &=& \le \#D \.\hat{\#v} \ri \hat{\#v} +
 \frac{1}{\sqrt{1 - \beta^2}} \,
\les \le \=I - \hat{\#v}\hat{\#v} \ri \. \#D + \frac{ \#v \times
\#H}{\co^2} \ris,  \l{Dp}
\end{eqnarray}
where $\beta = v / \co$.

In the remainder of this paper, the propagation of plane waves in
a medium described by the Minkowski constitutive relations \r{cr}
is investigated, with particular emphasis on the orientation of
the  phase velocity relative to the time--averaged Poynting
vector. The reader is referred to standard works
\c{Chen}--\c{ChUn} for  background details on
planewave propagation within the context of special theory of relativity.

\subsection{Planewave propagation}

Plane waves in the chosen medium are described  by the field phasors
\begin{equation}
\left.
\begin{array}{l}
\#E = \#E_0 \exp \le i \#k \. \#r \ri \\ \vspace{-3mm} \\
\#H = \#H_0 \exp \le i \#k \. \#r \ri
\end{array}
\right\}.  \l{pw}
\end{equation}
Our attention is restricted to uniform plane
waves, as given by the wavevector $\#k = k \, \hat{\#k}$ where the
unit vector $\hat{\#k} \in \mathbb{R}^3$. The wavenumber $k$ is
generally complex--valued, i.e.,
\begin{equation} k = k_R + i k_I\,,
\end{equation}
where $k_R = \mbox{Re} \lec k \ric$ and $k_I = \mbox{Im} \lec k
\ric$.

To calculate the wavenumbers, solutions of the form \r{pw} are
sought to the frequency--domain Maxwell curl postulates combined
with the Minkowski constitutive relations. Thus, we find
\c{NPV_STR,Chen}
\begin{equation}
k= \ko \, \frac{-\beta\xi \, \hat{\#k}\.\hat{\#v}  \pm
\sqrt{\Delta}} {1-\xi\le \beta \, \hat{\#k}\.\hat{\#v} \ri^2}\,,
\l{k_sqrt}
\end{equation}
wherein
\begin{eqnarray}
\xi &=& \frac{ \eps_r\mu_r-1} {1-\beta^2}\,,\\
\Delta &=& 1 + \le \eps_r\mu_r- 1 \ri \delta\,,\\
\delta &=& \frac{1 - \le \beta \, \hat{\#k}\.\hat{\#v} \ri^2 }{1-
\beta^2} \ge 1\,.
\end{eqnarray}
Note that the medium with constitutive relations \r{cr} is
unirefringent since
 the two wavenumbers represented by \r{k_sqrt} are not
independent. The direction of planewave propagation determines the
 choice of sign for the square root term in \r{k_sqrt}.

 The amplitudes of the  electromagnetic field phasors
may expressed as \c{NPV_STR}
\begin{eqnarray}
\#E_0 &=& C_1\,\#e_1 + C_2 \,\#e_2 \,, \l{E_vec}\\
\#H_0 &=&  \frac{C_1}{\omega\muo \mu_r}\,\#e_2 -\omega\epso
\eps_r\,C_2 \,\#e_1 \,, \l{H_vec}
\end{eqnarray}
where $C_1$ and $C_2$ are arbitrary constants. The orthogonal
eigenvectors $\#e_1$ and $\#e_2$ are given by
\begin{eqnarray}
\#e_1& =& \#k\times\hat{\#v}\,, \l{e1_vector}
\\[5pt]
\#e_2 &=& \#a \times \#e_1 \,, \l{e2_vector}
\end{eqnarray}
with
\begin{equation}
\#a = \#k + \frac{\xi \le \omega - \#k \. \#v \ri}{\co^2} \,
\#v\,. \l{a_vec}
\end{equation}
 Hence,  the time--averaged Poynting vector
\begin{equation}
\langle \#P \rangle  = \frac{ | \#e_1 |^2  \exp \le -2 k_I
\hat{\#k}\.\#r \ri}{2} \, \le   \frac{| C_1 |^2}{\omega \muo}\,
\#p_1
 + |C_2|^2
\omega \epso  \#p_2 \ri\,, \l{P_general}
\end{equation}
is delivered by \r{E_vec}--\r{e2_vector}, where
\begin{equation}
\left.
\begin{array}{l}
\#p_1 = {\rm Re} \lec \displaystyle{ \frac{1}{\mu_r}\, \#a} \ric
\\ \vspace{-2mm} \\ \#p_2 = {\rm Re} \lec \eps^*_r \, \#a \ric
\end{array}
\right\}\,.
\end{equation}

 The  phase velocity
\begin{equation}
\#v_p = \co  \frac{\ko}{k_R} \hat{ \#k} \,, \l{vp_def}
\end{equation}
is categorized in terms of its orientation relative to the
time--averaged Poynting vector as follows:
\begin{itemize}
\item  positive phase velocity (PPV) is
characterized by $\#v_p \. \langle \#P \rangle > 0$,
\item negative
phase velocity (NPV) is characterized by $\#v_p \. \langle \#P
\rangle < 0$, and
\item orthogonal phase velocity (OPV) is characterized
by $\#v_p \. \langle \#P \rangle = 0 $.
\end{itemize}
As described elsewhere,
the phenomenon of OPV propagation arises for nondissipative mediums when
$\Delta < 0$ \c{OPV_STR}.

\section{Numerical illustrations}\label{num}

Let us now  explore numerically the occurrence of PPV, NPV, and
OPV propagation within the parameter space provided by $\lec
\eps_r, \mu_r, \#v \ric$. The velocity vector
$\#v$ is characterized by its relative magnitude $\beta$ and its
orientation angle $\theta = \cos^{-1} (\hat{\#k}\.\hat{\#v})$ relative to the direction of planewave propagation, in the remainder of this paper.

\subsection{Wavenumber}
We begin by considering  the wavevector $\#k$. Notice from
\r{k_sqrt} that the wavenumber $k$ is a function of the product
$\eps_r \mu_r$. In Figure~1, the real and imaginary parts of $k$
are plotted  for nondissipative mediums with  $\eps_r \mu_r \in
\le -10, 10 \ri$, for $\beta \in \lec 0.3, 0.6, 0.9 \ric$ and the
range of $\theta$ values given in Table~1. The imaginary part of
$k$ is null--valued across much~---~but not all~---~of the
range $\eps_r \mu_r > 0$.

Furthermore, at higher values of
$\beta$, $k_I \neq 0$ for larger positive values of $\eps_r
\mu_r$. This reflects the fact that $\Delta $ can be either positive-- or
negative--valued for $0 < \eps_r \mu_r < 1$, depending upon the
magnitude and direction of $\#v$. Many schemes for materials satisfying the condition $0 <
\eps_r\mu_r< 1$ have been formulated \c{Rachford,SH03,PO03},
although we must note that dissipation then is hard to avoid.

The real part
of $k$ in Figure~1 becomes unbounded for $\eps_r \mu_r
> 1$, at points  at which the denominator $(1 - \xi \beta^2
\cos^2 \theta)$ in \r{k_sqrt} vanishes. In view of our interest in
the transition between PPV and  NPV propagation, it is significant
that $k_R$ changes sign as $\eps_r \mu_r$ increases from $-10$ to
$+10$ only for  $ 90^\circ < \theta < 180^\circ$.

For dissipative mediums, $\eps_r \in \mathbb{C}$ and $\mu_r \in
\mathbb{C}$; hence,  $\eps_r = \eps^R_r + i \eps^I_r$ and $\mu_r
= \mu^R_r + i \mu^I_r$, where $\eps^{R,I}_r \in
\mathbb{R}$ and $ \mu^{R,I}_r \in
\mathbb{R}$. The wavenumber is  plotted
as a function of $\eps^R_r \in \le -10, 10 \ri$ in Figure~2~---~for
$\eps^I_r \in \lec 0.1 | \eps^R_r |, 2 | \eps^R_r | \ric$, $\mu_r
= \pm 1+ 0.1 i$, $\beta = 0.9$ and the range of $\theta$ values
given in Table~1. Generally, $k_I$
is nonzero for all values of $\eps^R_r$. As $\eps^R_r$
varies, strong resonances are observed in Figure~2 for both $k_R$ and $k_I$.
Notice in
Figure~2 that the real part of the wavenumber changes sign as
$\eps^R_r $ increases from $-10$ to $+10$
 only for  $ 90^\circ < \theta <
180^\circ$. This behaviour is similar to that observed for the
nondissipative scenario illustrated in Figure~1.

\subsection{Phase velocity}
Let us now turn to nature of the phase velocity, as delineated by
$\#v_p \. \langle \#P \rangle$. In Figure~3 the distribution of
PPV, NPV, and OPV propagation is mapped out across the $\beta
\theta$ plane for nondissipative mediums with $\eps_r \mu_r \in
\lec 1/3, 2/3, 10/3 \ric$. The  OPV regime arises for $\eps_r
\mu_r < 1$ and for large values of $\beta$, spanning a range of
values of $\theta$ which are symmetrically distributed about
$\theta = 90^\circ$. The  NPV regime is restricted to $\theta >
90^\circ$ for both $\eps_r \mu_r < 1$ and $\eps_r \mu_r > 1$. We
note that at $\eps_r \mu_r = 1$, the regions of NPV and OPV both
vanish, in consonance with the Lorentz invariance of vacuum
\c{Pappas}. Planewave propagation for the nondissipative scenario
corresponding to $\eps_r \mu_r < 0$ (not shown in Figure~1) is OPV
in nature for all values of $\beta \in (0,1)$, since $\Delta < 0$.

 For dissipative mediums, the
distribution of the NPV and PPV regimes in the $\beta \theta$
plane is displayed in Figure~4, for $\eps^R_r \in \lec \pm 1/3,
\pm 2/3, \pm 10/3 \ric$, $\eps^I_r \in \lec 0.1 | \eps^R_r |, 2 |
\eps^R_r | \ric$, and $\mu_r = \pm 1+ 0.1 i$. When $\eps^R_r > 0$
and $\mu^R_r > 0$, propagation is predominantly PPV
in nature, with the  NPV regime limited to  $\theta > 90^\circ$
 for large values of $\beta$. Similarly, for $\eps^R_r < 0$ and
$\mu^R_r < 0$, propagation is predominantly of the NPV
type, with PPV propagation arising only for $\theta < 90^\circ$
at large values of $\beta$. If $\eps^R_r$ and $\mu^R_r$ have
different signs,  and the imaginary parts of $\eps_r$ and $\mu_r$
are relatively small, then the distribution of NPV and PPV
propagation is evenly divided in the $\beta \theta$ plane along
the line $\theta = 90^\circ$. This even split between the
NPV and PPV regimes breaks down if the imaginary parts of
$\eps_r$ and/or $\mu_r$ become sufficiently large.

We note, in particular, from Figure~4 that OPV propagation does
not arise for dissipative mediums. Furthermore, there are no
regions in the $\beta \theta$ plane for which the sign of $\#v_p
\. \langle \#P \rangle $ is  indeterminate; i.e., both $\#v_p \. \#p_1
$ and $\#v_p \. \#p_2 $ always have the same sign.

In order
to explore the orientation of the vectors $\#p_1$ and
$\#p_2$ relative to $\#v_p$,  let us introduce the
angles
\begin{equation}
\phi_\ell  = \cos^{-1} \le \, \frac{\#v_p \. \#p_\ell
}{|\#v_p||\#p_\ell| } \, \ri, \qquad \qquad (\ell = 1,2)\,.
\end{equation}
In Figure~5, contour plots of the angles $\phi_1$ and $\phi_2$ are shown in
the $\beta \theta$ plane for $\eps^R_r =  \pm 2/3$,  $\eps^I_r \in
\lec 0.1 | \eps^R_r |, 2 | \eps^R_r | \ric$, and $\mu_r = \pm 1+
0.1 i$. We see that $\phi_{1} < 90^\circ$ and $\phi_{2} < 90^\circ$ for all values of
$\beta$ and $\theta$. Thus, the projection of the time--averaged
Poynting vector onto $\hat{\#k}$ is always positive. Therefore, NPV propagation arises when $k_R$ becomes
negative--valued.

The vectors $\#p_1$ and $\#p_2$ are close to being perpendicular
to $\hat{\#k}$  for large values of $\beta$ when $\theta$ takes
values close to $90^\circ$, in Figure~5. This is most noticeable
when the imaginary parts of $\eps_r$ and $\mu_r$ are relatively
small. We infer that \emph{nearly} OPV propagation is achieved for
weakly dissipative mediums for large values of $\beta$  when
$\theta$ is close to $ 90^\circ$. It is clear from Figure~5 that
the orientations of $\#p_1$ and $\#p_2$ with respect  to
$\hat{\#k}$ are generally very similar. In particular, for weakly
dissipative mediums, the orientations of $\#p_1$ and $\#p_2$
relative to $\hat{\#k}$
 are almost the same across the entire  $\beta \theta$ plane.

The occurrences of NPV and PPV propagation are further explored in
Figure~6, wherein the respective regimes are delineated as
functions $\eps^R_r \in \le -10, 10 \ri$ and $\eps^I_r \in \le 0,
10 \ri $ for $\theta \in \lec 30^\circ, 150^\circ \ric$, $\beta
\in \lec 0.1, 0.3, 0.5 \ric$, and $\mu_r = \pm 1 + 0.1 i$. When
$\theta > 90^\circ$ and $\beta $ is small, the predominant type of
propagation is PPV for  values of $\eps^R_r$ much less than zero
but NPV for values of $\eps^R_r$ much greater than zero. As
$\beta$ increases, the NPV  regime expands, to the extent that
when $\beta = 0.5$ most of the $\eps^R_r \eps^I_r$ plane supports
NPV propagation. The reverse situation is observed when $\theta <
90^\circ$: planewave propagation becomes increasingly PPV in
nature as $\beta$ increases.

The relative orientation angles $\phi_{\ell}$ of $\#p_{\ell}$, $(\ell=1, 2)$,
corresponding to the NPV and PPV distributions presented in
Figure~6, are provided in Figure~7. Both $\phi_1$ and
$\phi_2$ are acute across the entire $\eps^R_r \eps^I_r$ plane.
Thus,  our observation from Figure~5 that the projection of the
time--averaged Poynting vector onto $\hat{\#k}$ is always
positive, is further confirmed in Figure~7. Unlike the situation
in Figure~5,  the two vectors $\#p_1$ and $\#p_2$  clearly have
quite different orientations  when viewed as functions of
$\eps^R_r$ and $\eps^I_r$. Since $\beta \leq 0.5$ and $\theta$ is
not close to $90^\circ$, the propagation represented in Figure~7
does not approximate to OPV in nature.

\section{Conclusions}\label{con}

From our numerical studies, for a simply moving, dielectric--magnetic
medium that is isotropic in the co--moving reference frame,
we conclude the following.

\begin{itemize}
\item[(i)] Whether the nature of planewave propagation is PPV or  NPV
(or OPV in the case of nondissipative mediums)  depends strongly
upon the inertial frame of reference that the observer is situated in.

\item[(ii)] The projection of the
time--averaged Poynting vector onto
 the direction of planewave propagation, as
given by $\langle \#P \rangle \. \hat{\#k}$, is  non--negative.
PPV propagation is characterized by $k_R
> 0$; NPV propagation arises when   $k_R$ becomes
negative--valued.

\item[(iii)] OPV propagation only occurs for nondissipative
mediums; for such mediums it occurs when the imaginary part of the
wavenumber is nonzero.
 Weakly dissipative mediums support nearly OPV propagation
for large values of $\beta$, when $\theta$ is close to $90^\circ$.

\item[(iv)] The orientations of the
component vectors $\#p_1$ and $\#p_2$ of the time--averaged
Poynting vector are almost the same when viewed as functions of
$\beta$ and $\theta$, but the  orientation of $\#p_1$ relative to
$\#p_2$ is  sensitively dependent upon the constitutive parameters
$\eps_r$ and $\mu_r$.

\end{itemize}

\newpage

\begin{figure}[ht!] \centering \psfull  \epsfig{file=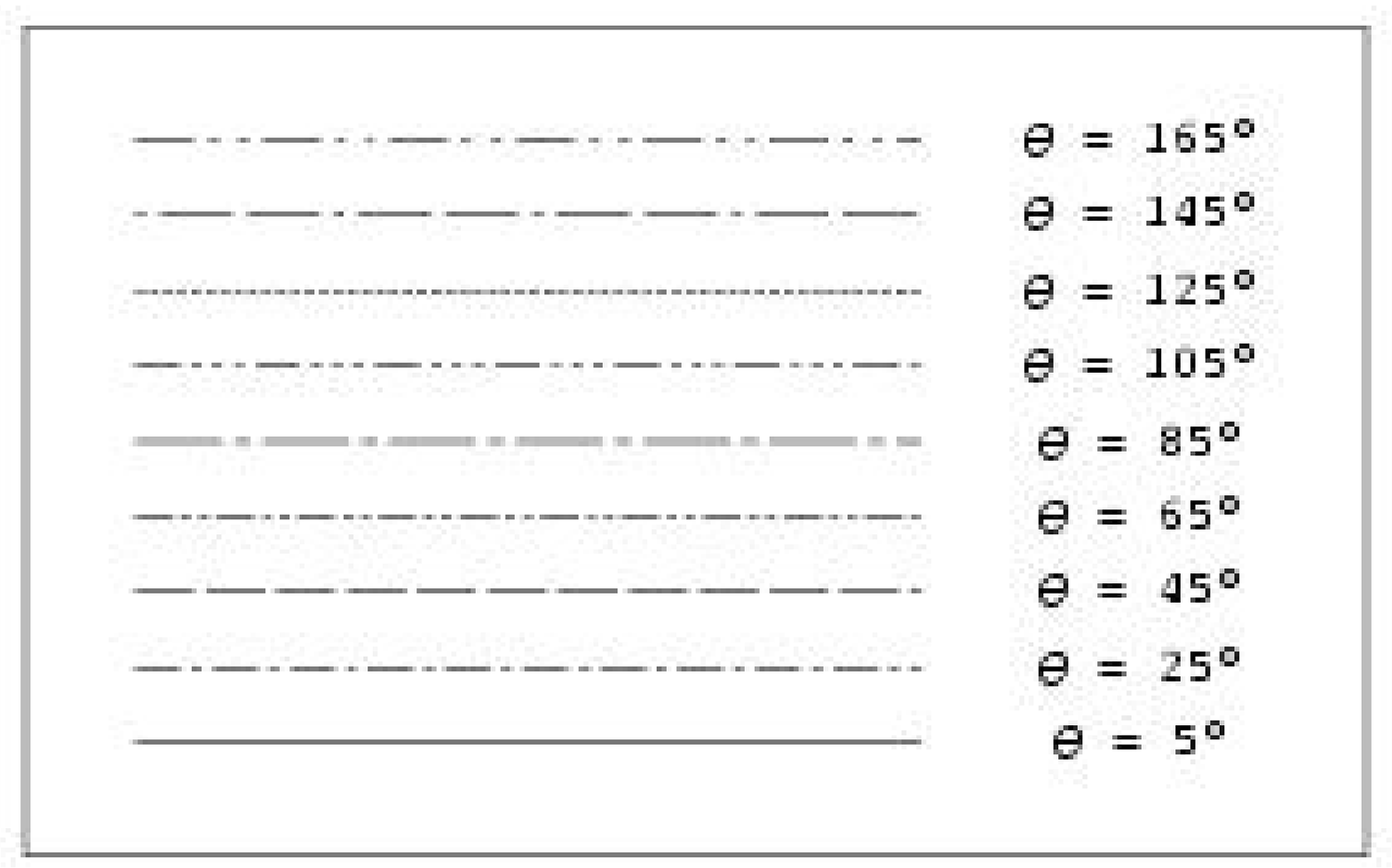,width=5.2in}
\end{figure}
\begin{center}
 Table~1. Key for the values of $\theta $ used in Figures~1 and 2.
\end{center}

\newpage

\setcounter{figure}{0}

\begin{figure}[!ht]
\centering \psfull \epsfig{file=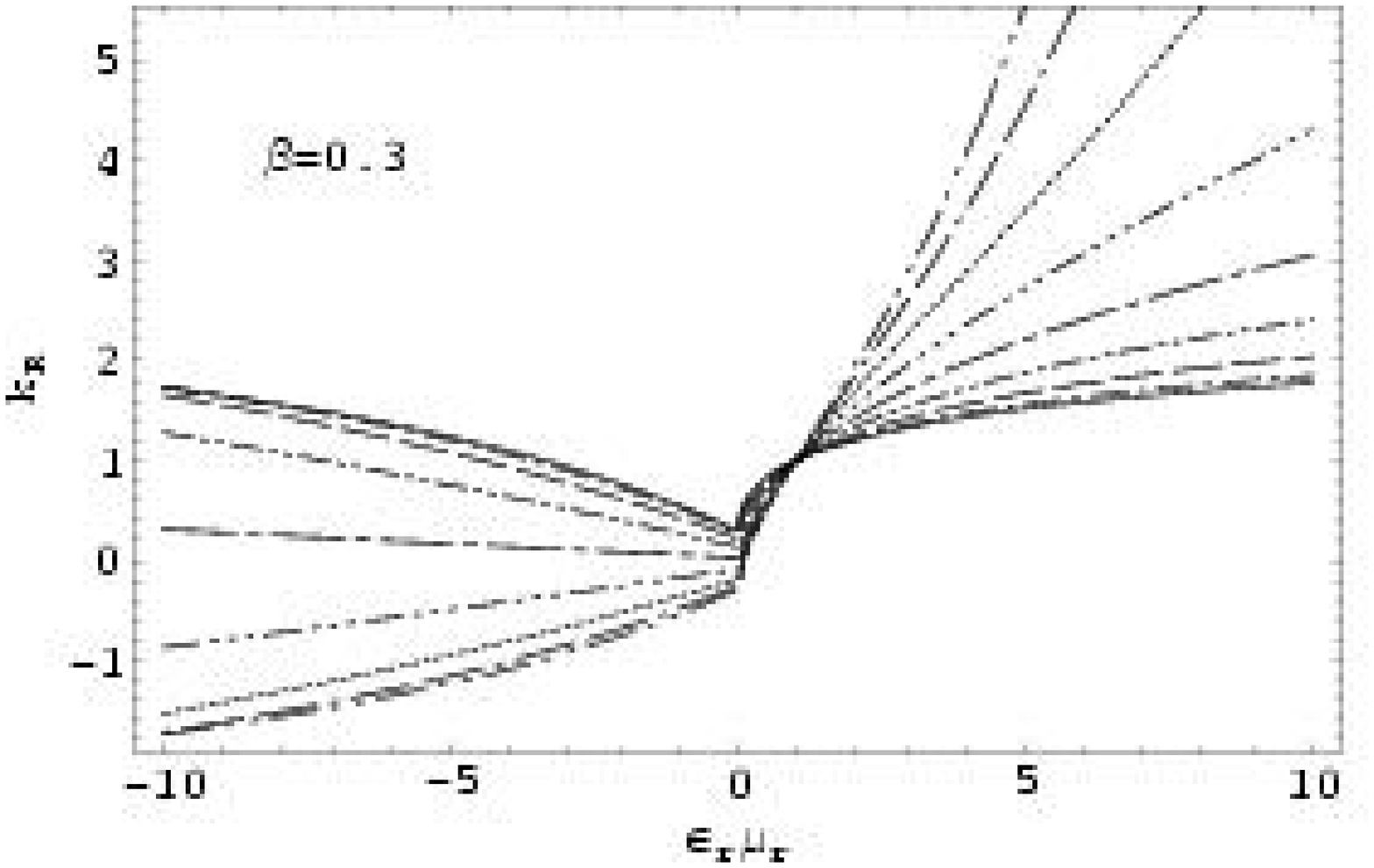,width=3.0in} \hfill
\epsfig{file=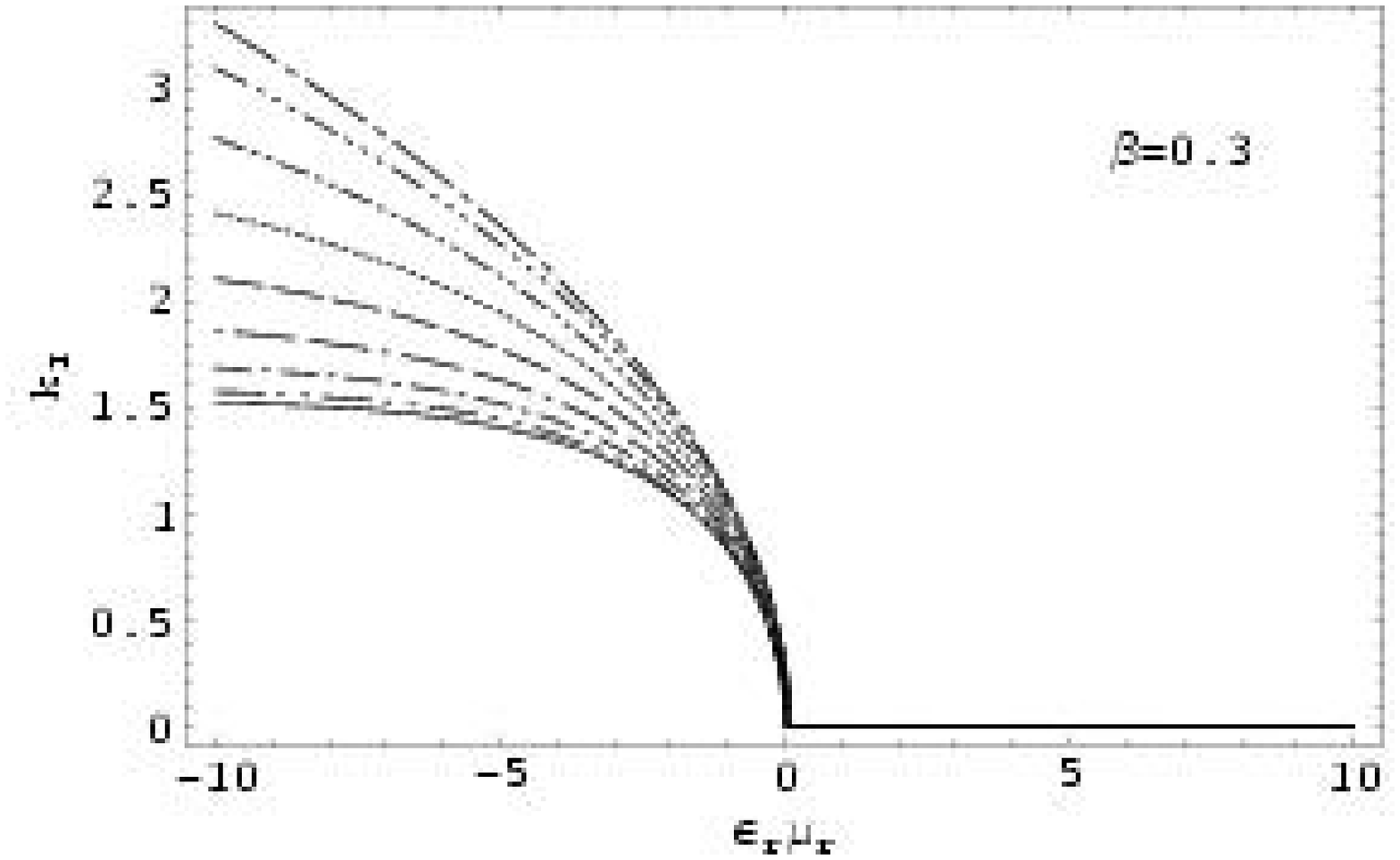,width=3.0in}
\\ \vspace{5mm}
\epsfig{file=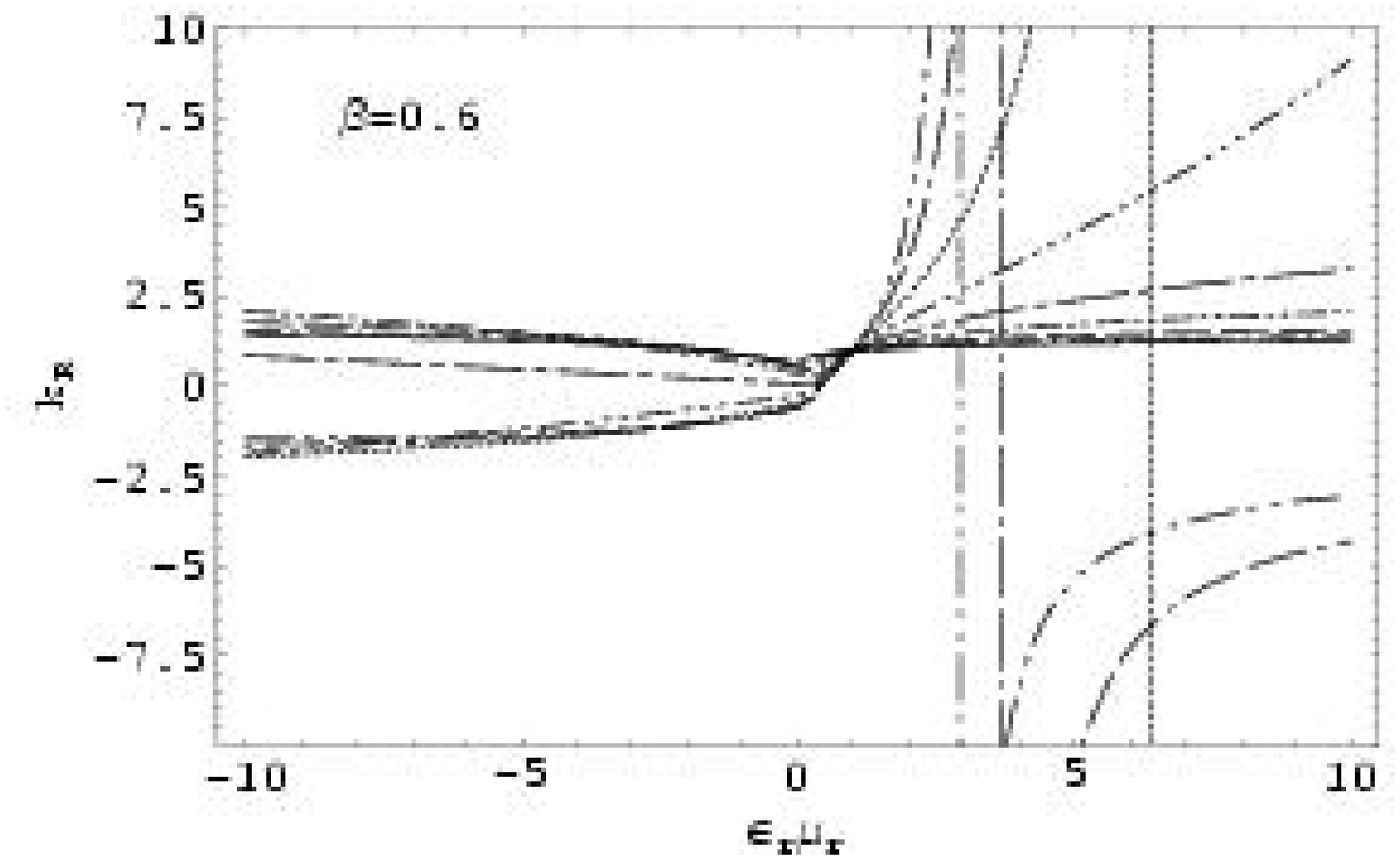,width=3.0in} \hfill
\epsfig{file=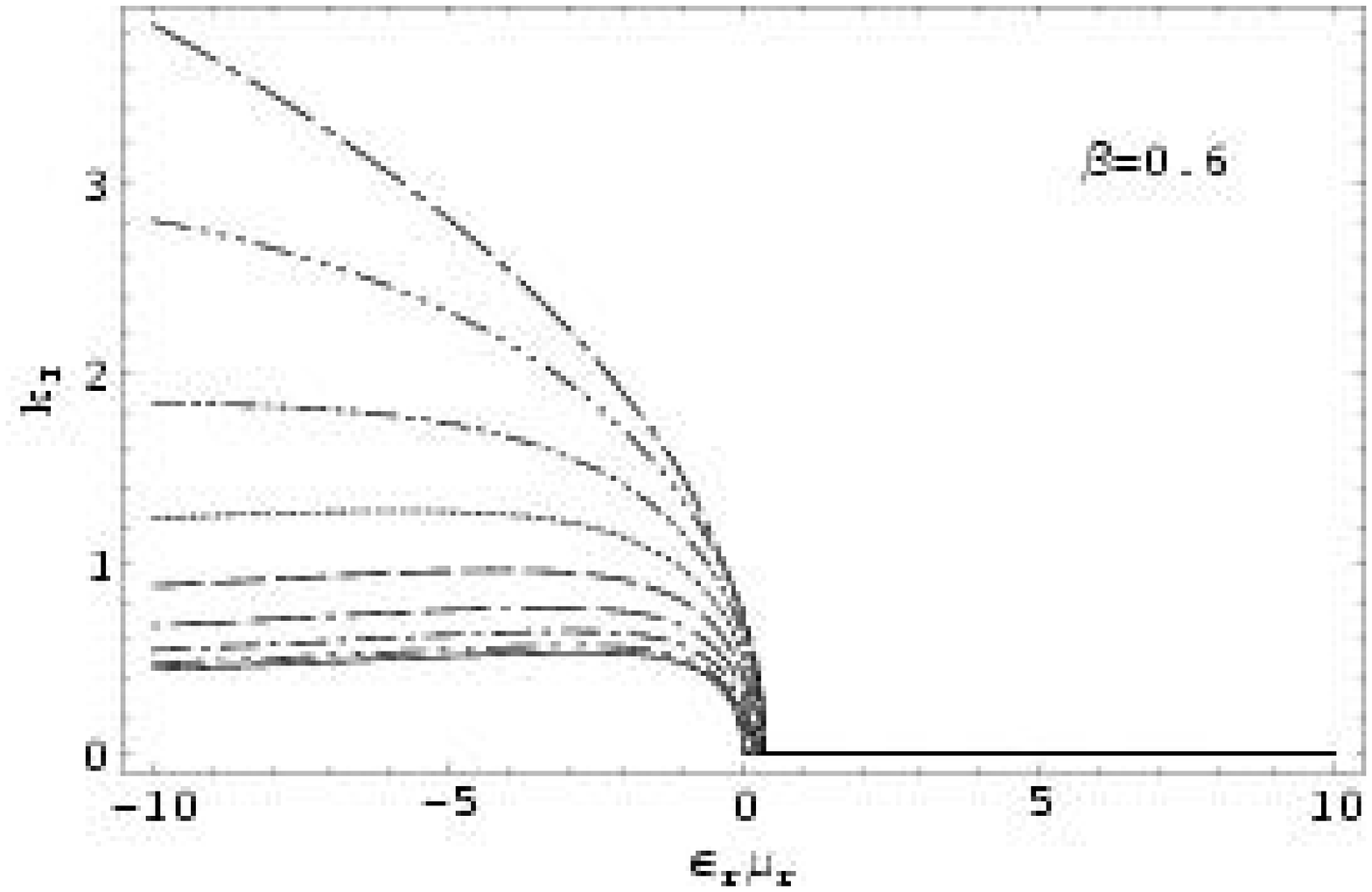,width=3.0in}
\\ \vspace{5mm}
\epsfig{file=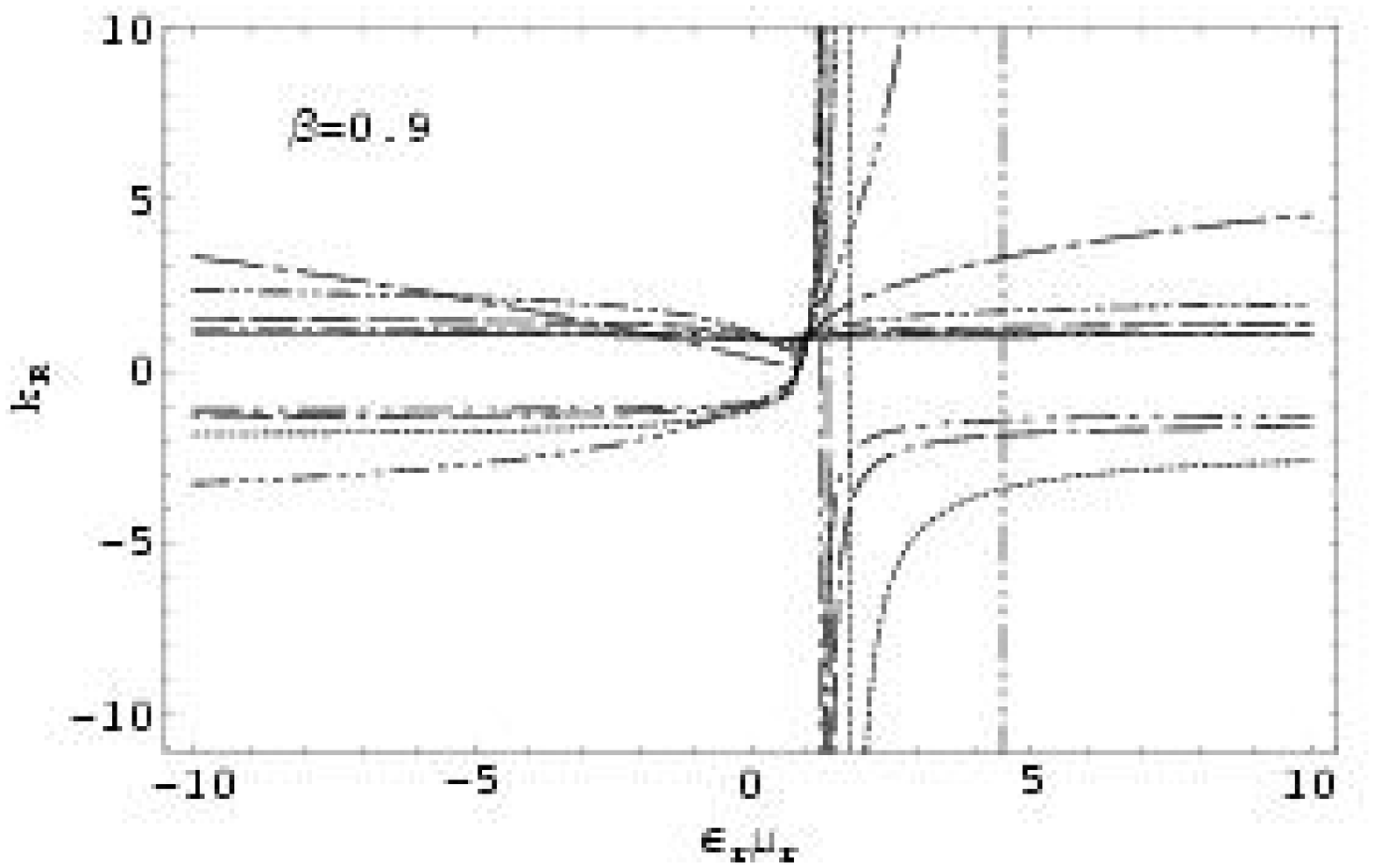,width=3.0in} \hfill
\epsfig{file=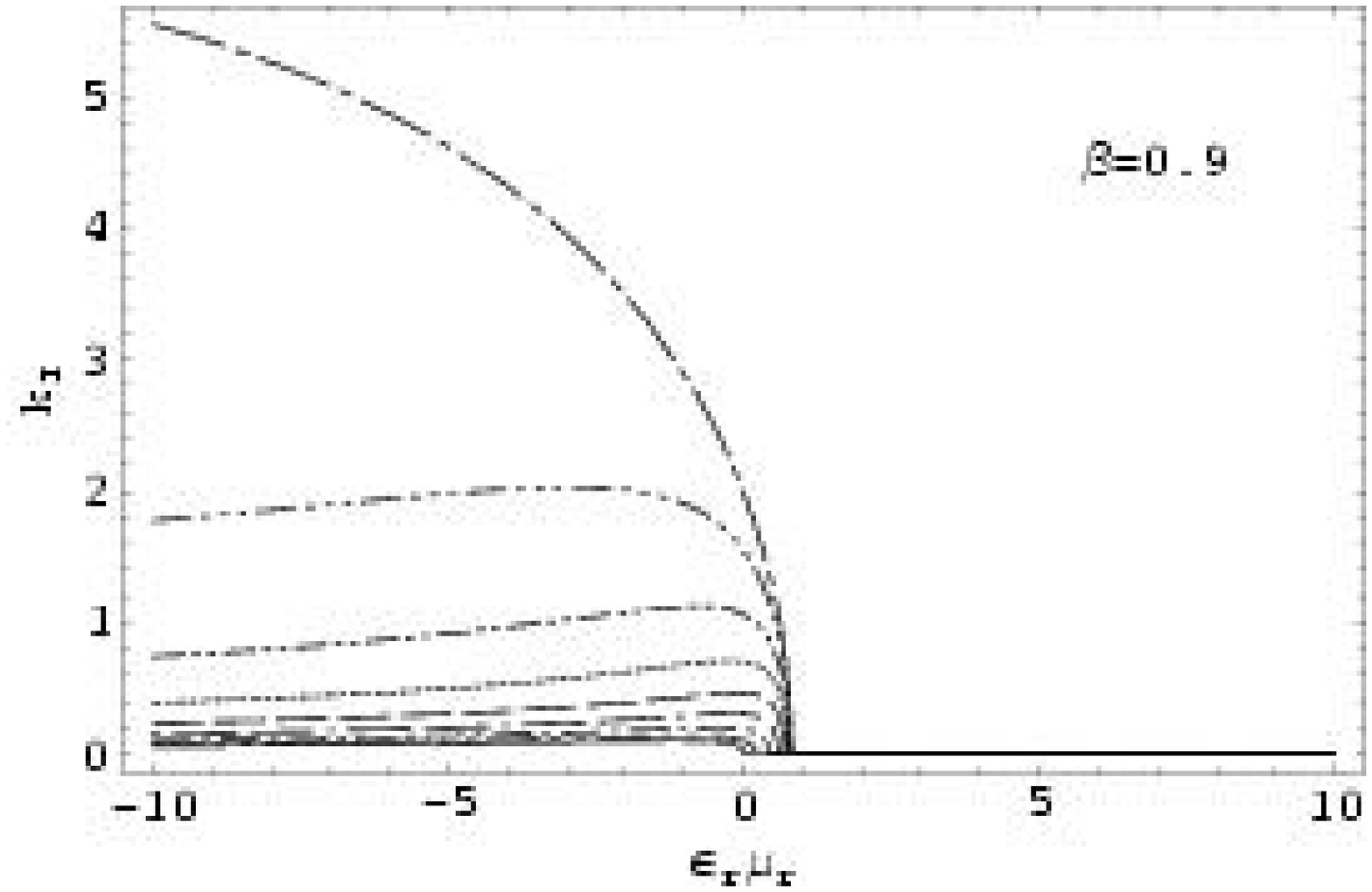,width=3.0in}
  \caption{\label{fig1}
The real (left) and imaginary (right) parts of the  wavenumber
$k$ for nondissipative mediums, normalized with respect to $ \ko$,
plotted against $\epsilon_r \mu_r$  for $\beta \in \lec 0.3, 0.6,
0.9 \ric$. The values of $\theta$ are identified in Table~1.
 }
\end{figure}

\newpage

\begin{figure}[!ht]
\centering \psfull \epsfig{file=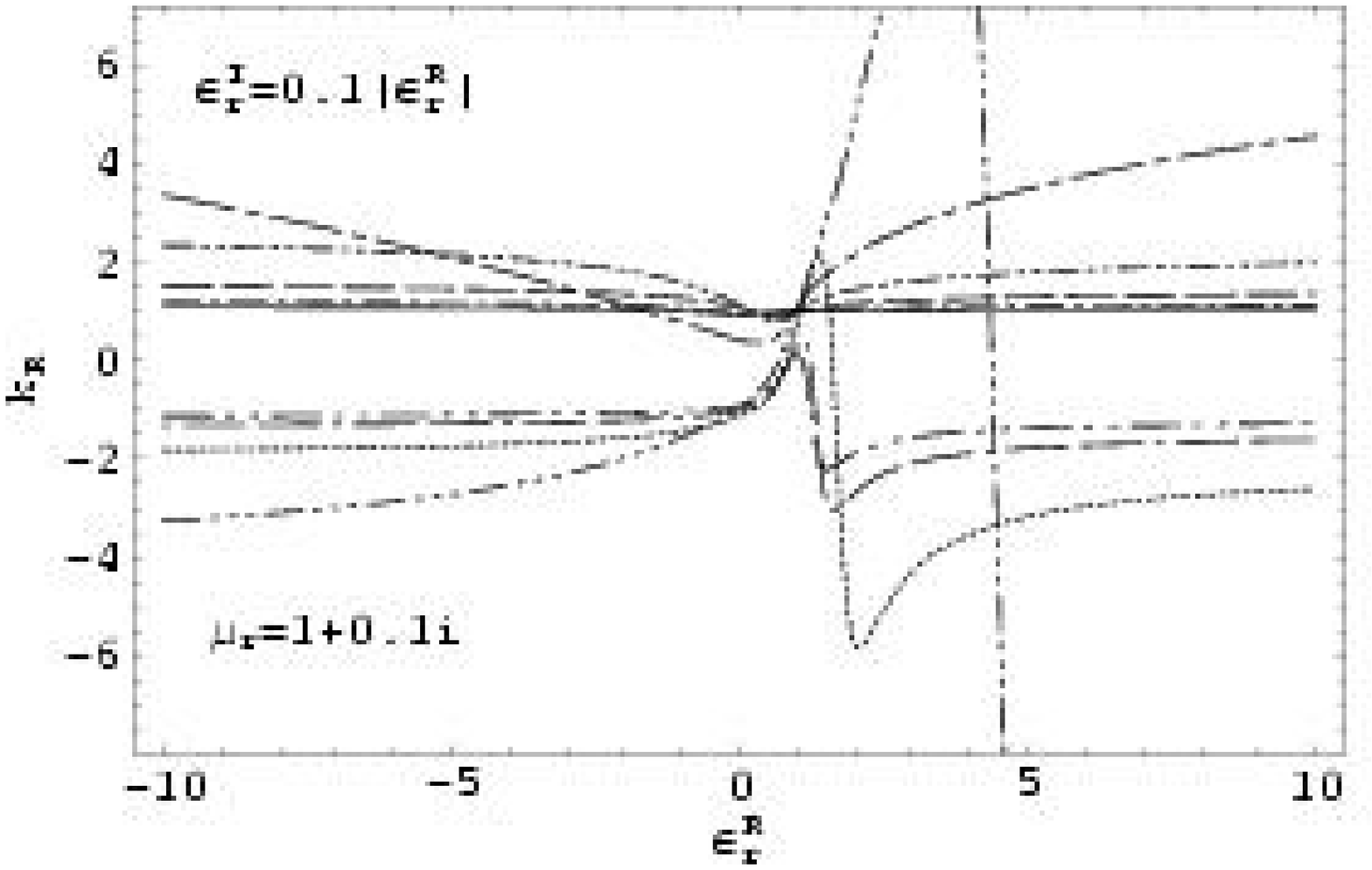,width=3.0in} \hfill
\epsfig{file=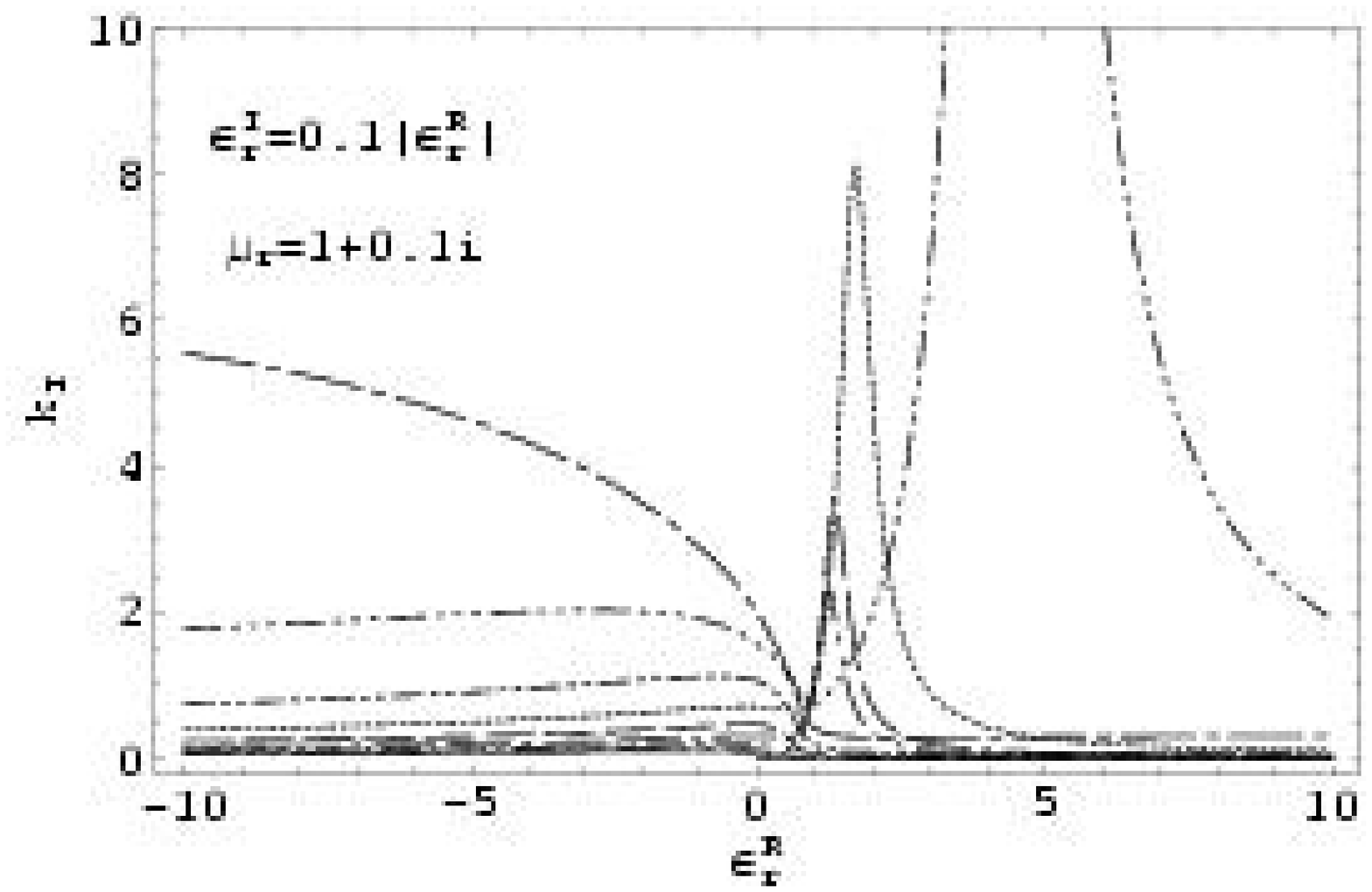,width=3.0in}
\\ \vspace{5mm}
\epsfig{file=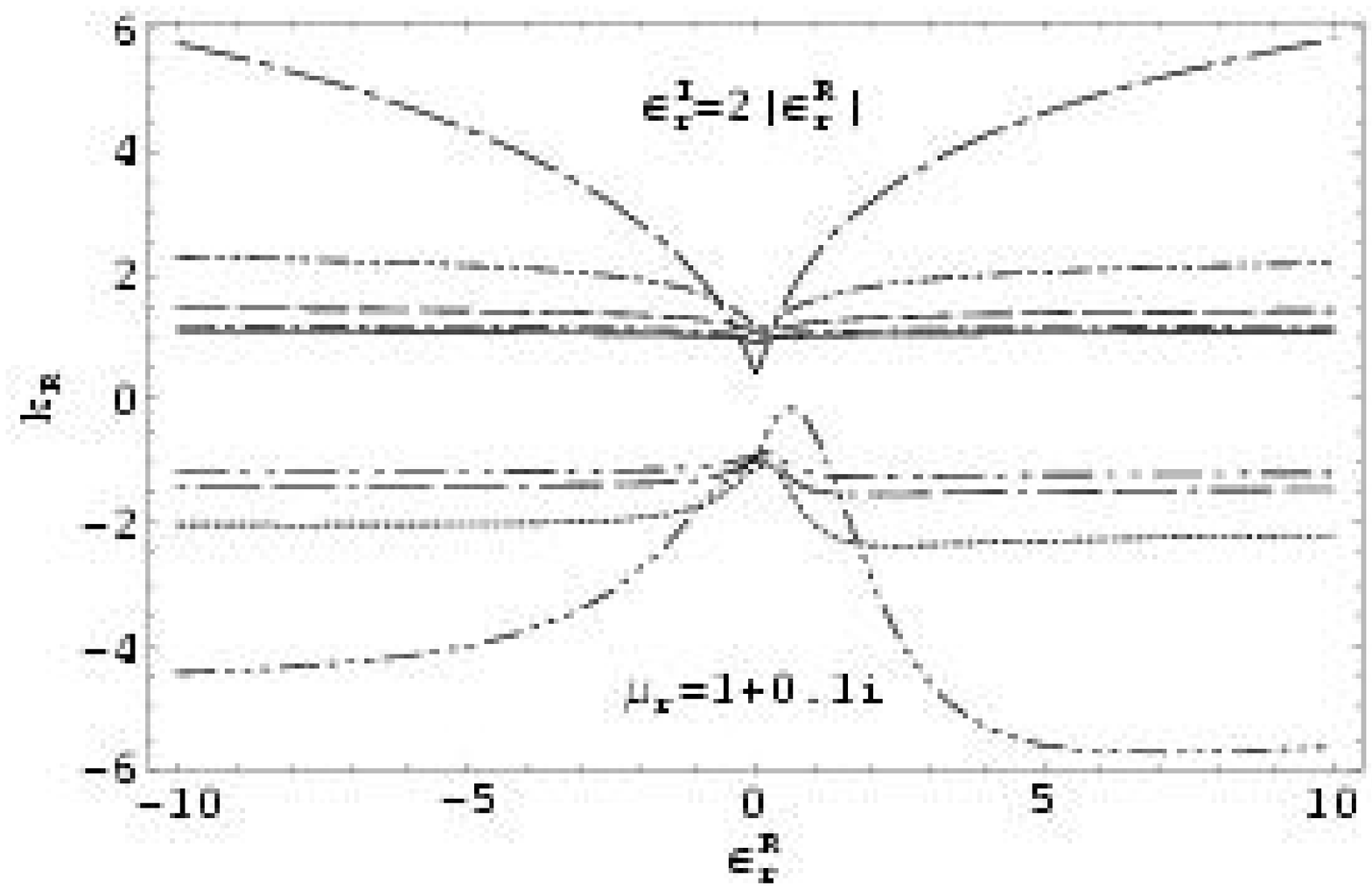,width=3.0in} \hfill
\epsfig{file=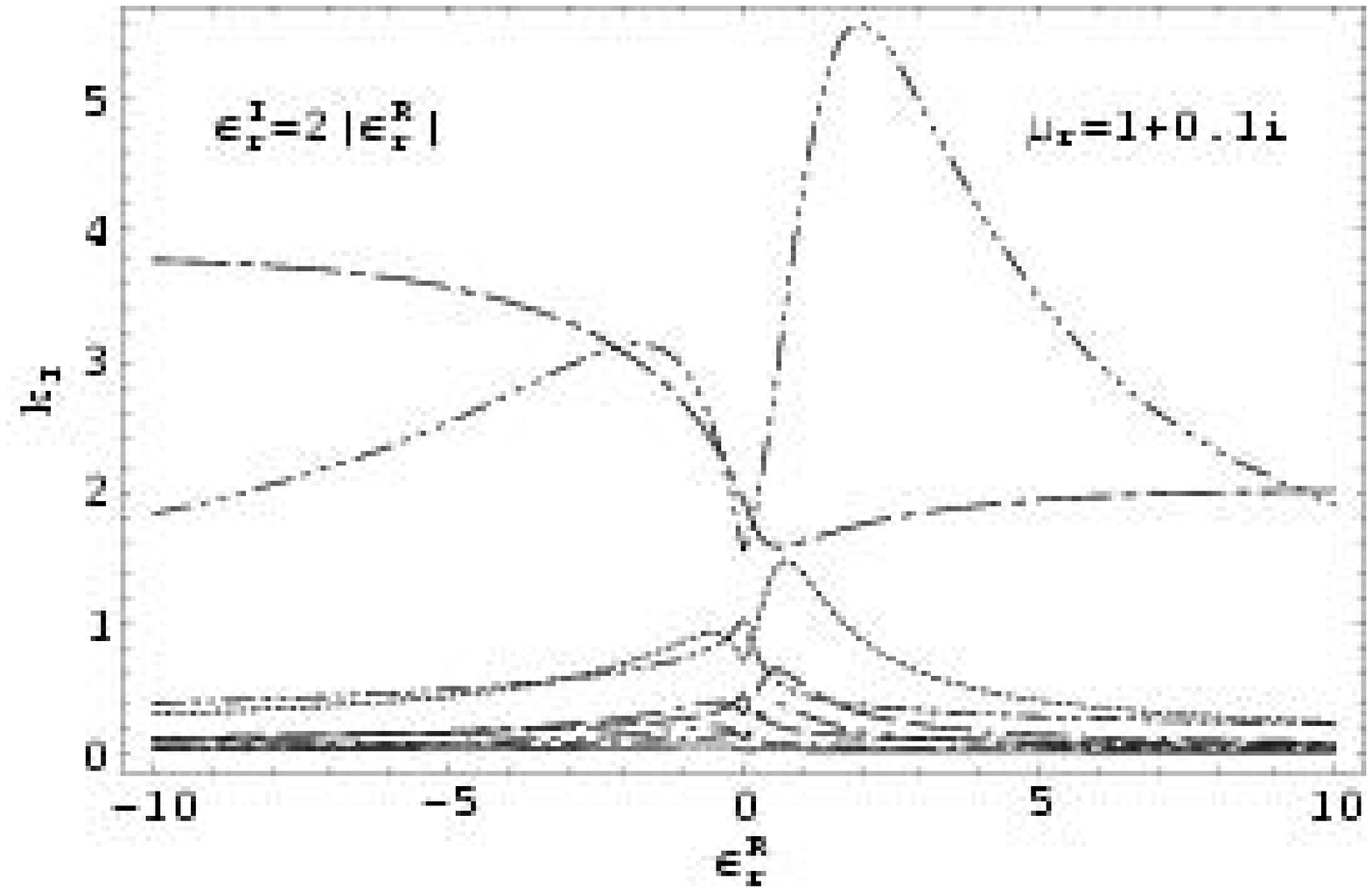,width=3.0in}
\\ \vspace{5mm}
\epsfig{file=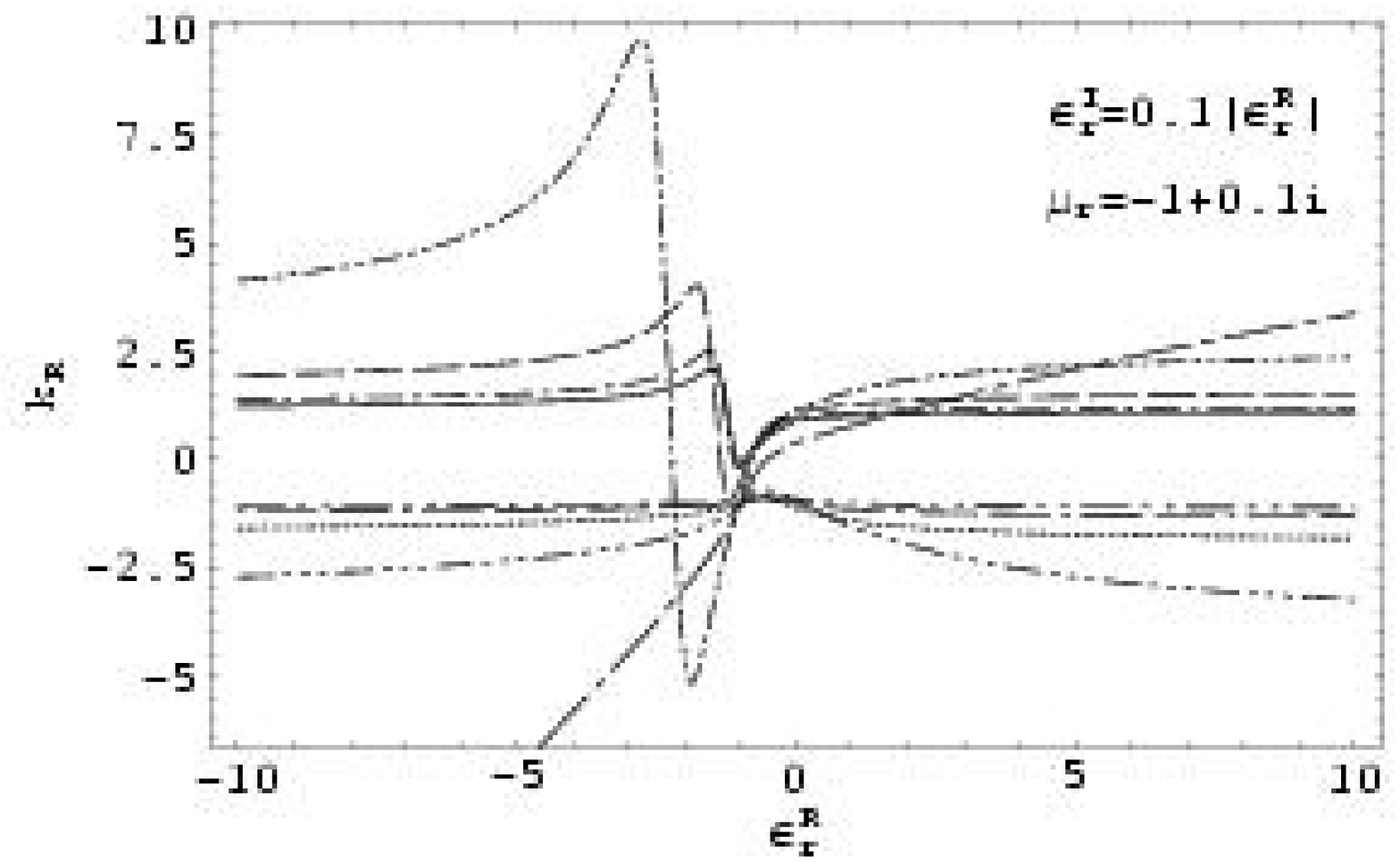,width=3.0in} \hfill
\epsfig{file=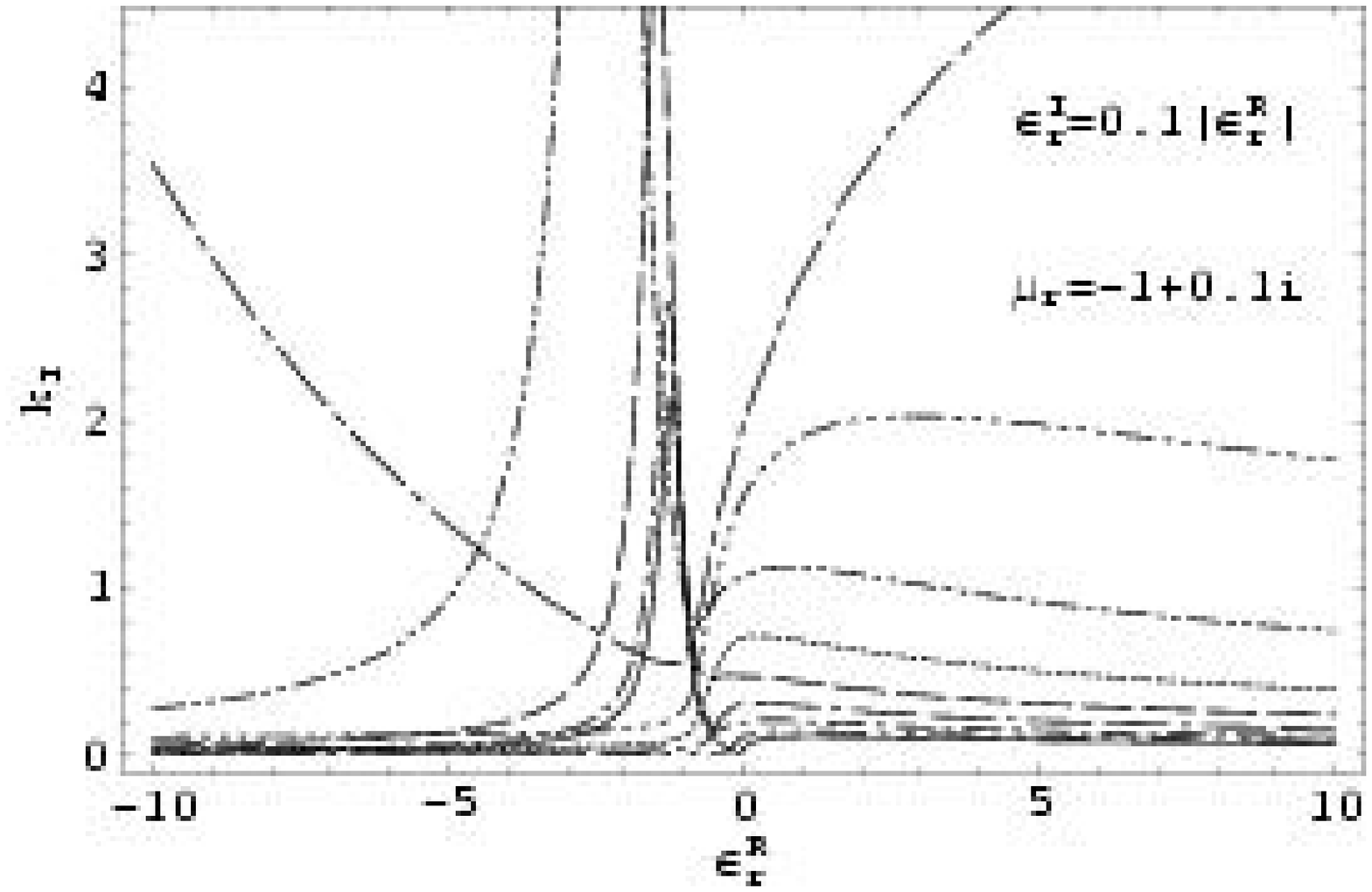,width=3.0in}
  \caption{\label{fig2}
The real (left) and imaginary (right) parts of the  wavenumbers
$k$ for dissipative mediums, normalized with respect to $ \ko$,
plotted against $\epsilon^R_r $  for  $\epsilon^I_r \in \lec 0.1 |
\eps^R_r |, 2 | \eps^R_r | \ric$, $\mu_r = \pm 1+ 0.1 i$, and
 $\beta = 0.9 $. The values of
$\theta$ are identified in Table~1.
 }
\end{figure}

\newpage

\begin{figure}[!ht]
\centering \psfull \epsfig{file=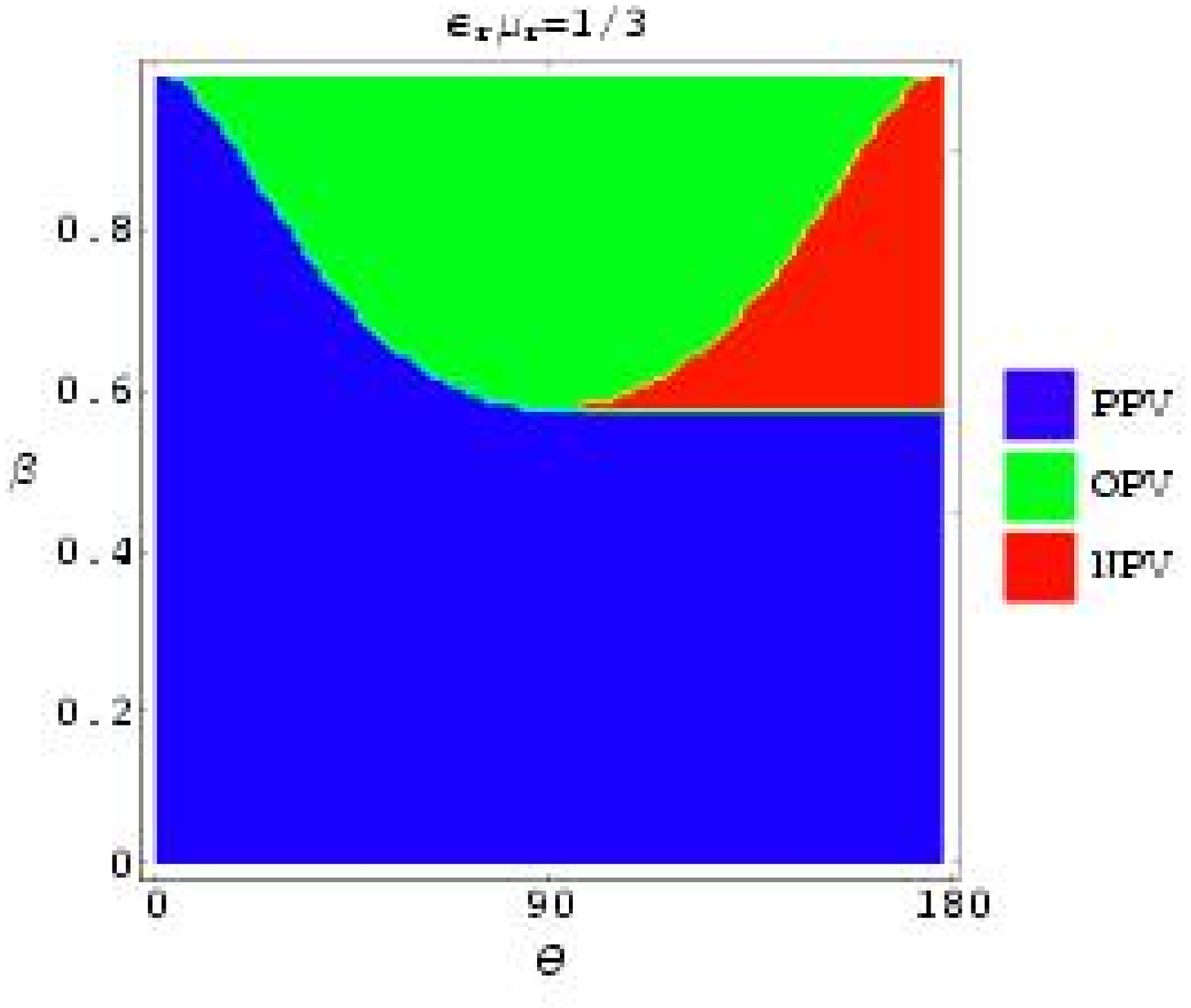,width=3.0in}
\\ \vspace{2mm}
\epsfig{file=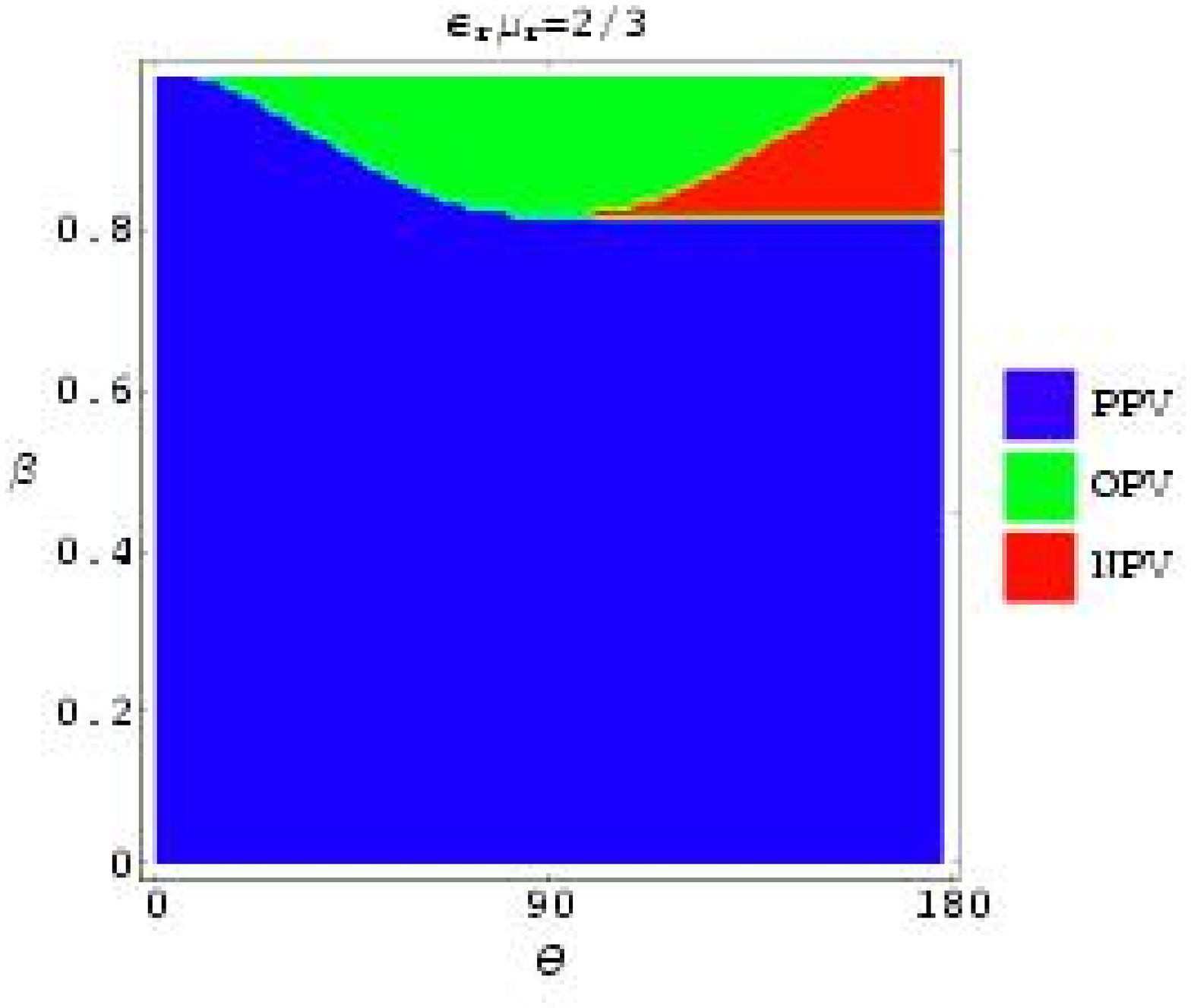,width=3.0in}
\\ \vspace{2mm}
\epsfig{file=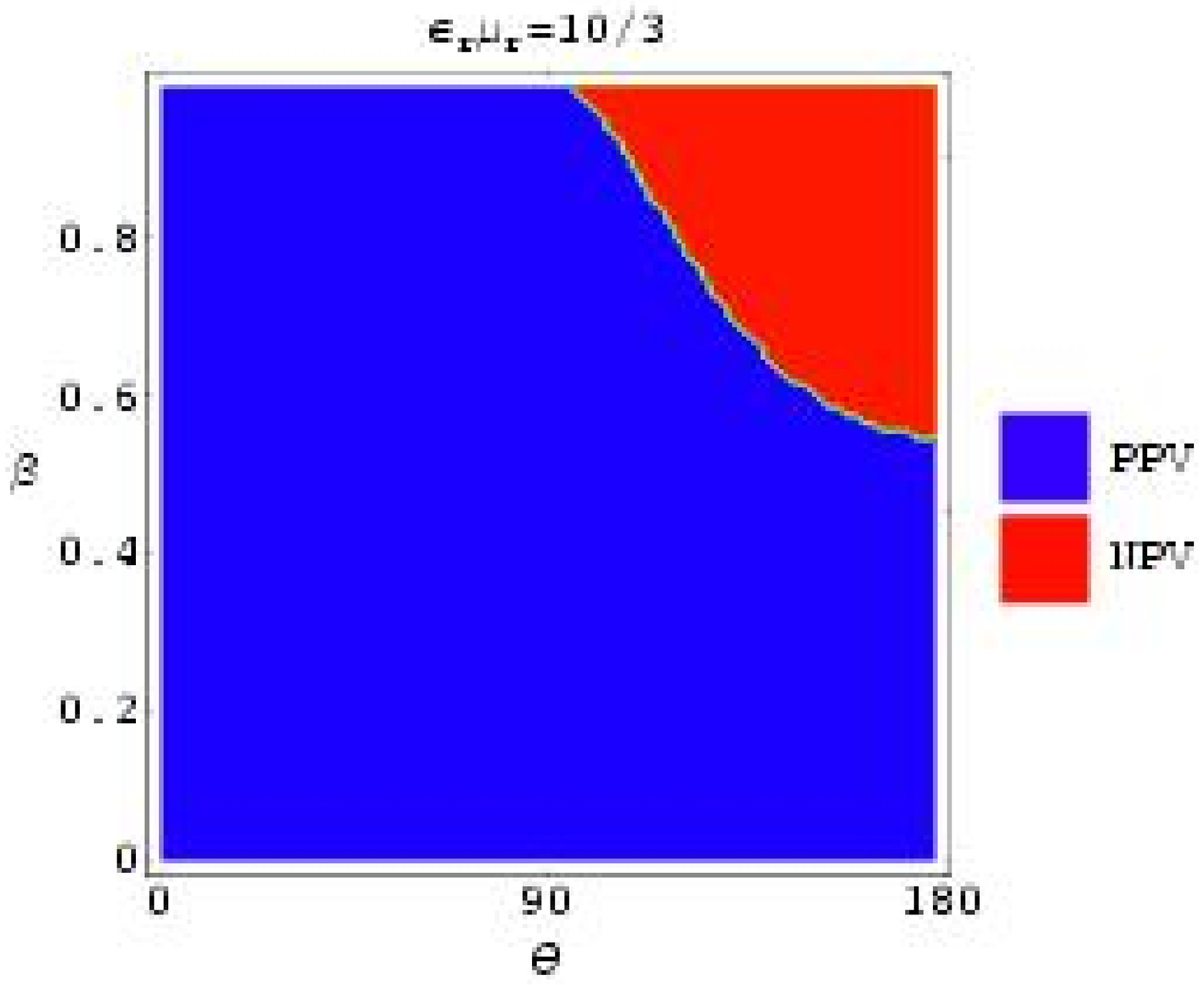,width=3.0in}
  \caption{\label{fig3}
  The distribution of positive phase velocity (PPV), negative
  phase velocity (NPV), and orthogonal phase velocity (OPV),
in relation to $\beta \in (0,1)$ and $\theta \in (0^\circ,
180^\circ)$
 when
 $
\epsilon_r \mu_r= 1/3$, $2/3$, and $10/3$. The labels `NPV' and
`PPV' shown hold for $\eps_r>0$ and $ \mu_r> 0$, but must be
 reversed for
$\eps_r<0$ and $ \mu_r< 0$.
 }
\end{figure}

\newpage

\begin{figure}[!ht]
\centering \psfull \epsfig{file=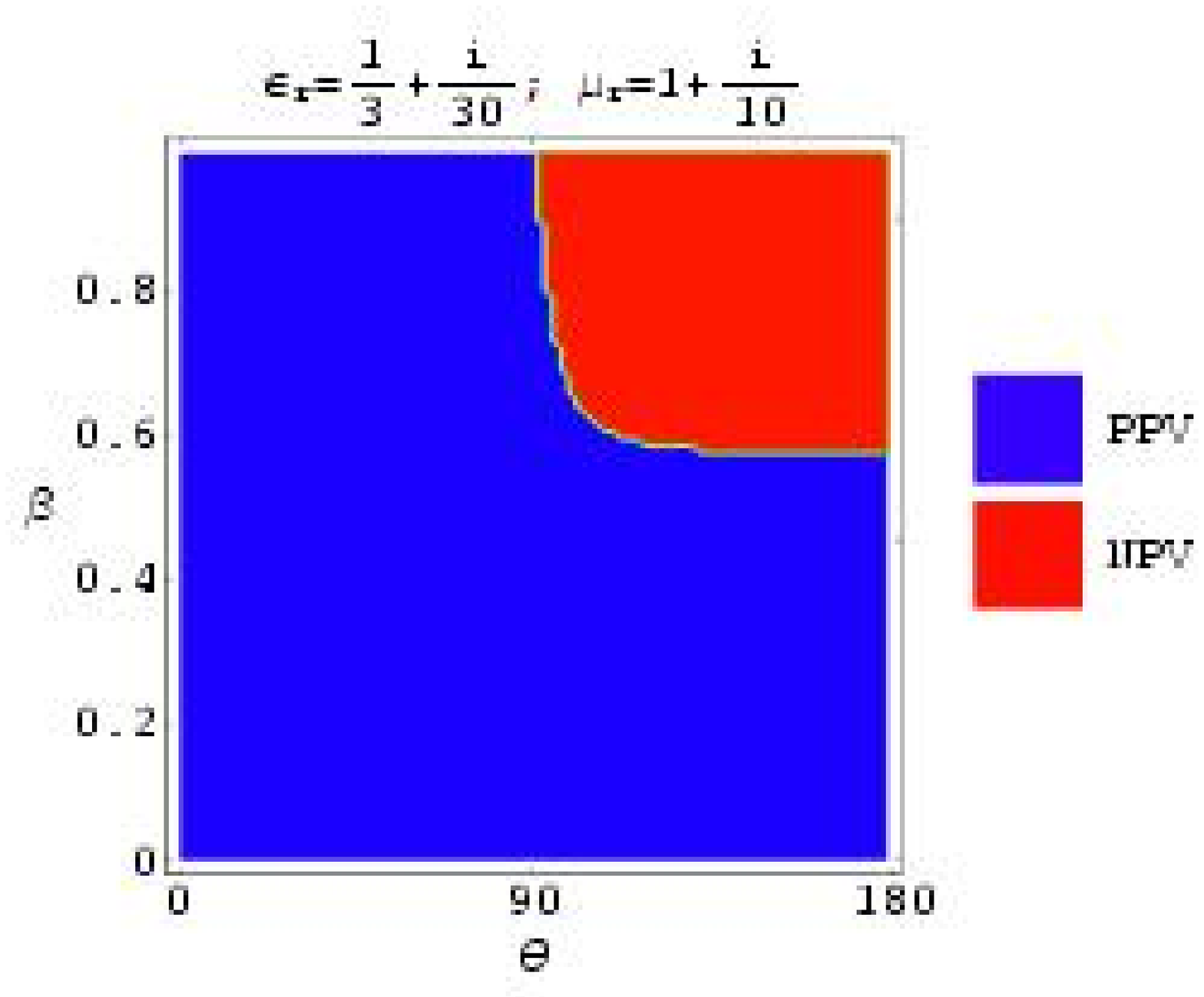,width=2.1in}
\epsfig{file=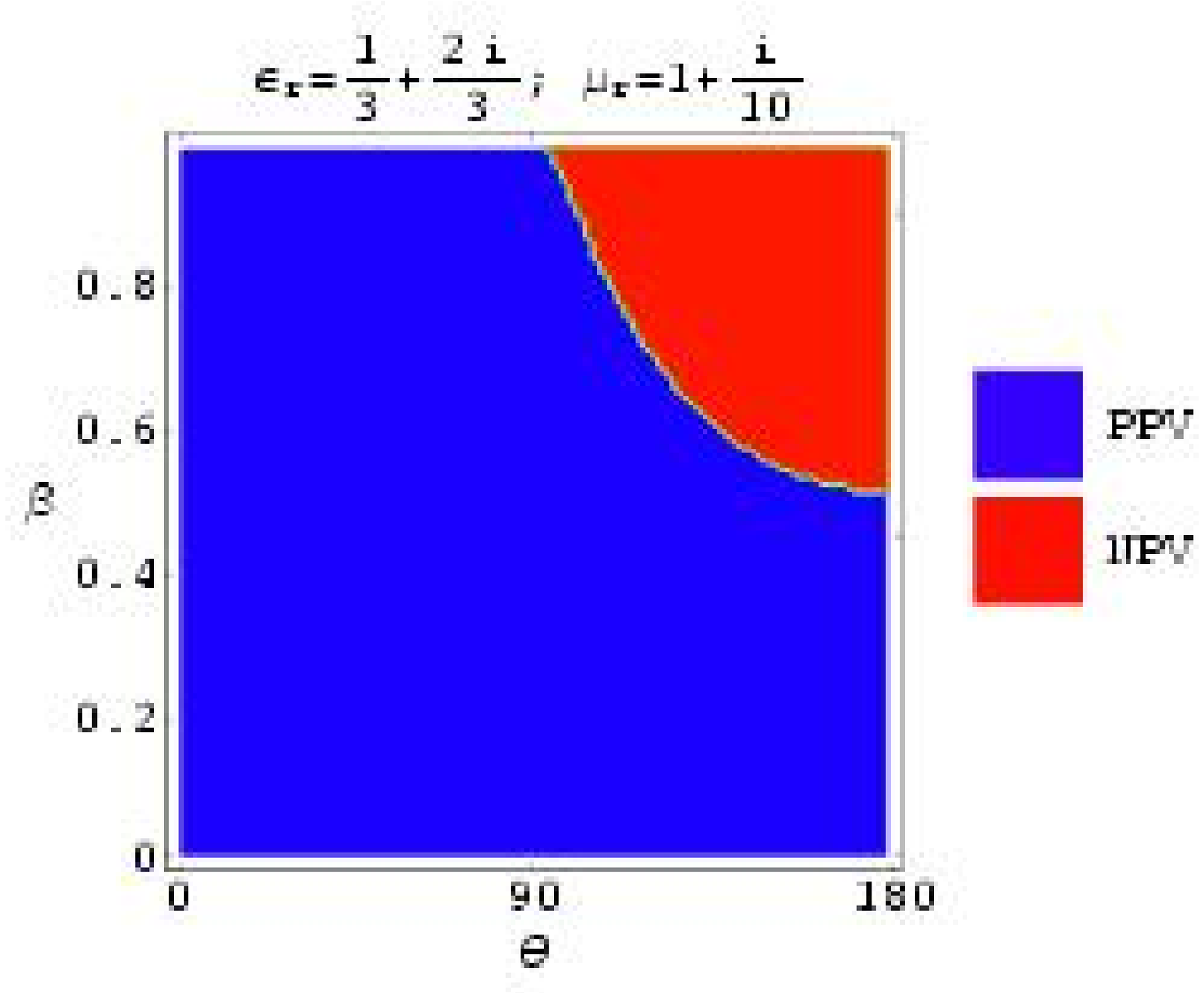,width=2.1in}
\epsfig{file=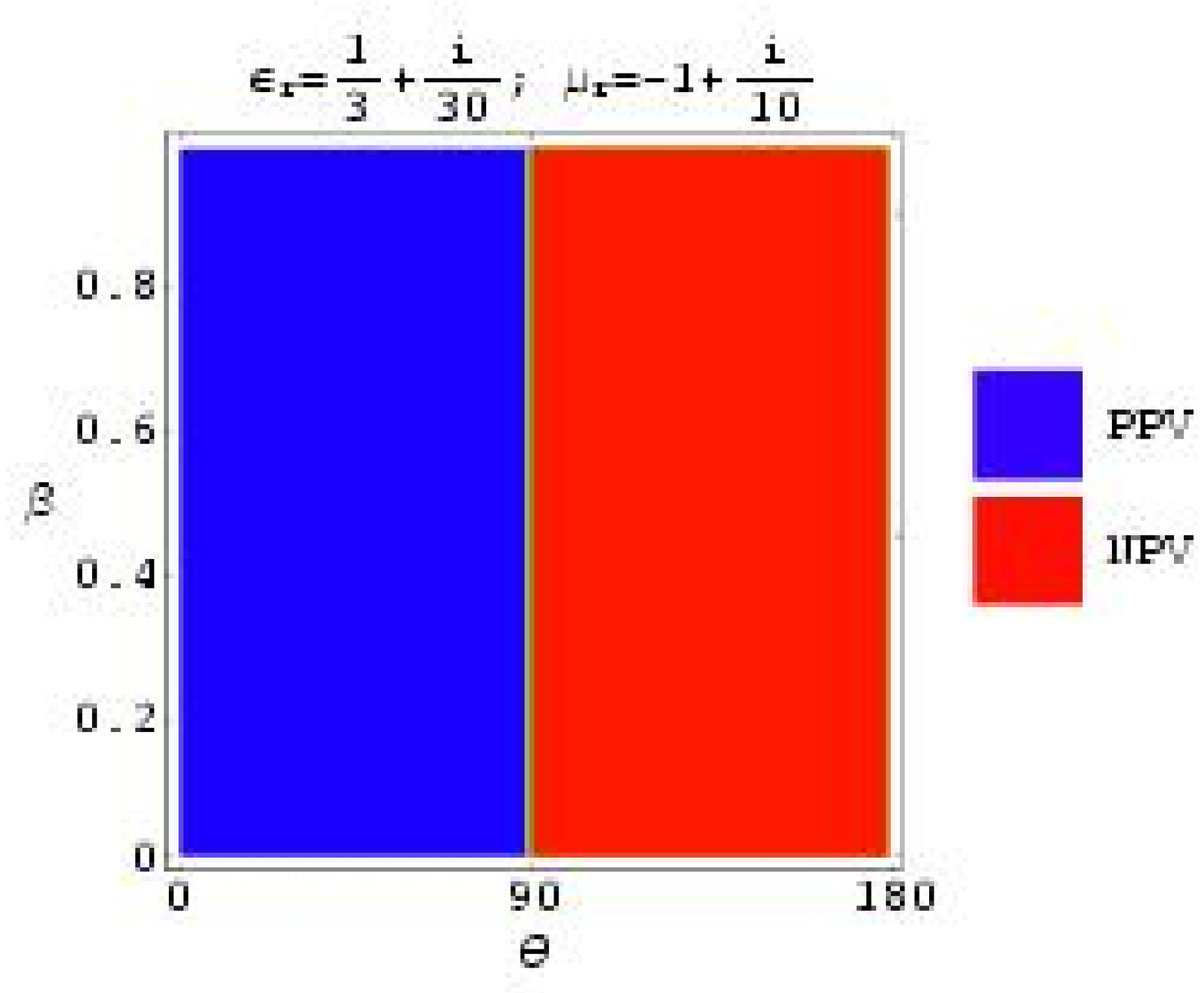,width=2.1in}
\\ \vspace{5mm}
\epsfig{file=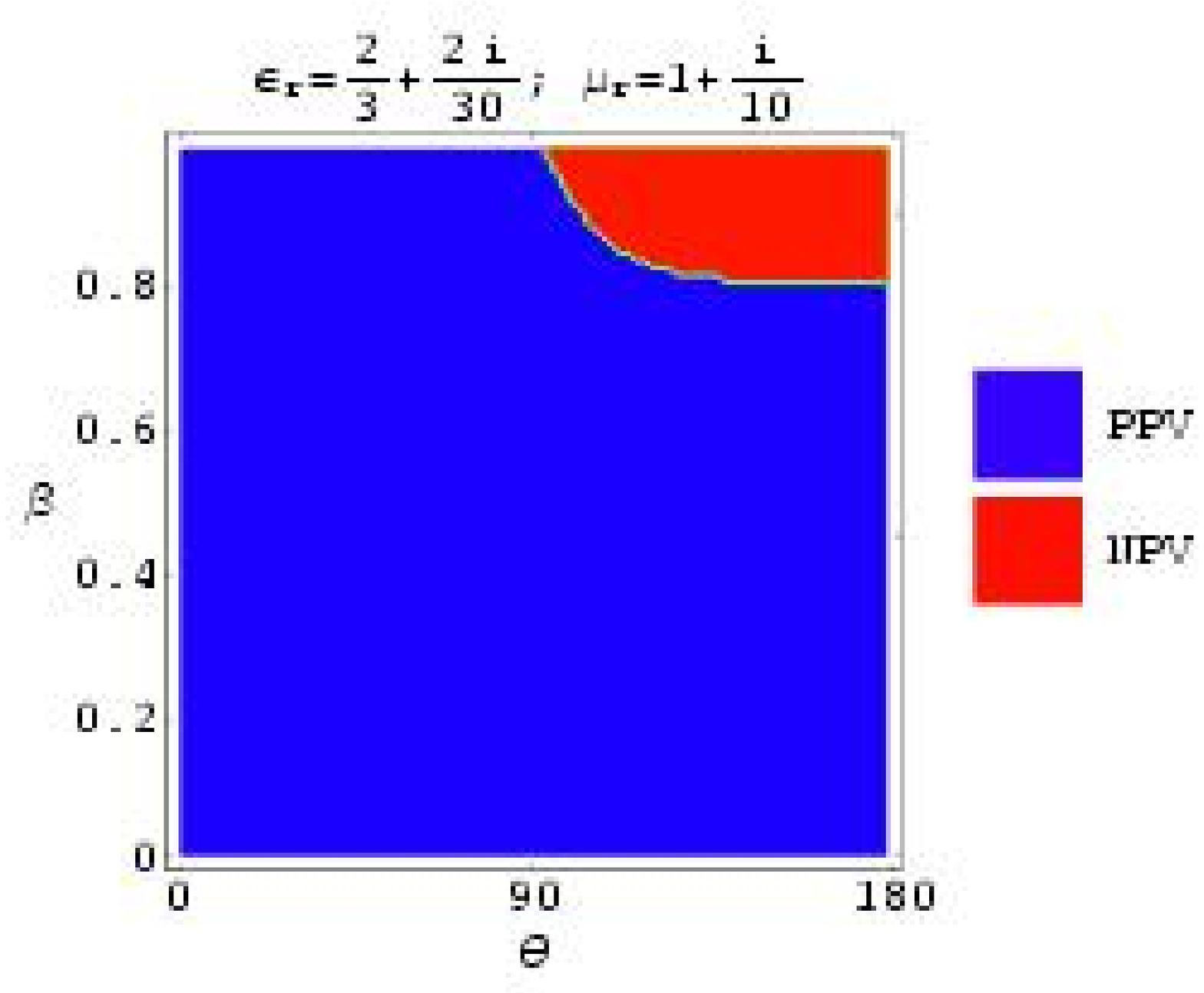,width=2.1in}
\epsfig{file=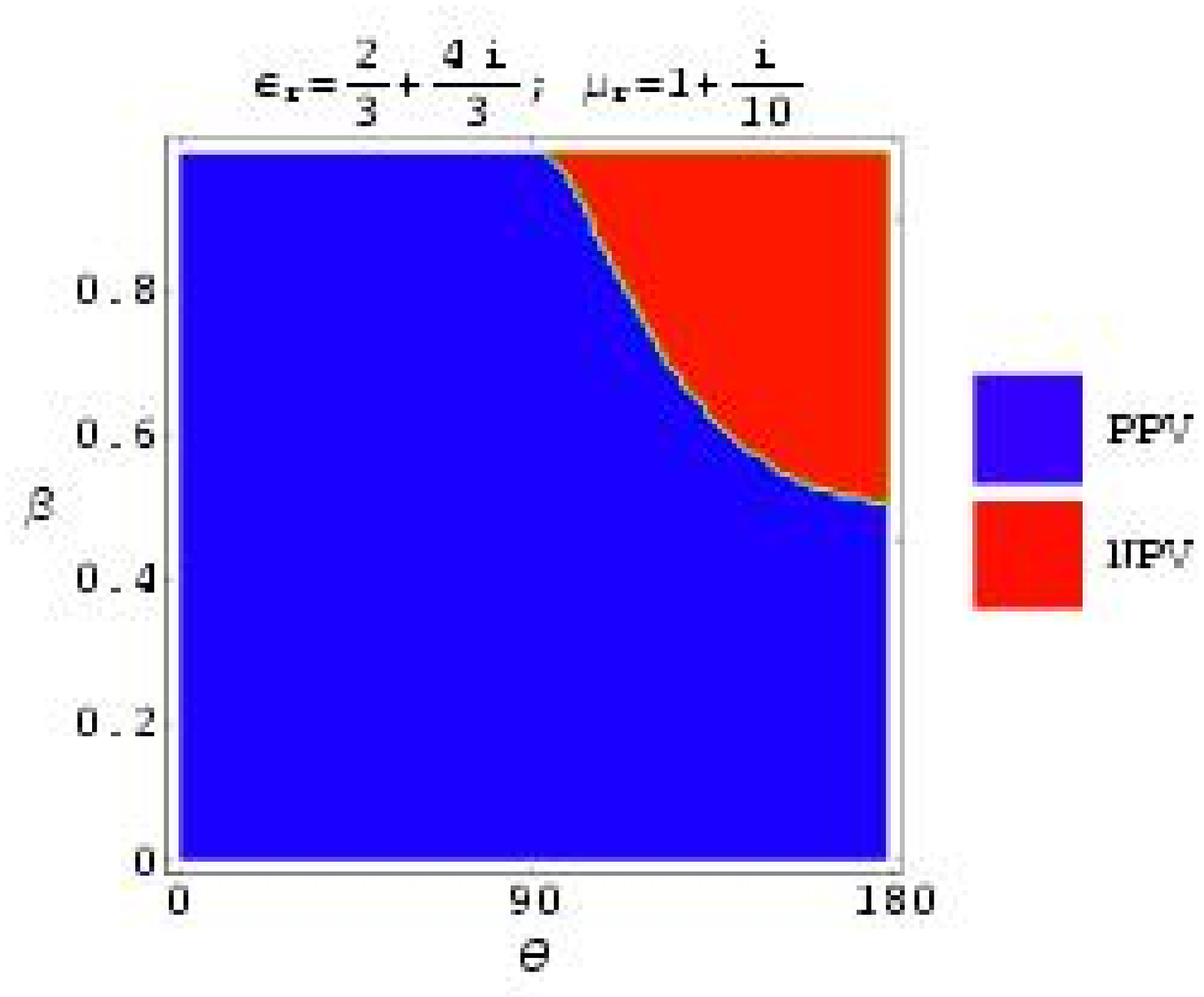,width=2.1in}
\epsfig{file=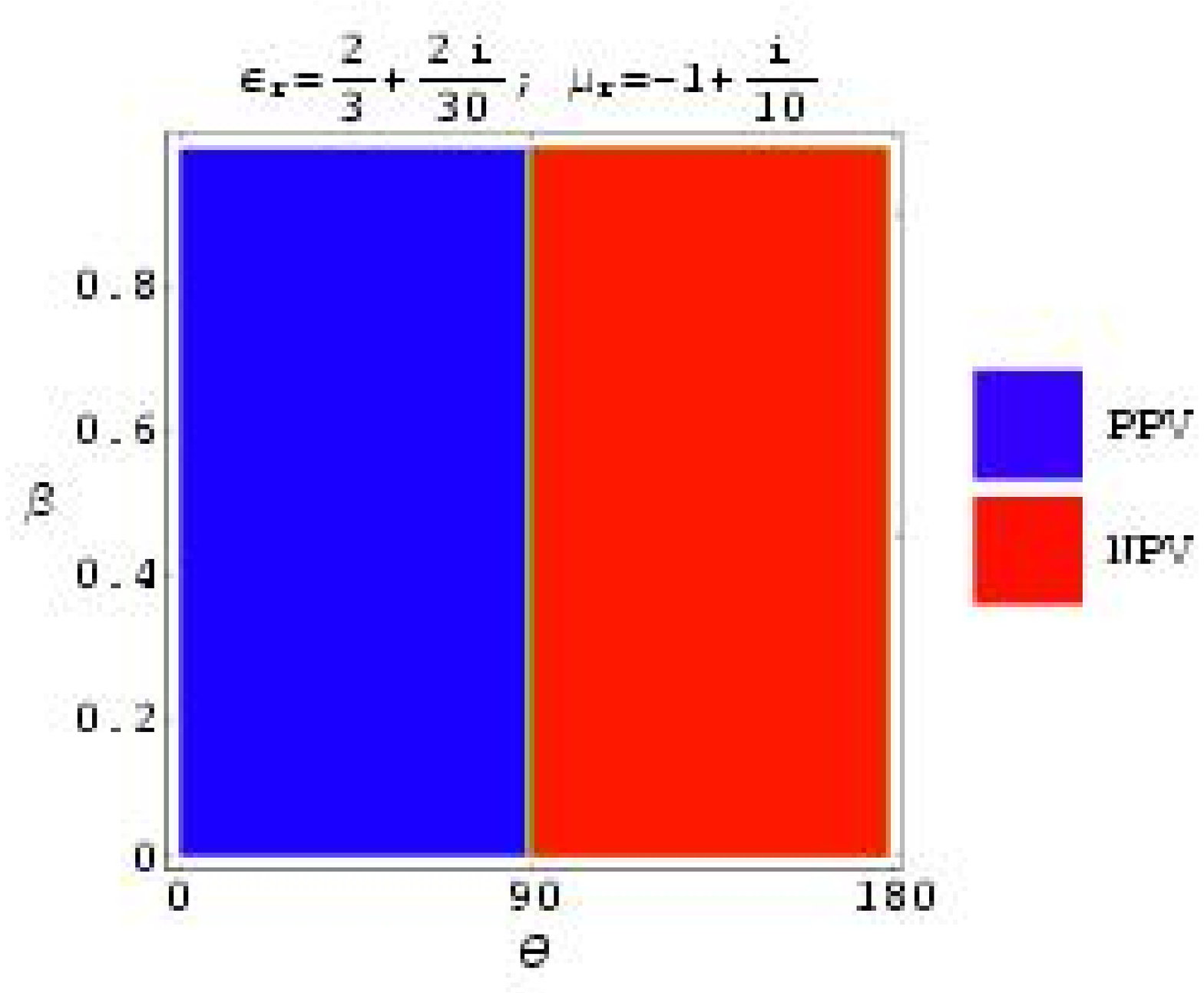,width=2.1in}
\\ \vspace{5mm}
\epsfig{file=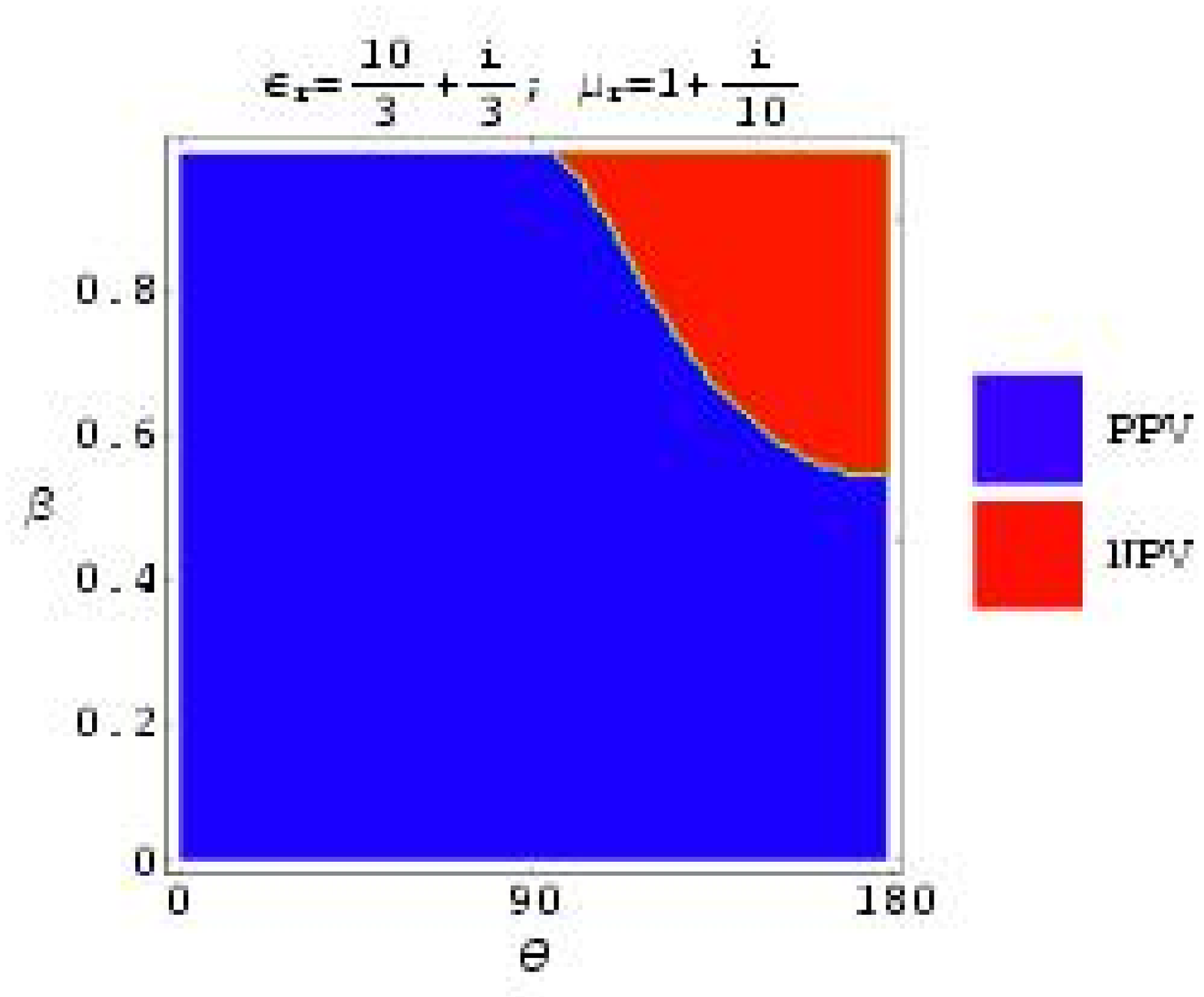,width=2.1in}
\epsfig{file=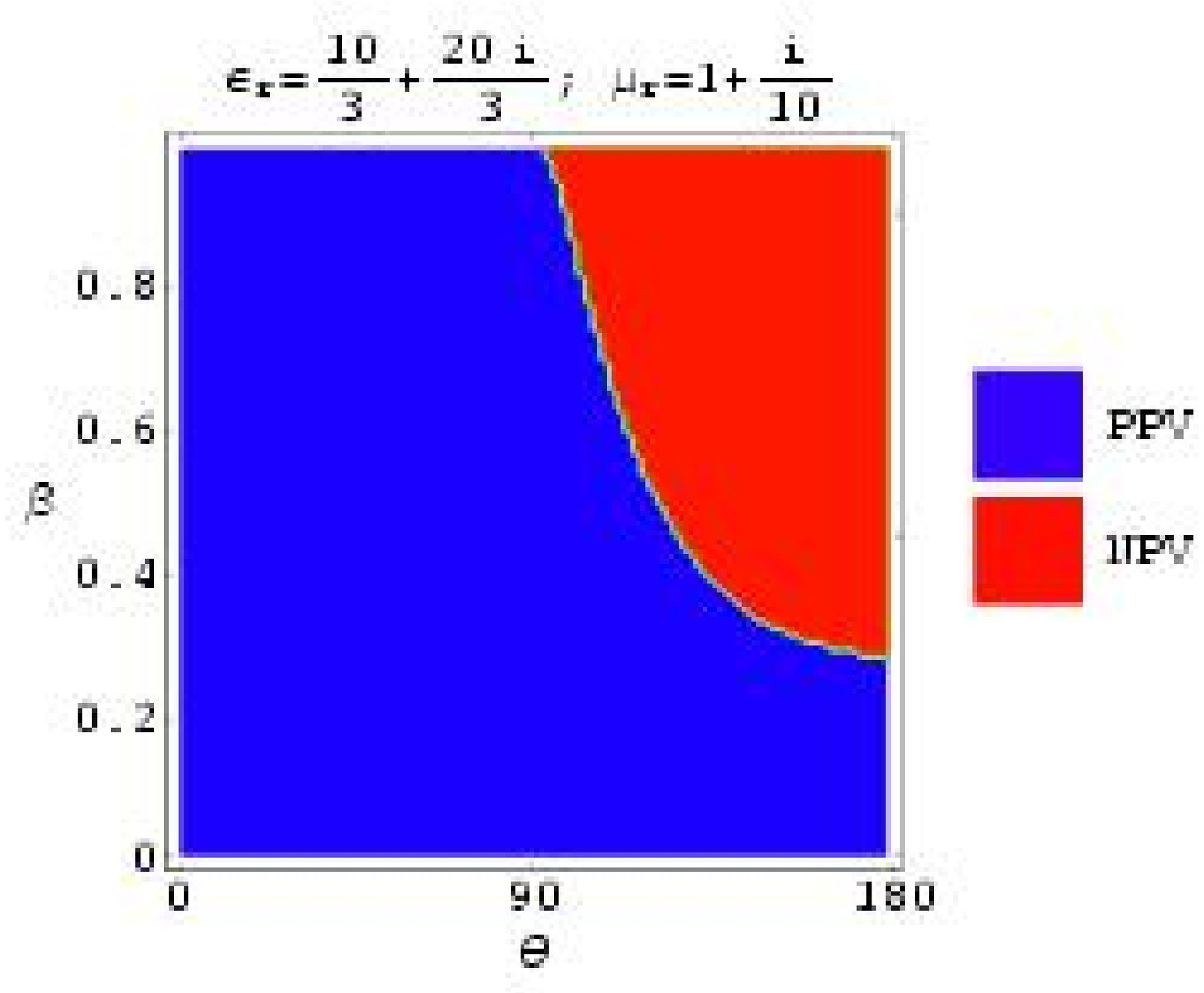,width=2.1in}
\epsfig{file=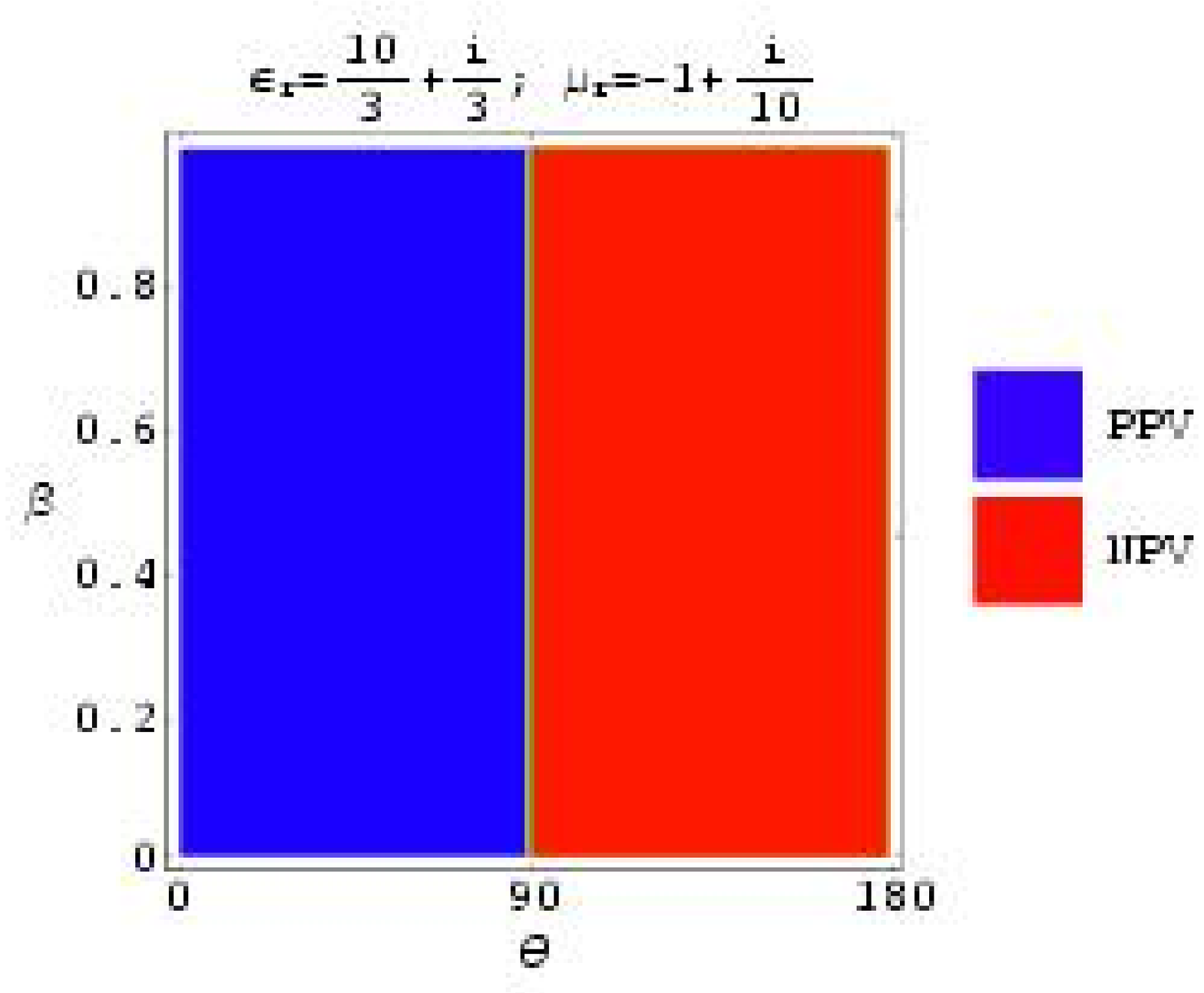,width=2.1in}
  \caption{\label{fig4a}
  The distribution of positive phase velocity (PPV) and negative
  phase velocity (NPV)
in relation to $\beta \in (0,1)$ and $\theta \in (0^\circ,
180^\circ)$,
for
 $\eps^R_r \in \lec \pm 1/3,
\pm 2/3, \pm 10/3 \ric$, $\eps^I_r \in \lec | 0.1 \eps^R_r |, 2 |
\eps^R_r | \ric$, and $\mu_r = \pm 1+ 0.1 i$.
 }
\end{figure}

\newpage

\setcounter{figure}{3}

\begin{figure}[!ht]
\centering \psfull \epsfig{file=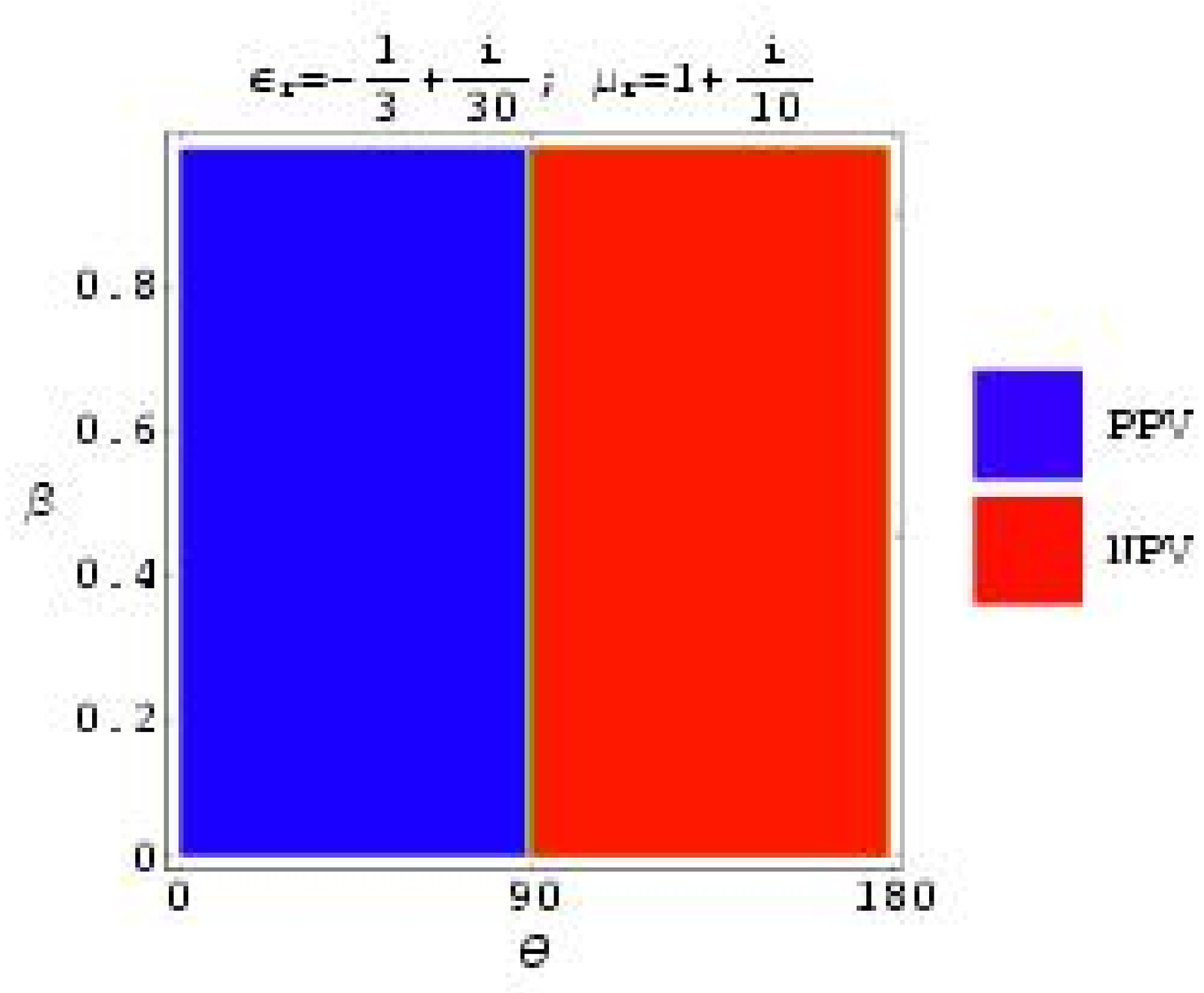,width=2.1in}
\epsfig{file=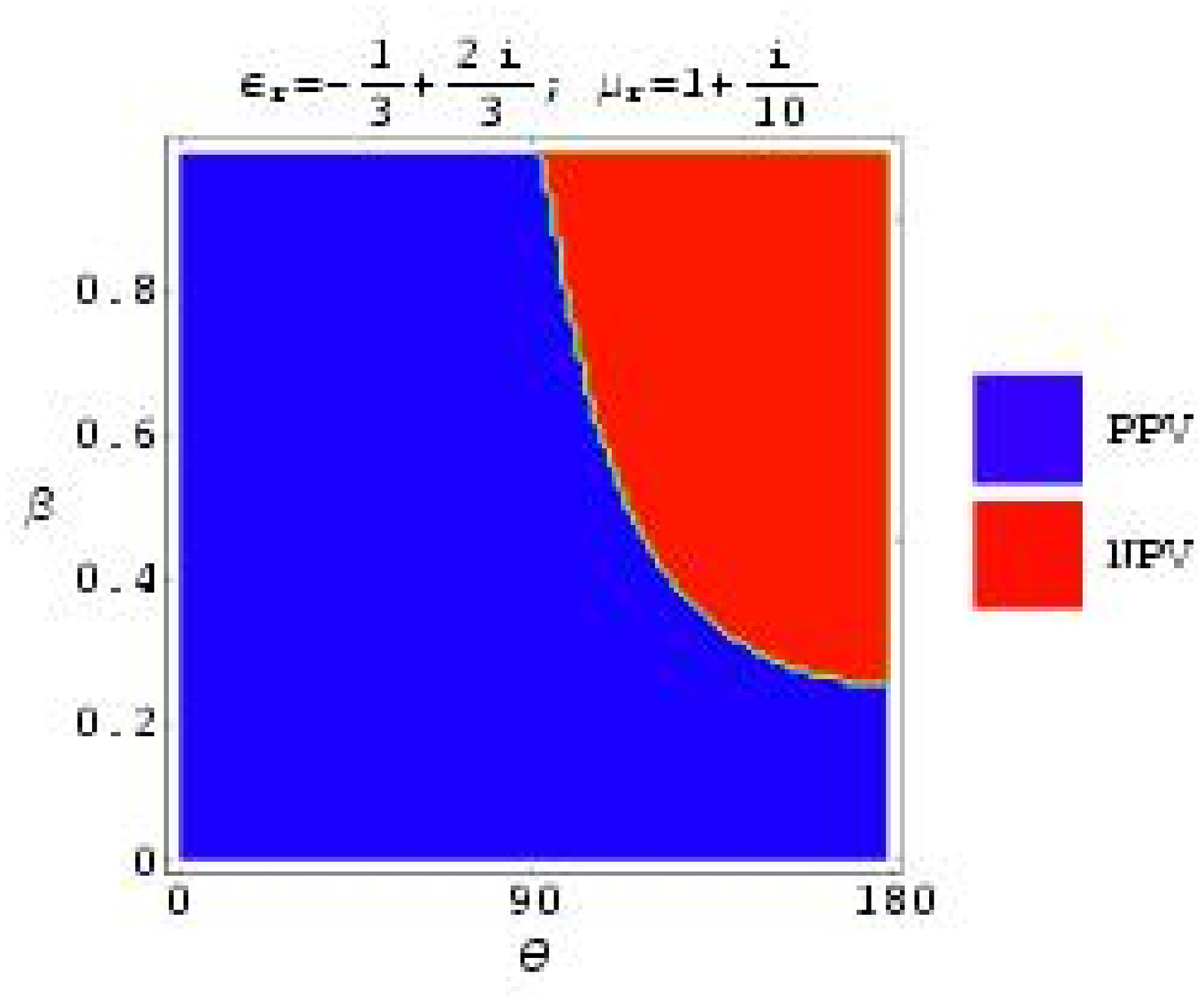,width=2.1in}
\epsfig{file=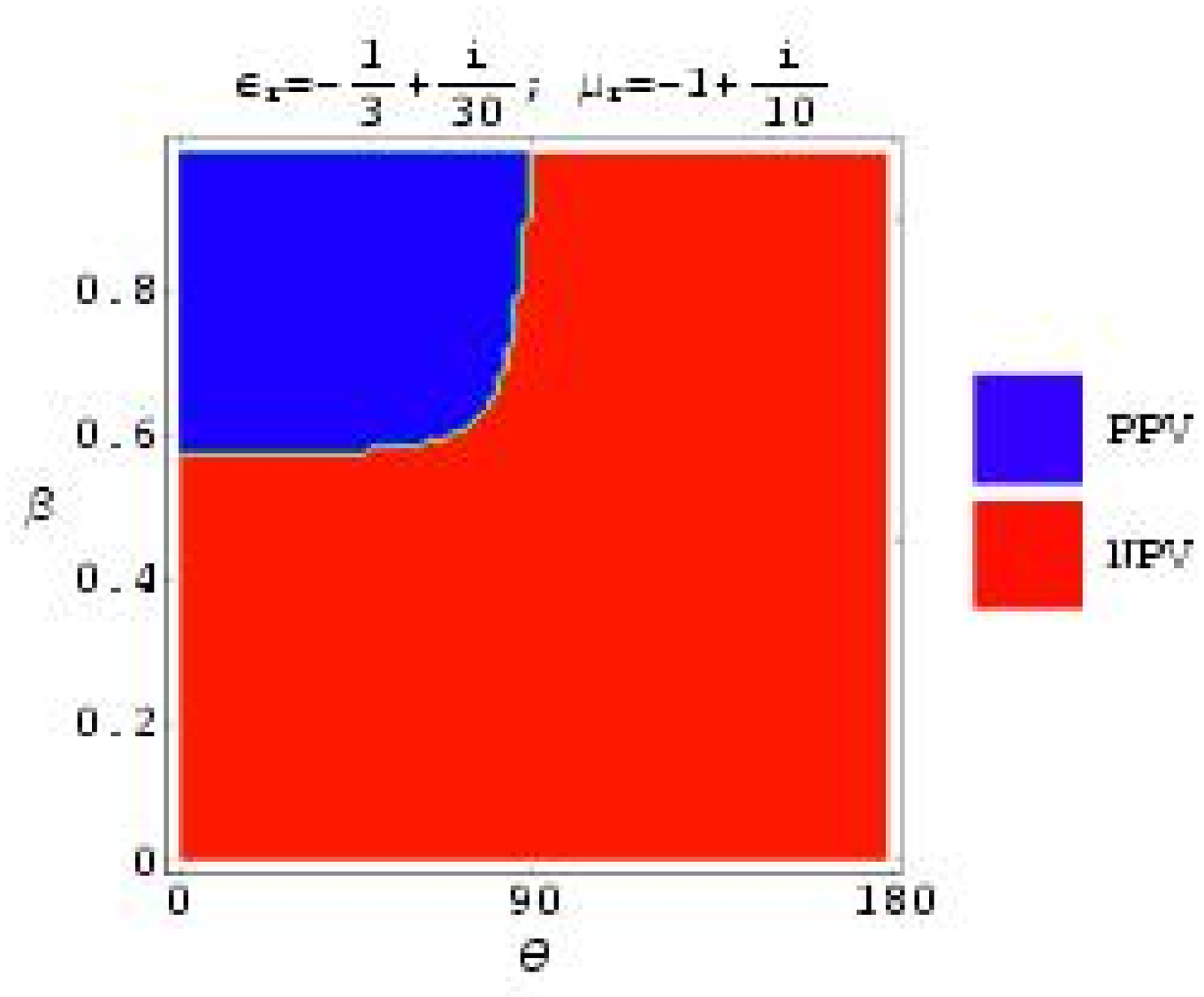,width=2.1in}
\\ \vspace{5mm}
\epsfig{file=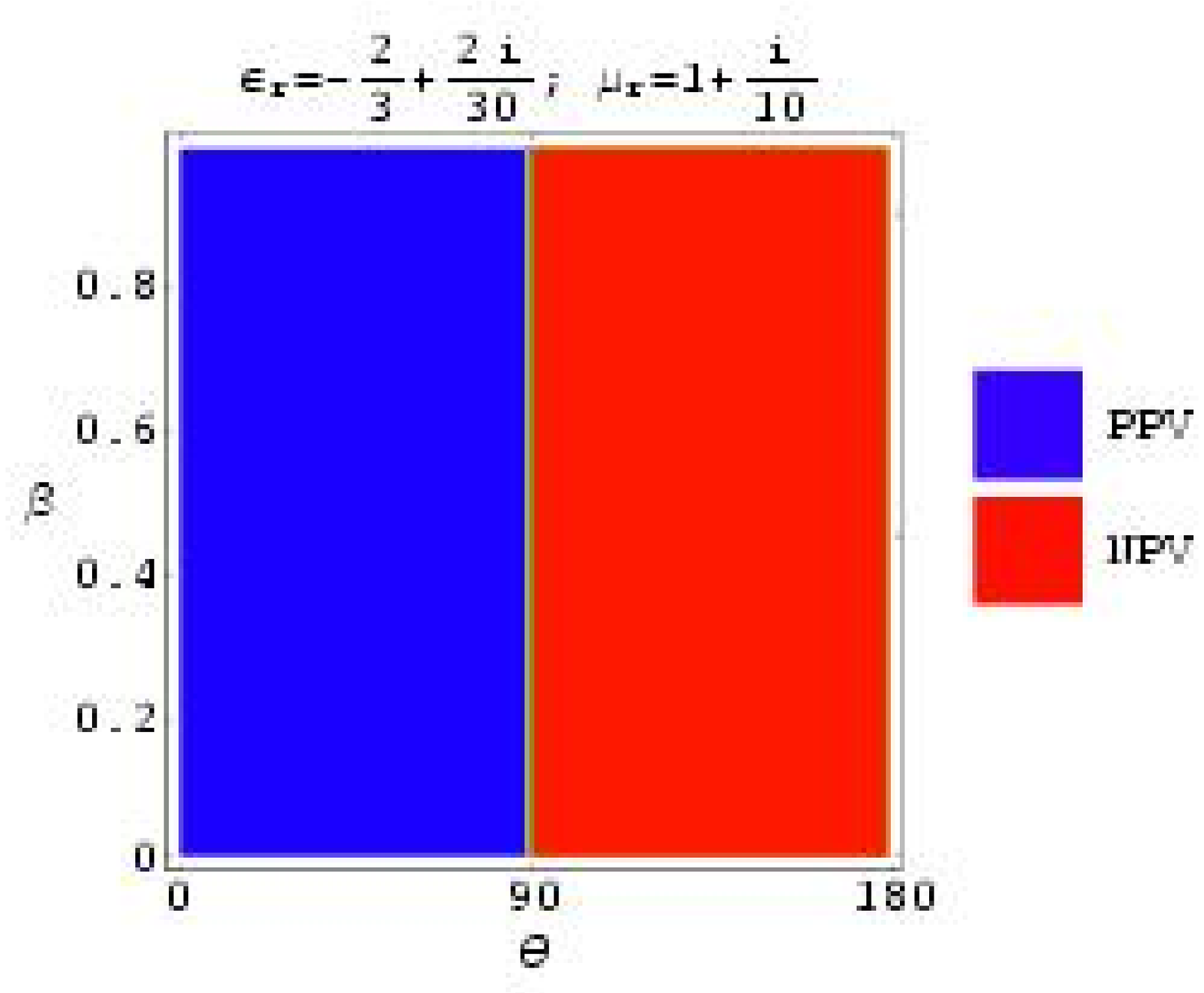,width=2.1in}
\epsfig{file=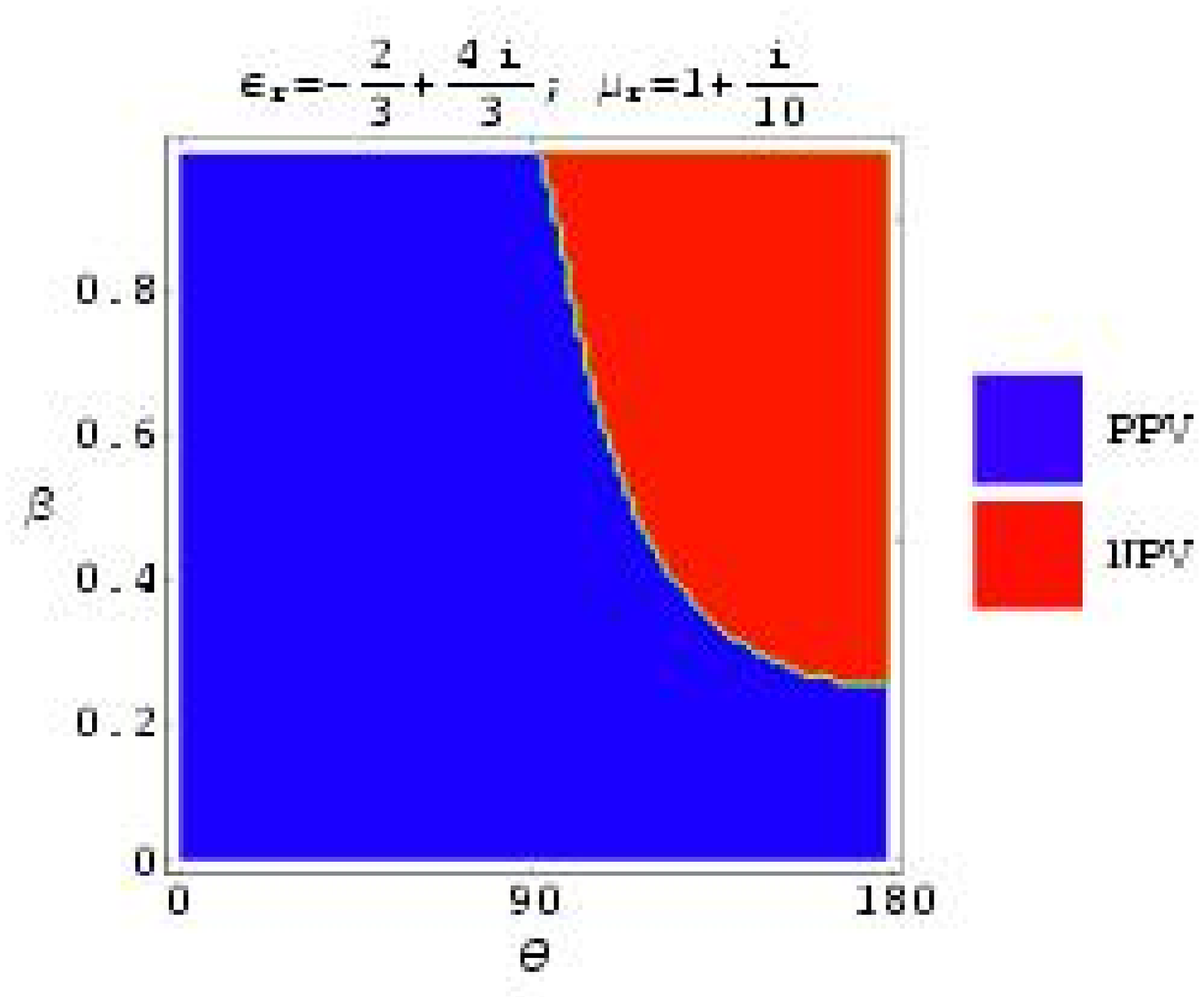,width=2.1in}
\epsfig{file=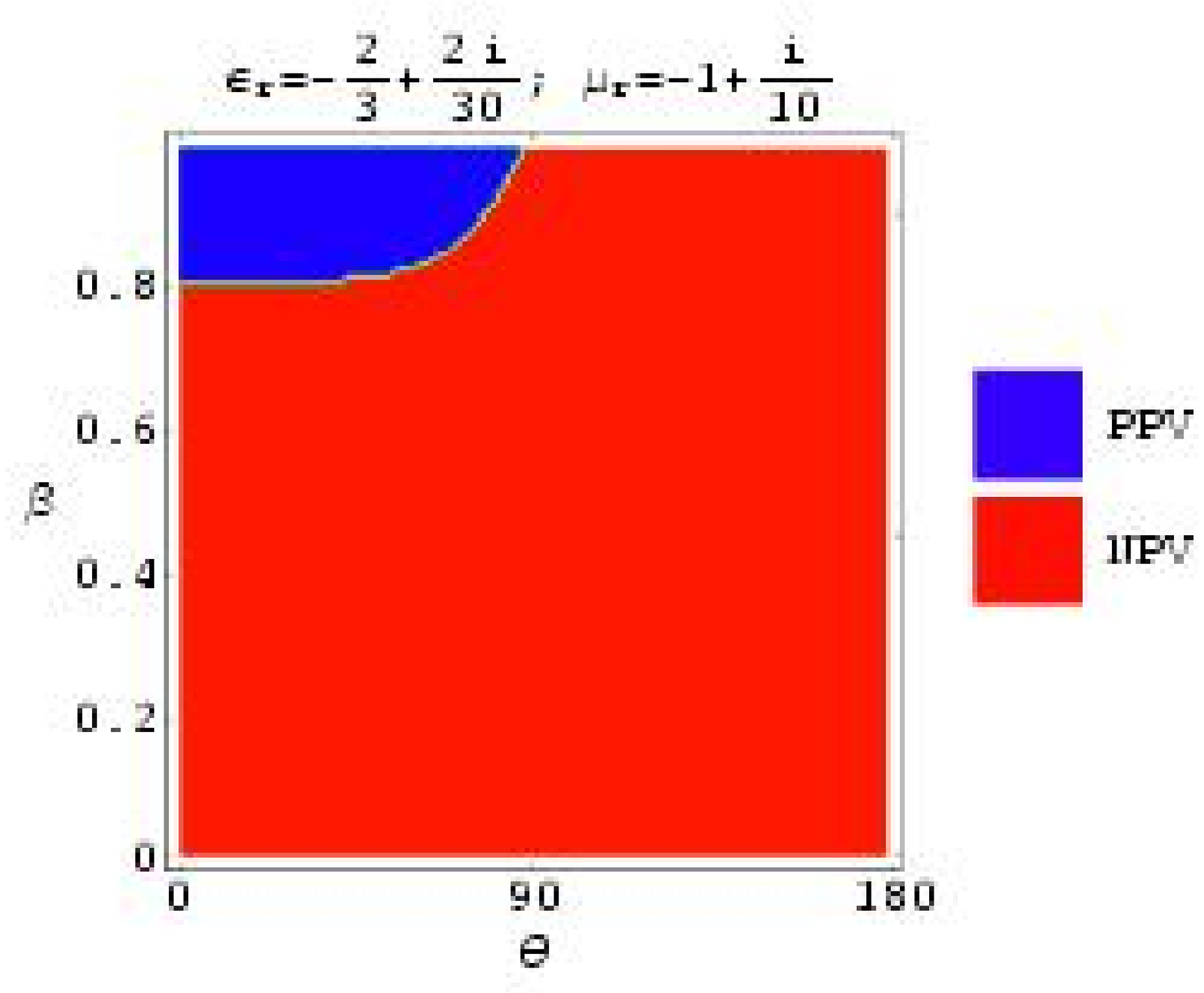,width=2.1in}
\\ \vspace{5mm}
\epsfig{file=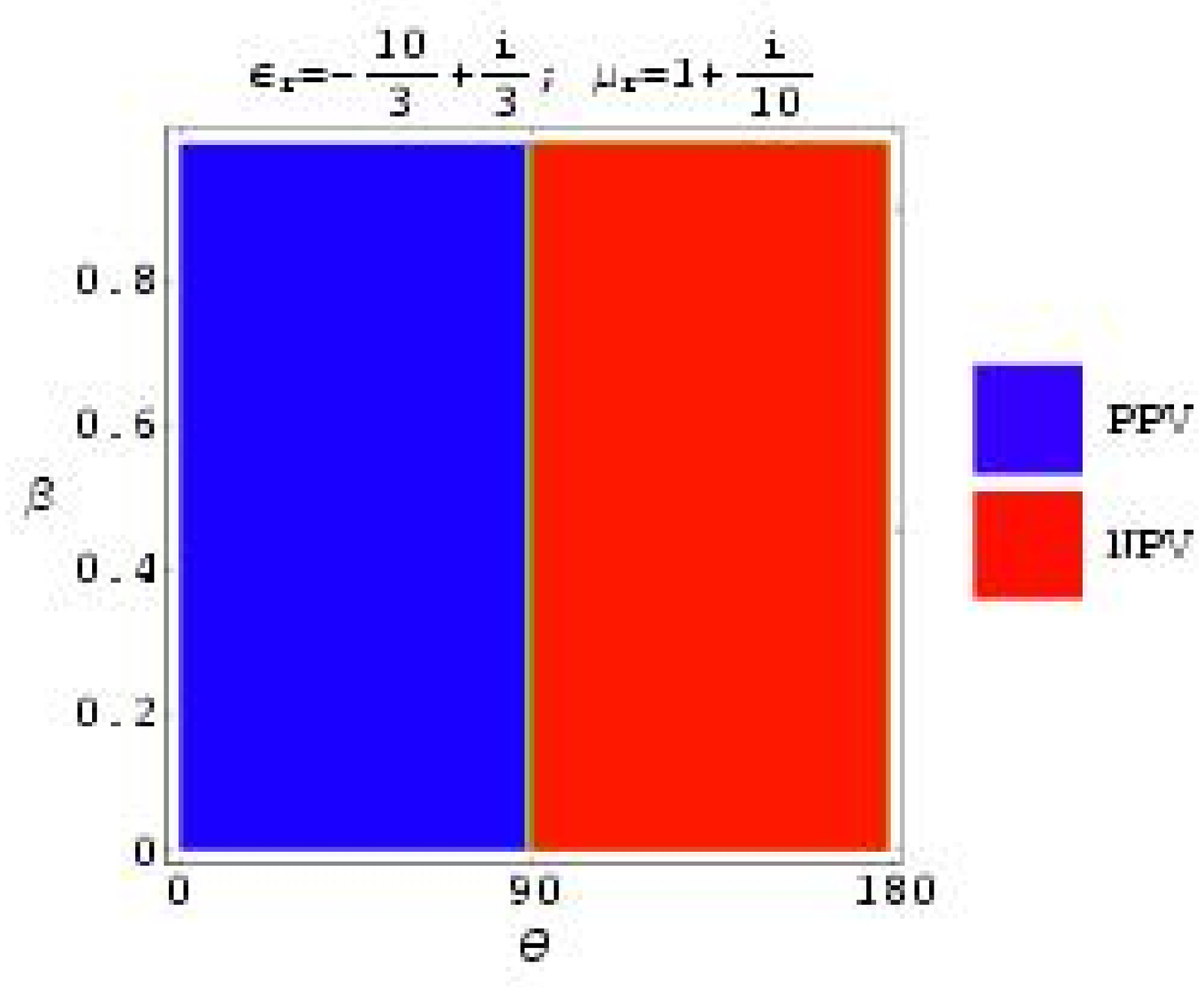,width=2.1in}
\epsfig{file=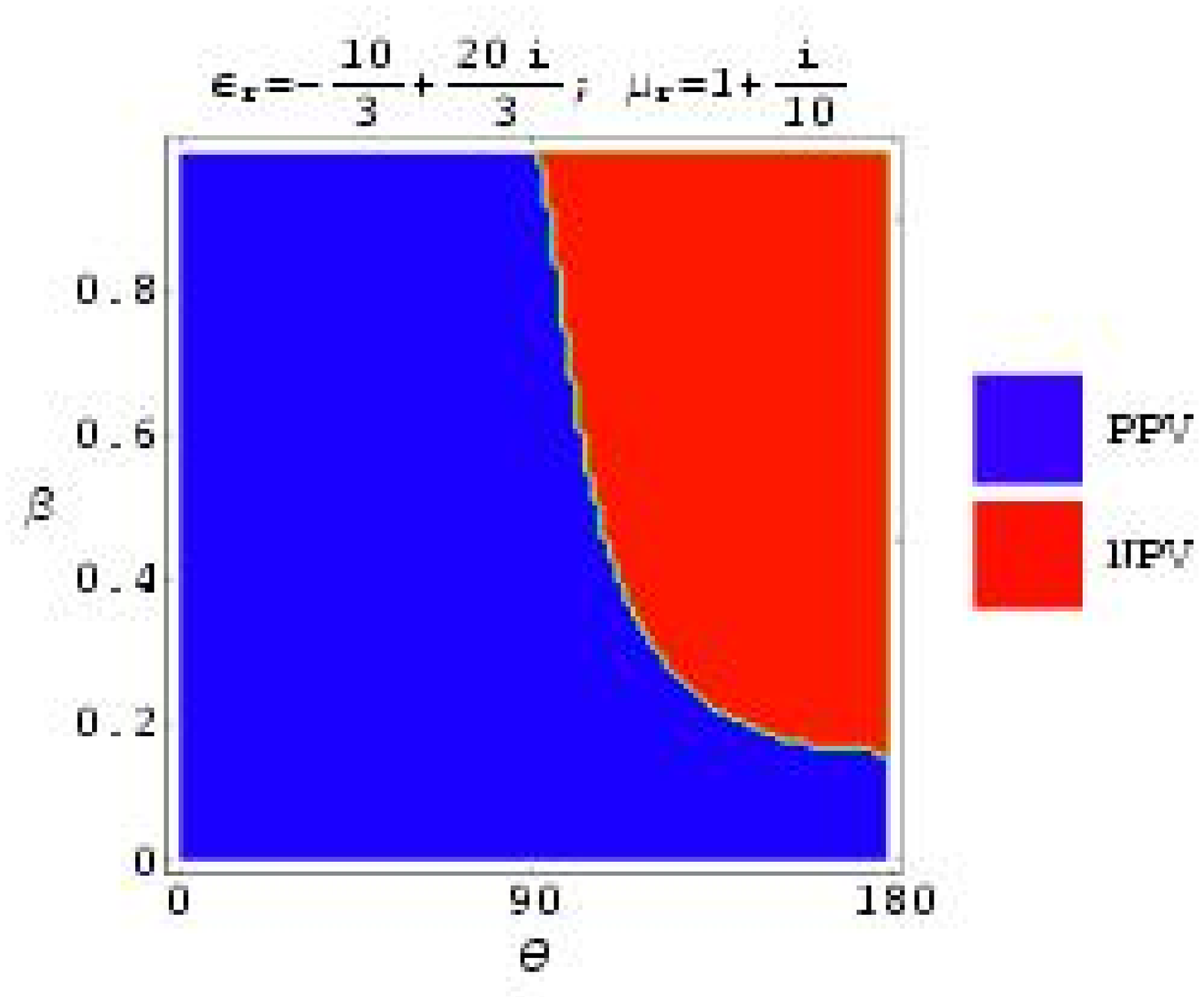,width=2.1in}
\epsfig{file=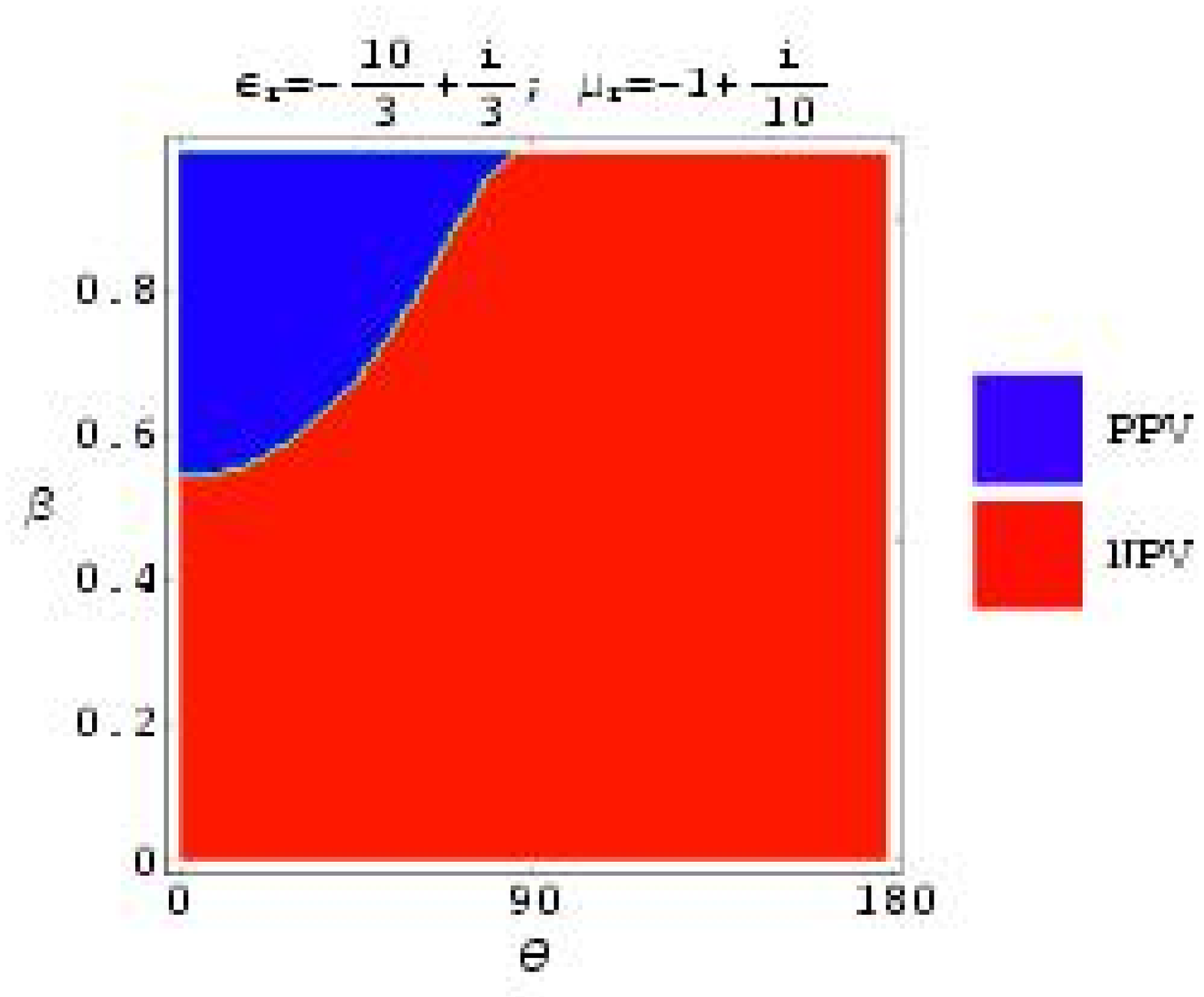,width=2.1in}
  \caption{\label{fig4b} Continued.
 }
\end{figure}

\newpage

\setcounter{figure}{4}

\begin{figure}[!ht]
\centering \psfull \epsfig{file=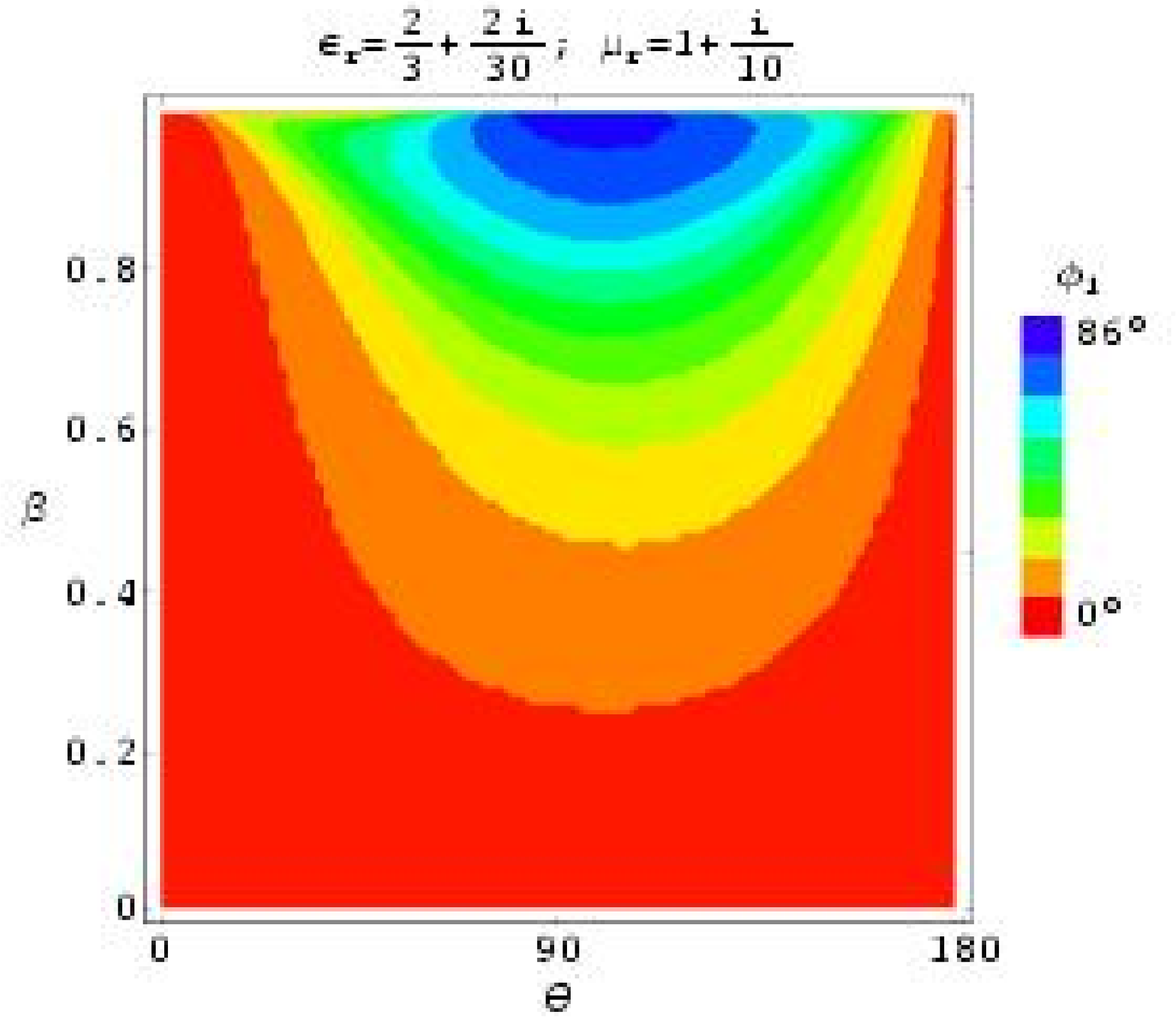,width=2.1in}
\epsfig{file=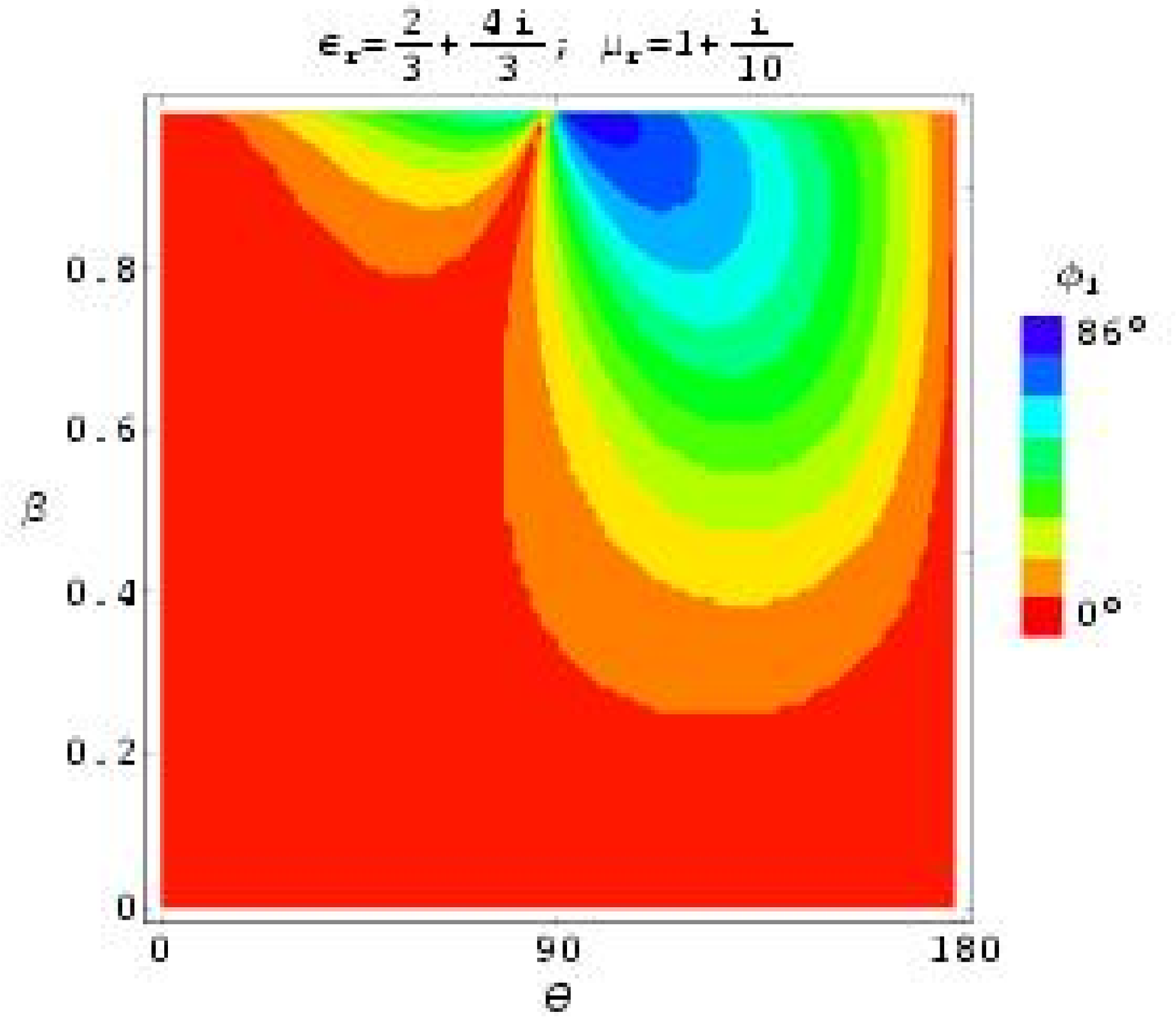,width=2.1in}
\epsfig{file=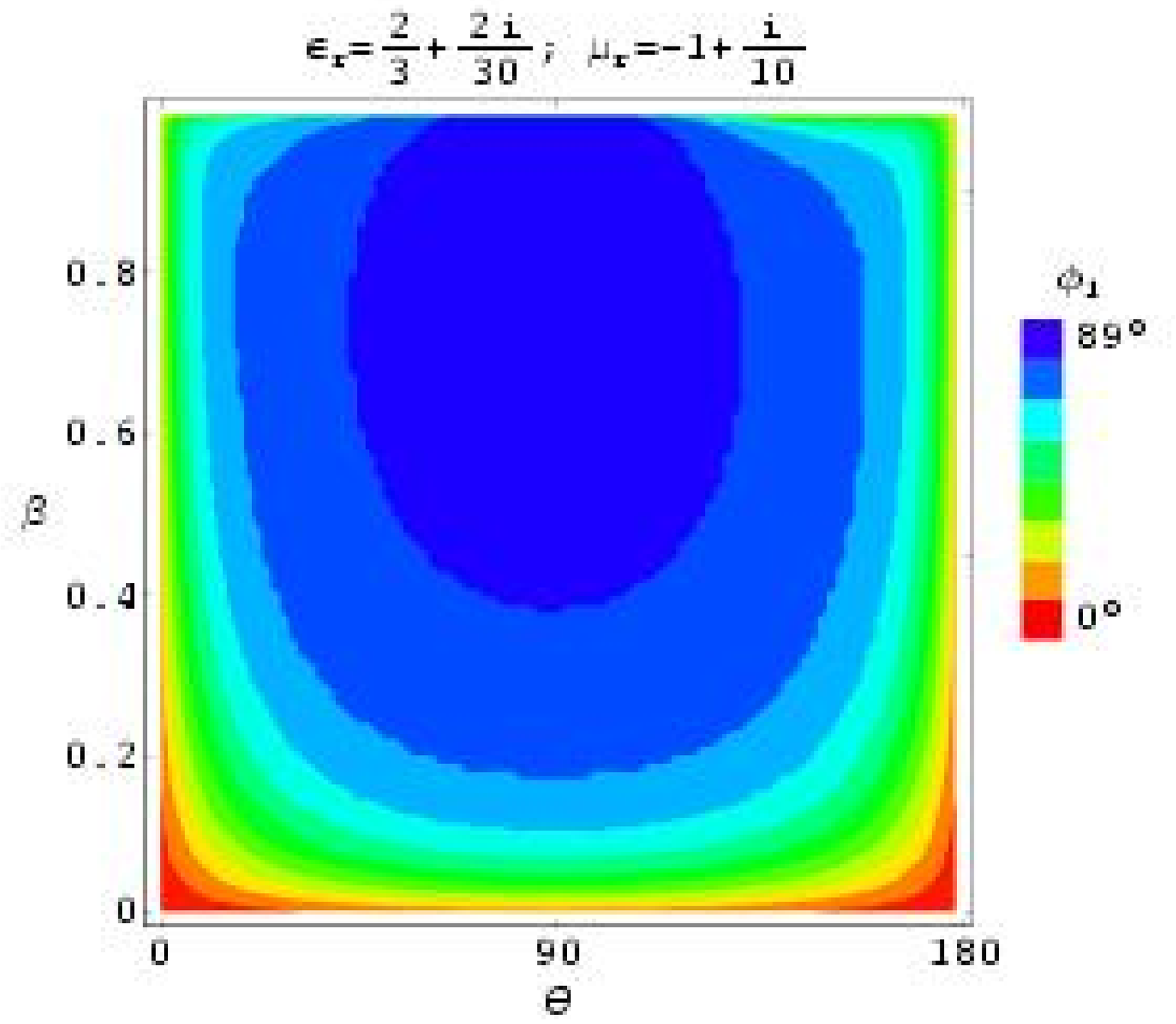,width=2.1in}
\\ \vspace{5mm}
\epsfig{file=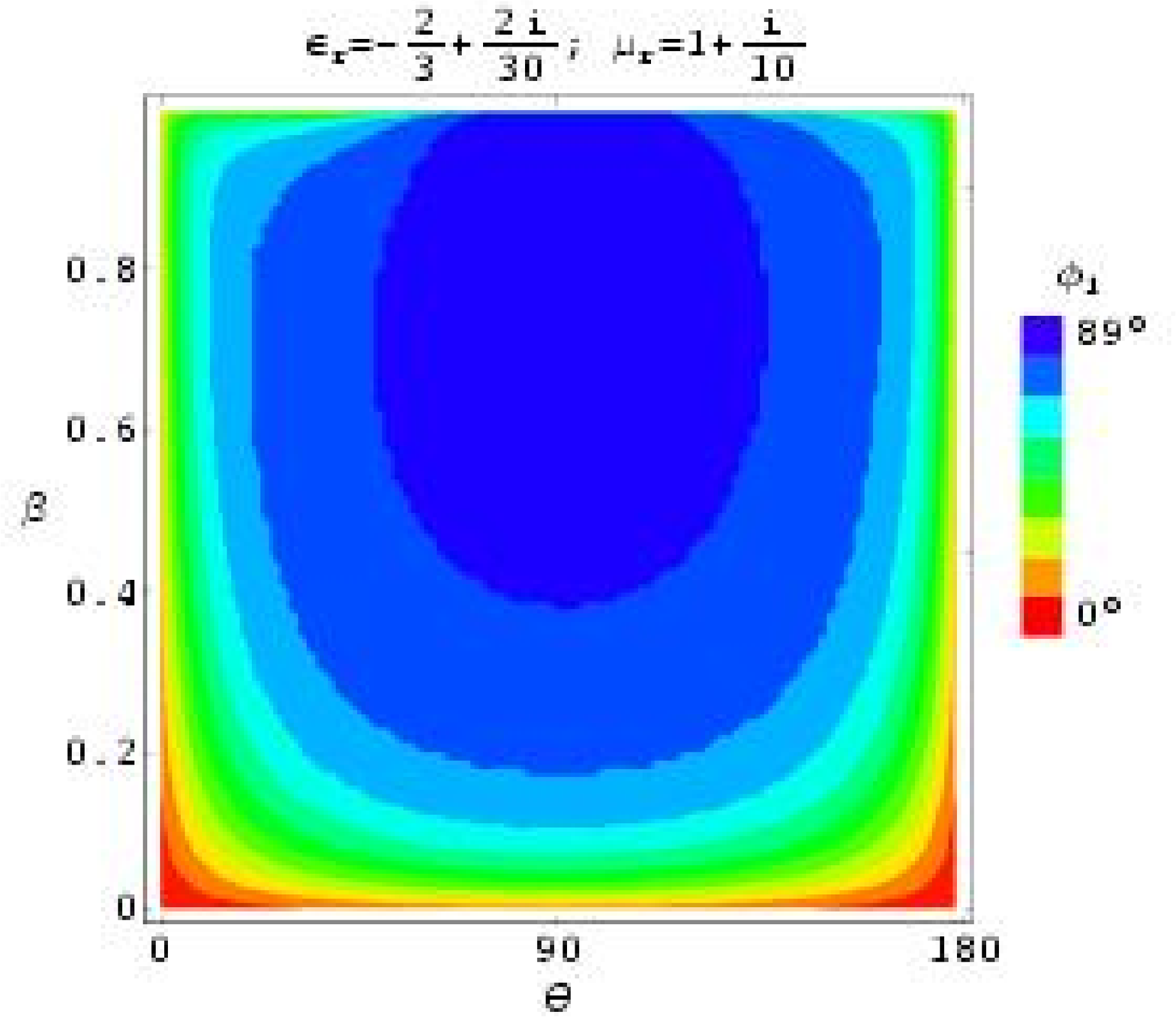,width=2.1in}
\epsfig{file=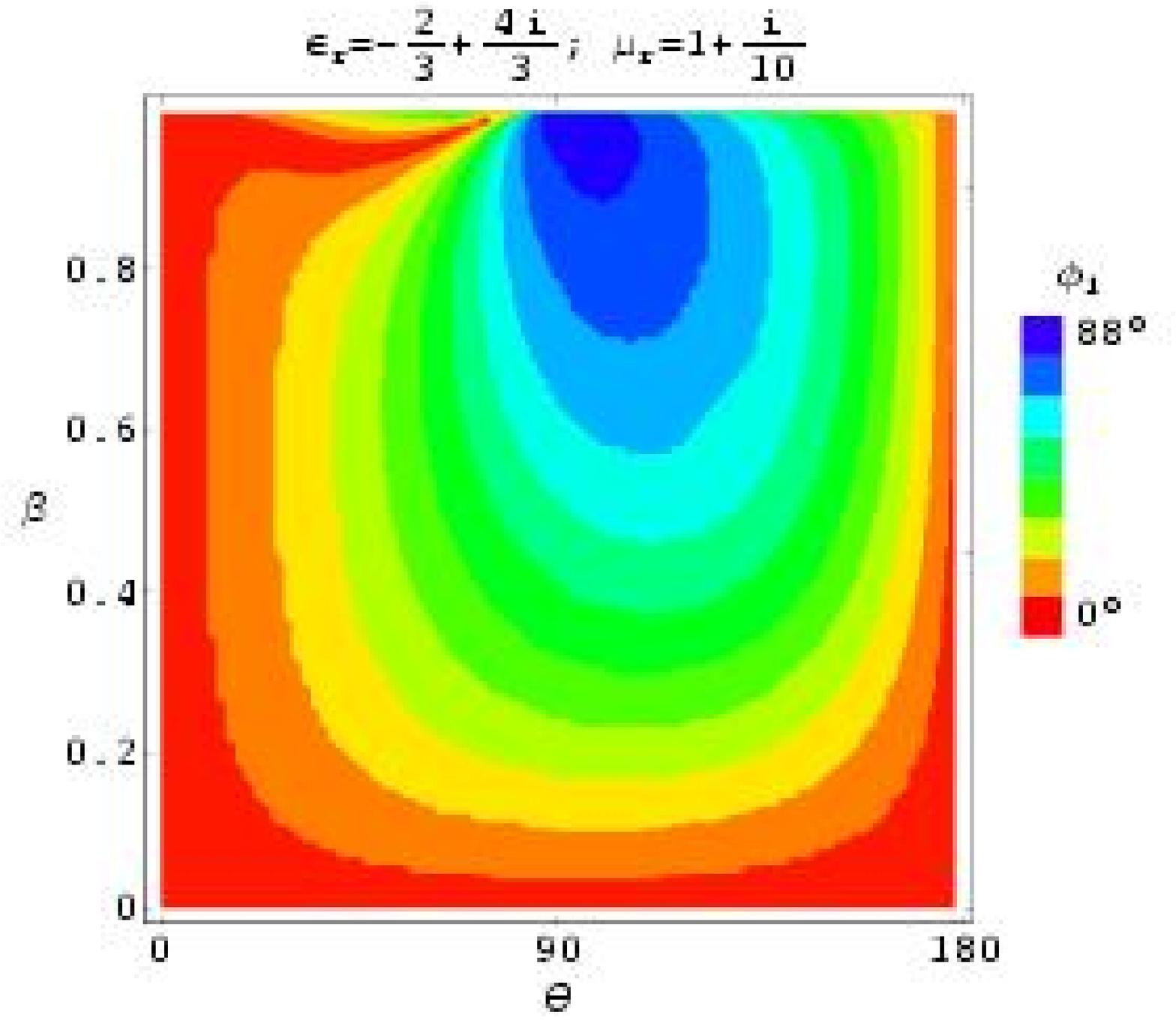,width=2.1in}
\epsfig{file=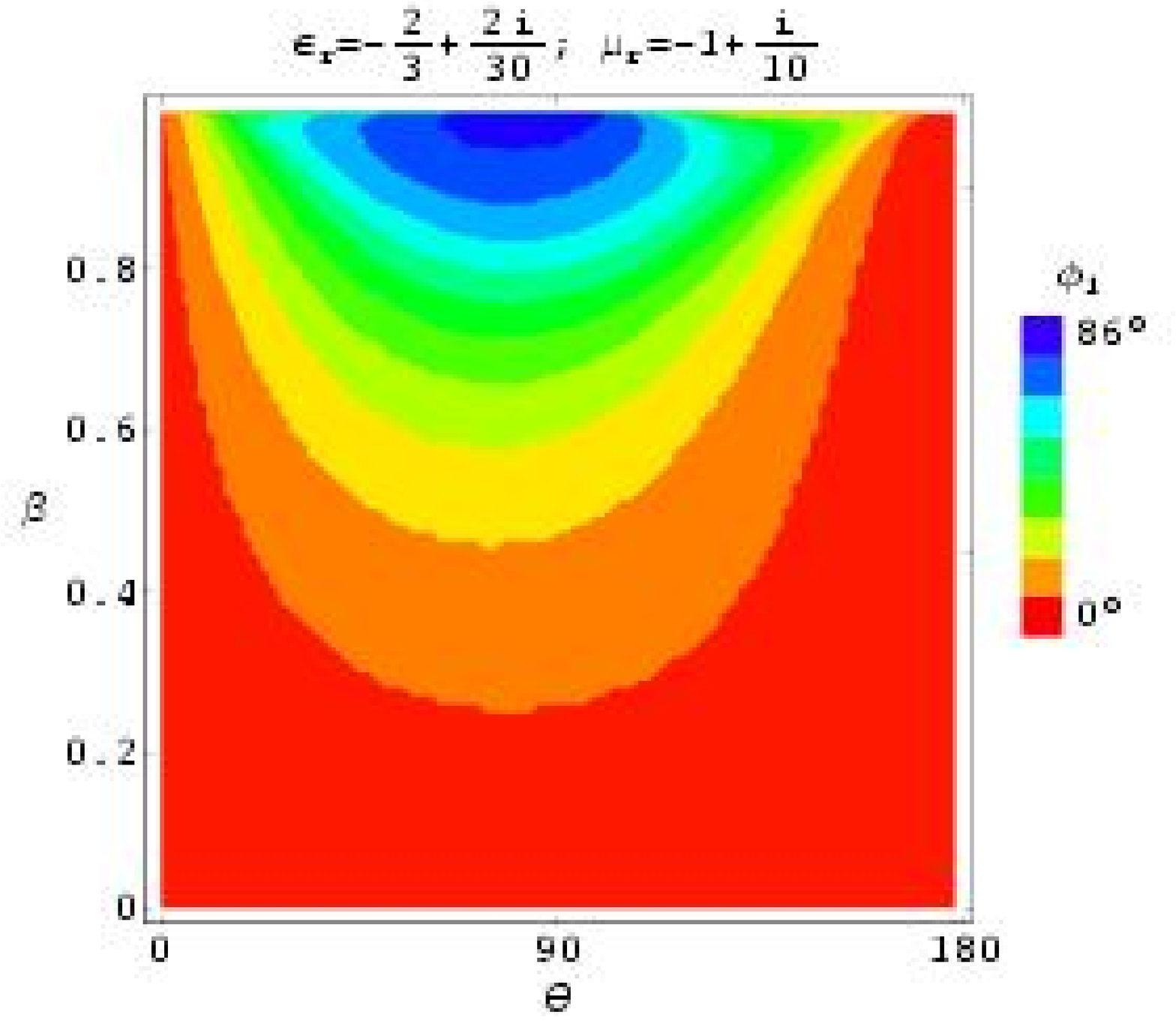,width=2.1in}
  \caption{\label{fig5a}
  The angles $\phi_1$ and $\phi_2$
    as functions of $\beta \in (0,1)$ and $\theta \in (0^\circ,
180^\circ)$, for $\eps^R_r =  \pm 2/3$,  $\eps^I_r \in \lec 0.1 |
\eps^R_r |, 2 | \eps^R_r | \ric$, and $\mu_r = \pm 1+ 0.1 i$.
 }
\end{figure}

\newpage

\setcounter{figure}{4}

\begin{figure}[!ht]
\centering \psfull \epsfig{file=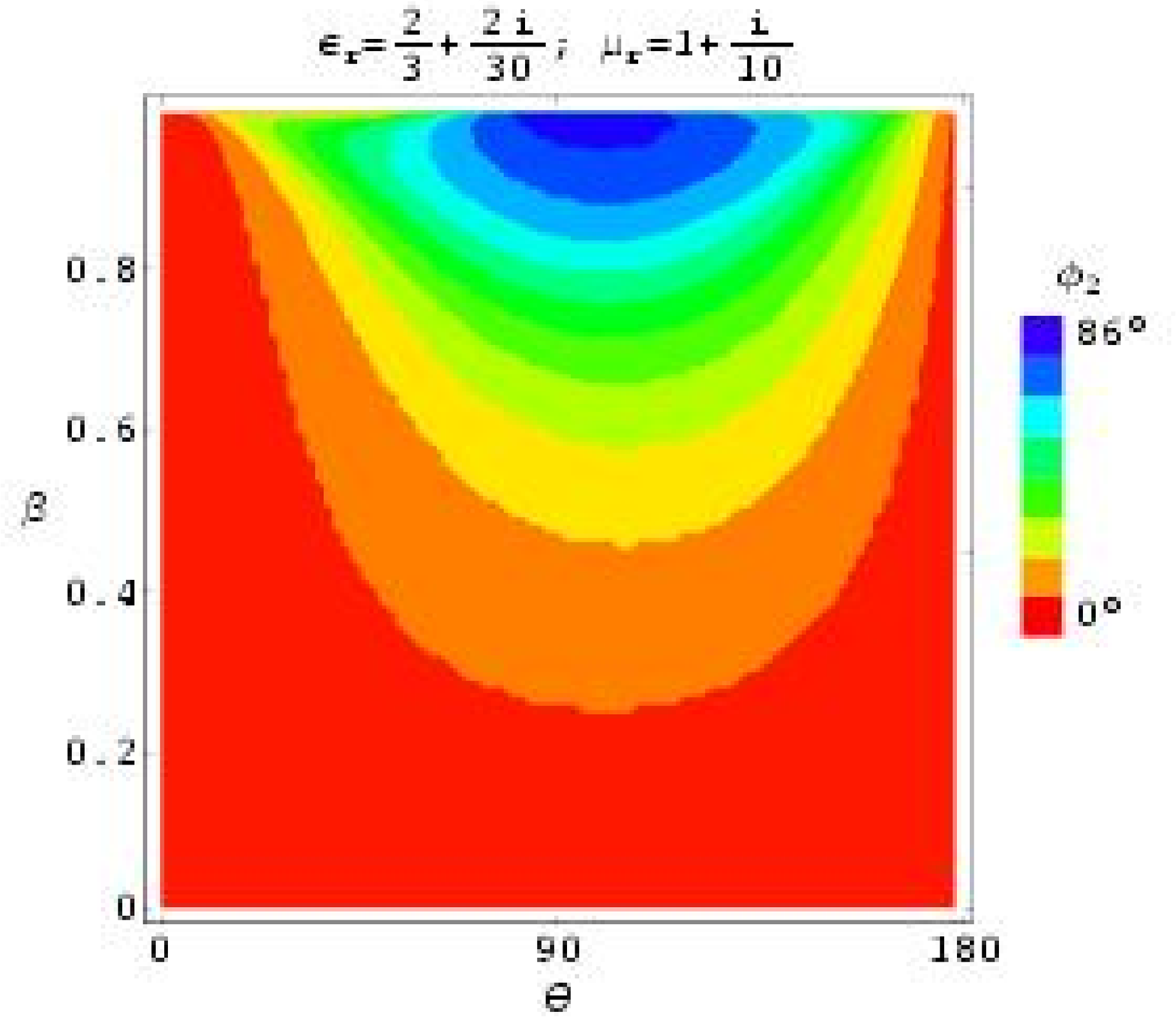,width=2.1in}
\epsfig{file=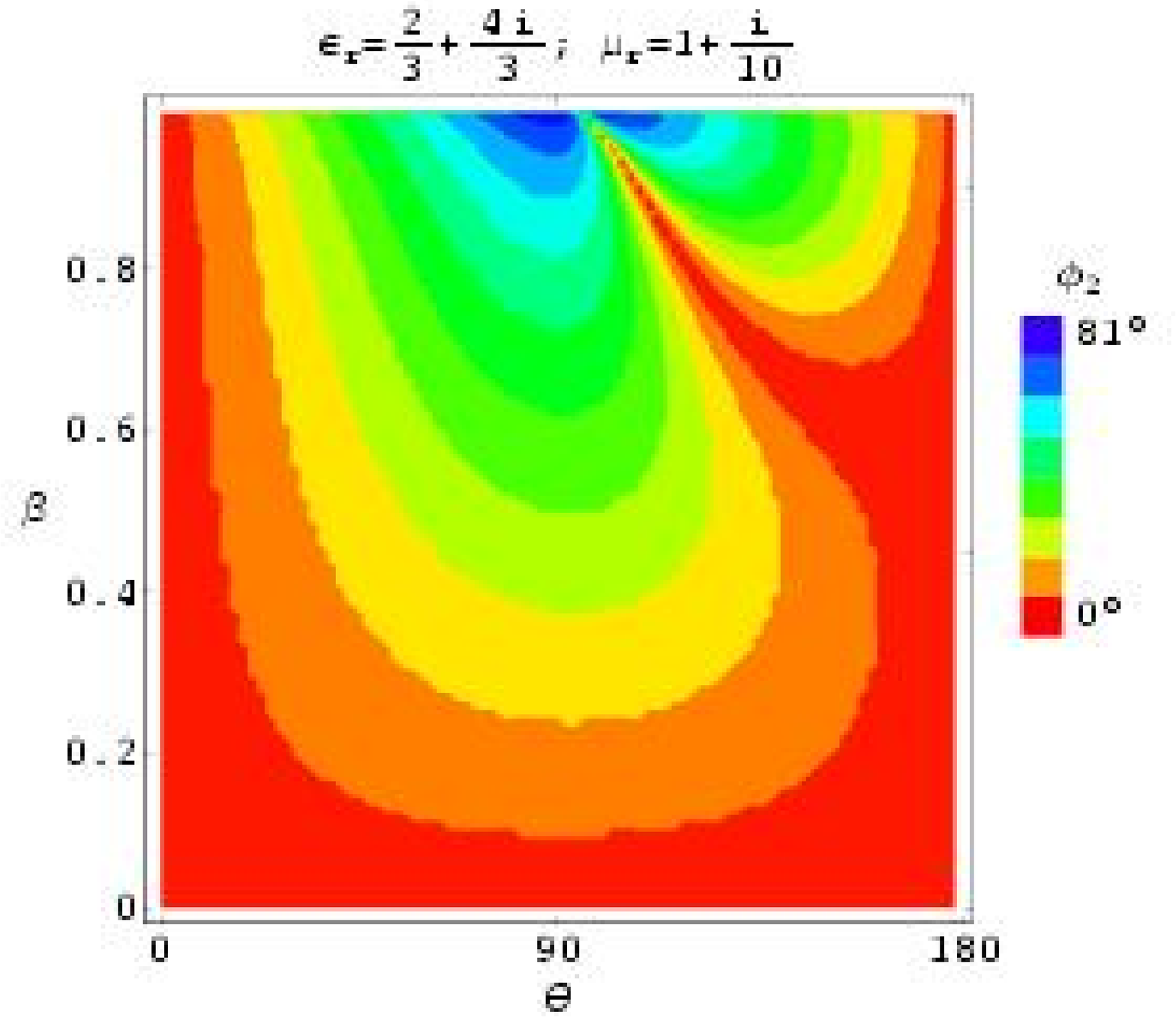,width=2.1in}
\epsfig{file=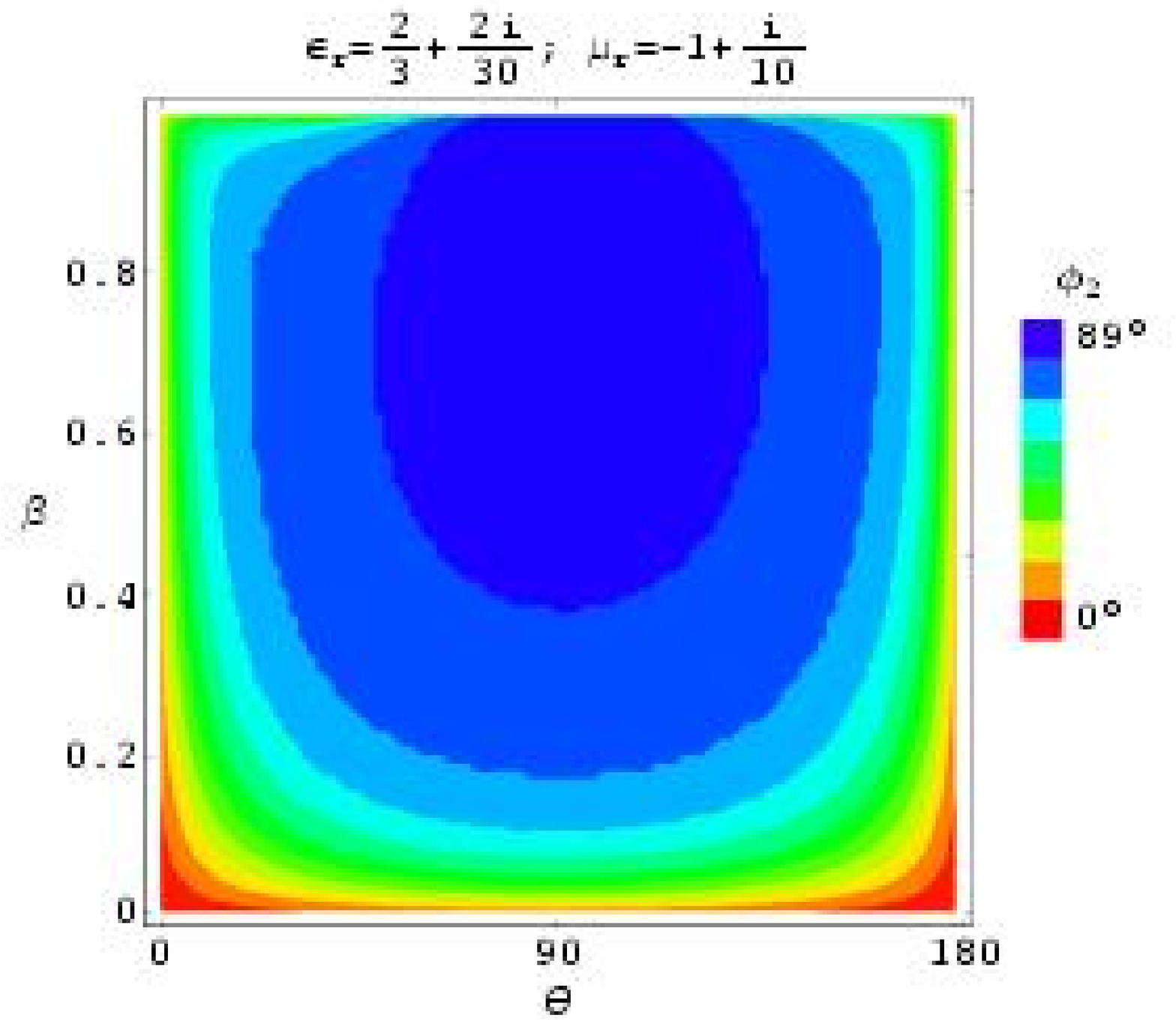,width=2.1in}
\\ \vspace{5mm}
\epsfig{file=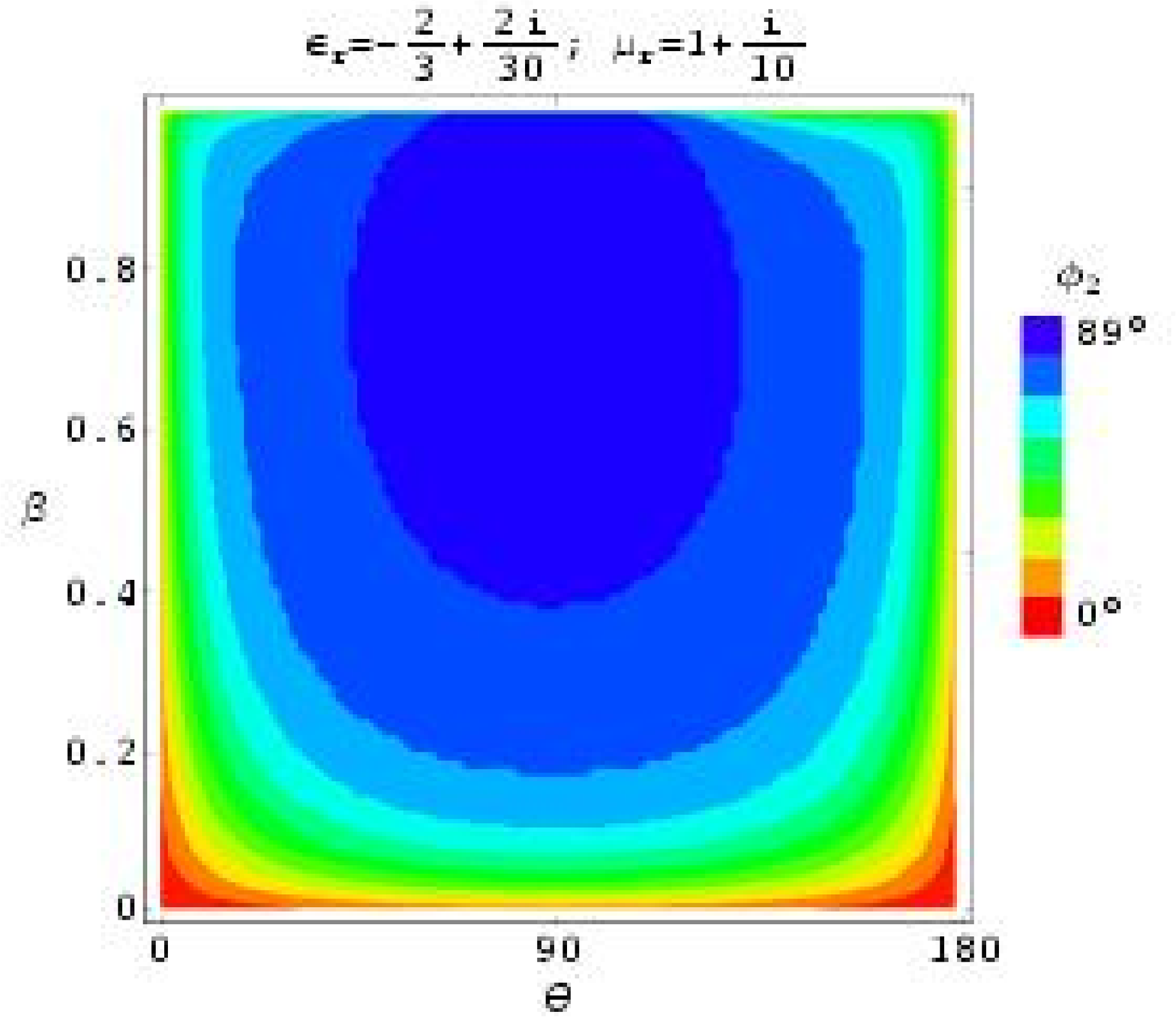,width=2.1in}
\epsfig{file=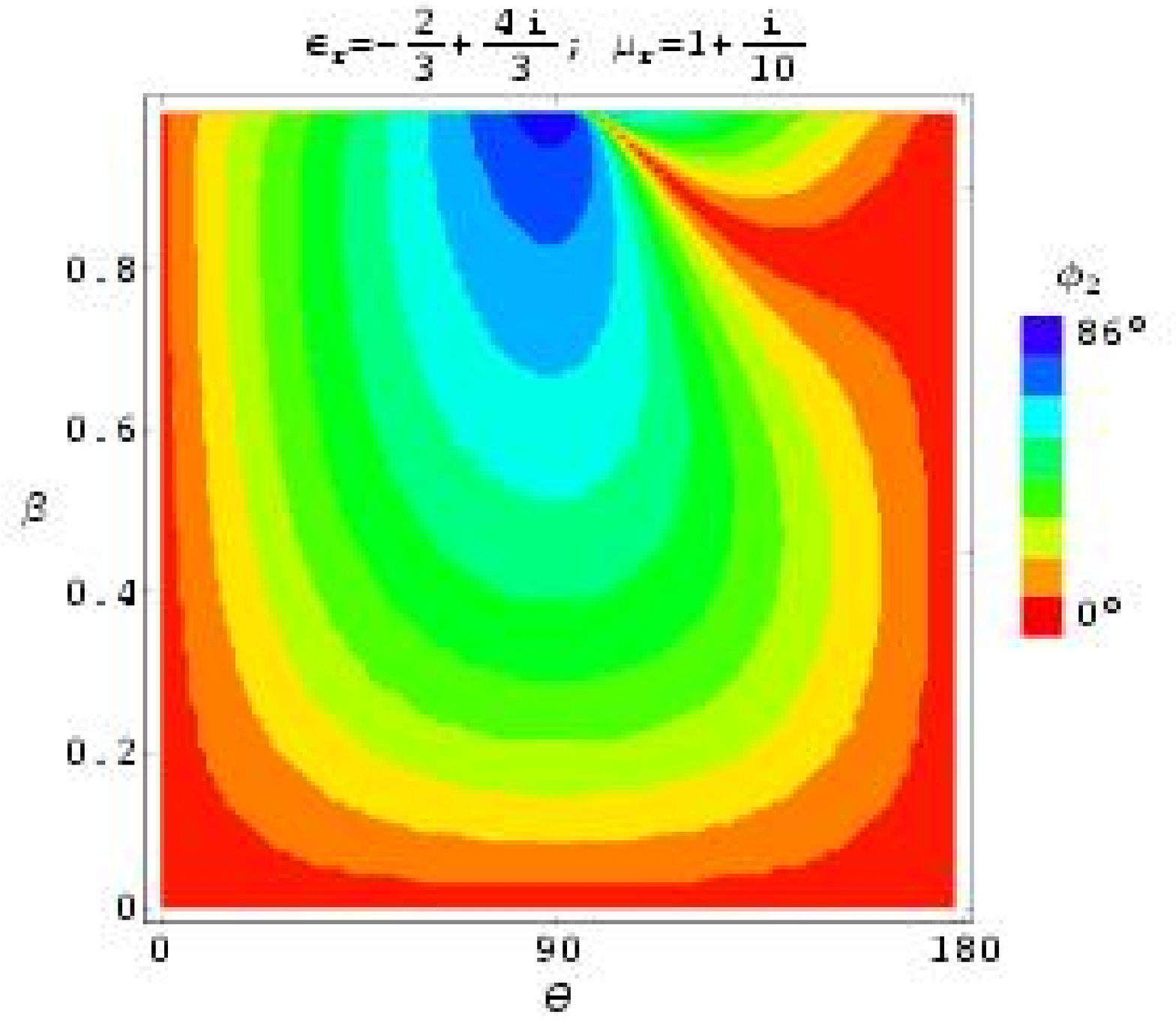,width=2.1in}
\epsfig{file=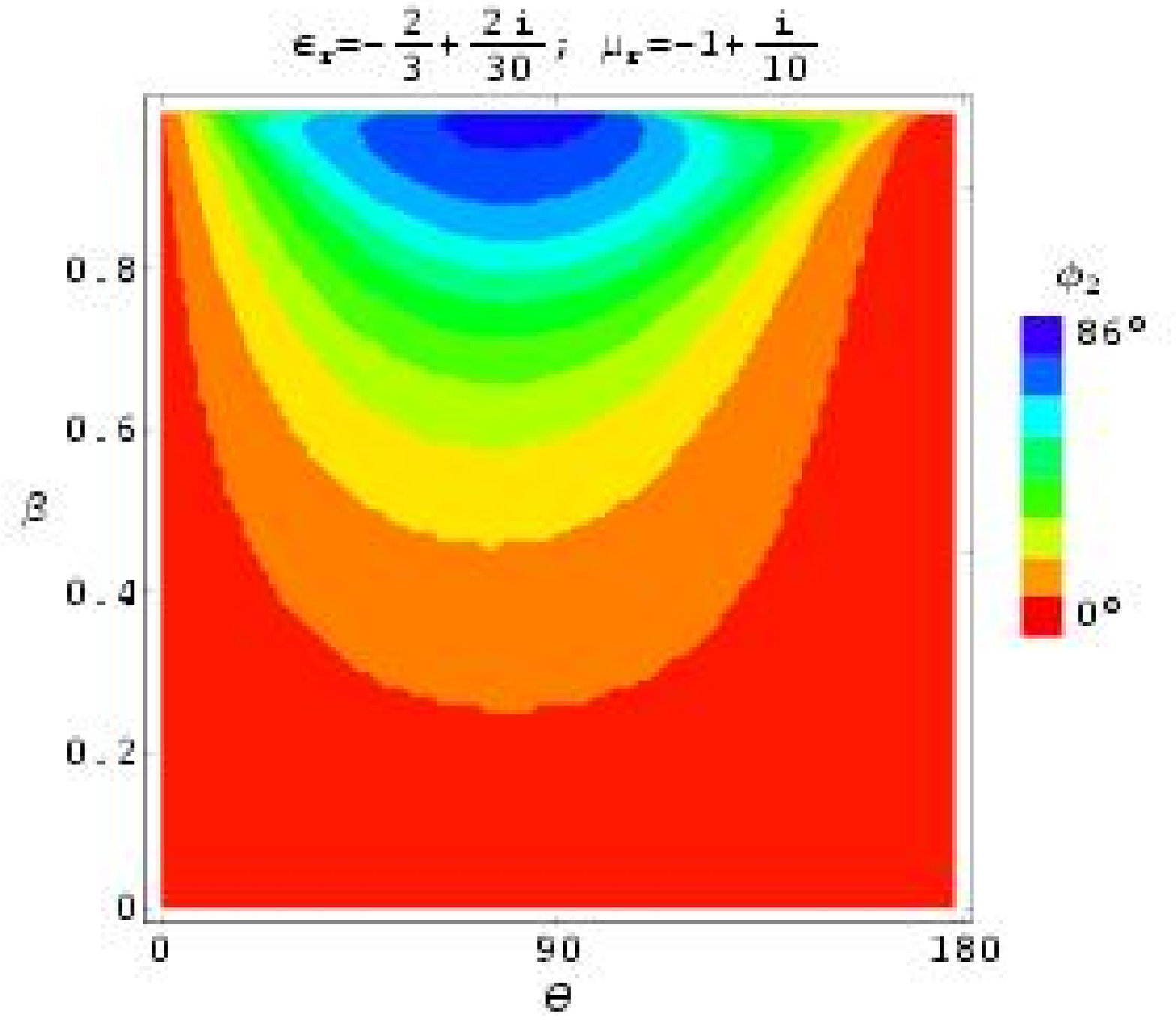,width=2.1in}
  \caption{\label{fig5b} Continued.
 }
\end{figure}

\newpage

\setcounter{figure}{5}

\begin{figure}[!ht]
\centering \psfull \epsfig{file=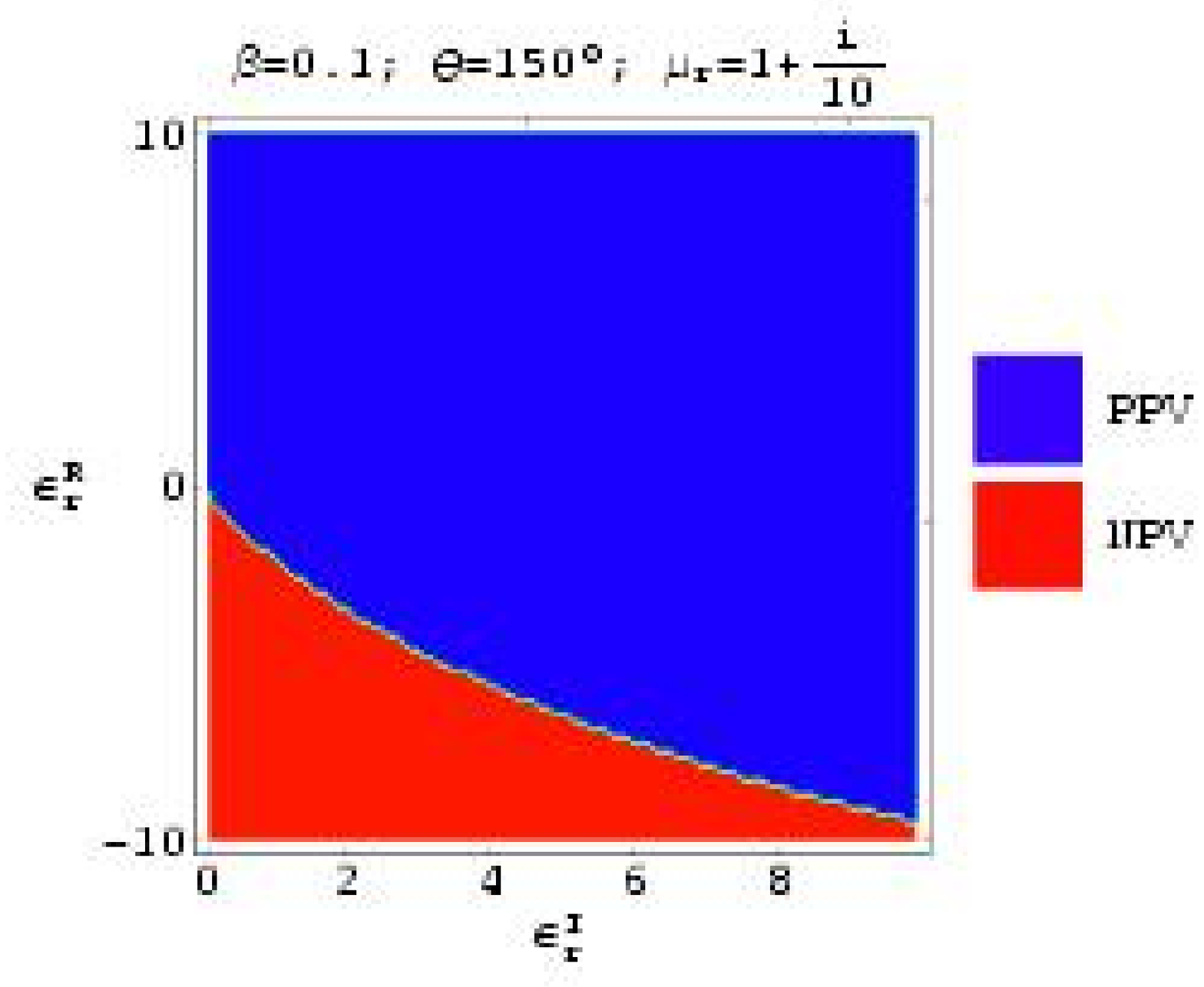,width=2.1in}
\epsfig{file=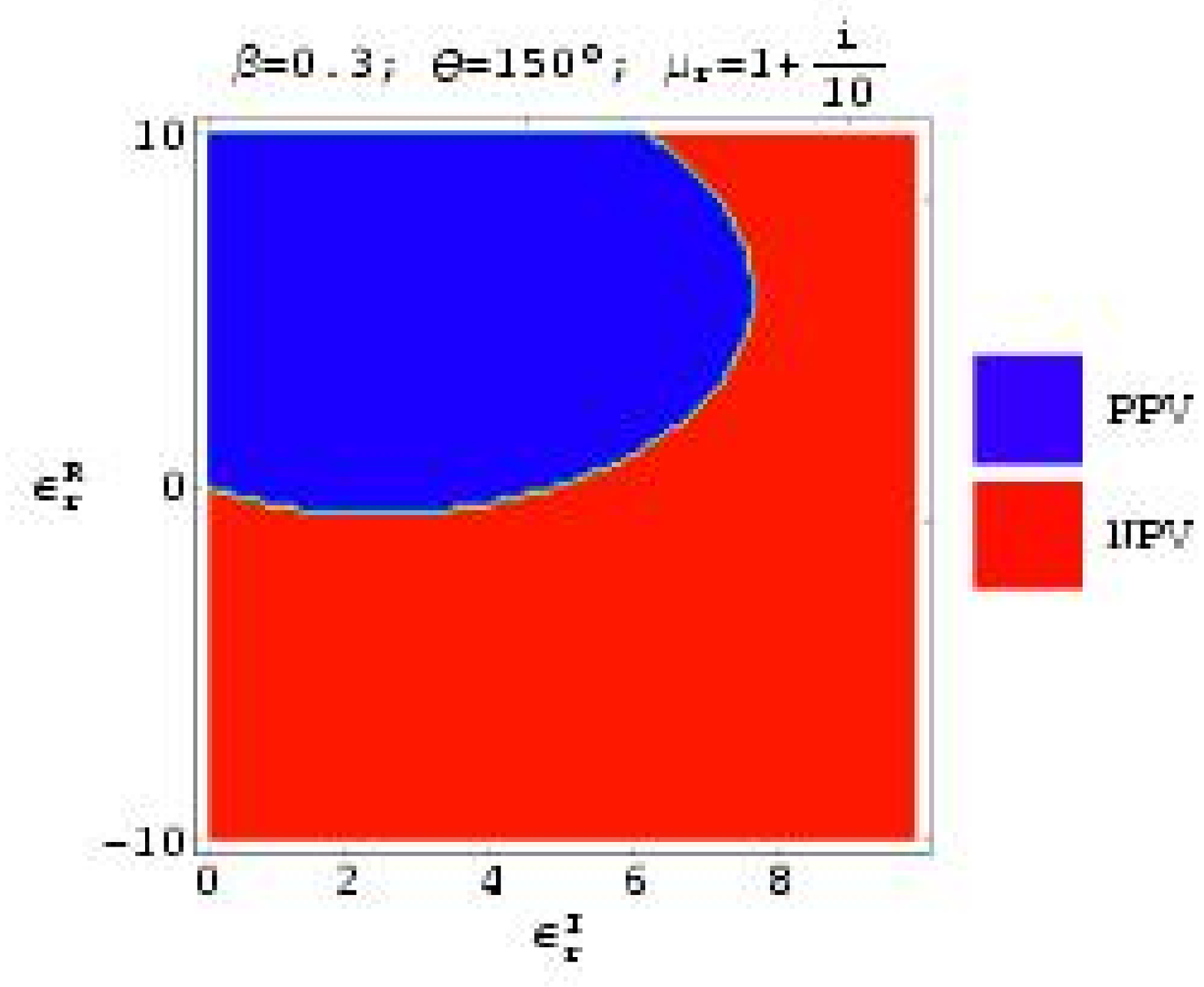,width=2.1in}
\epsfig{file=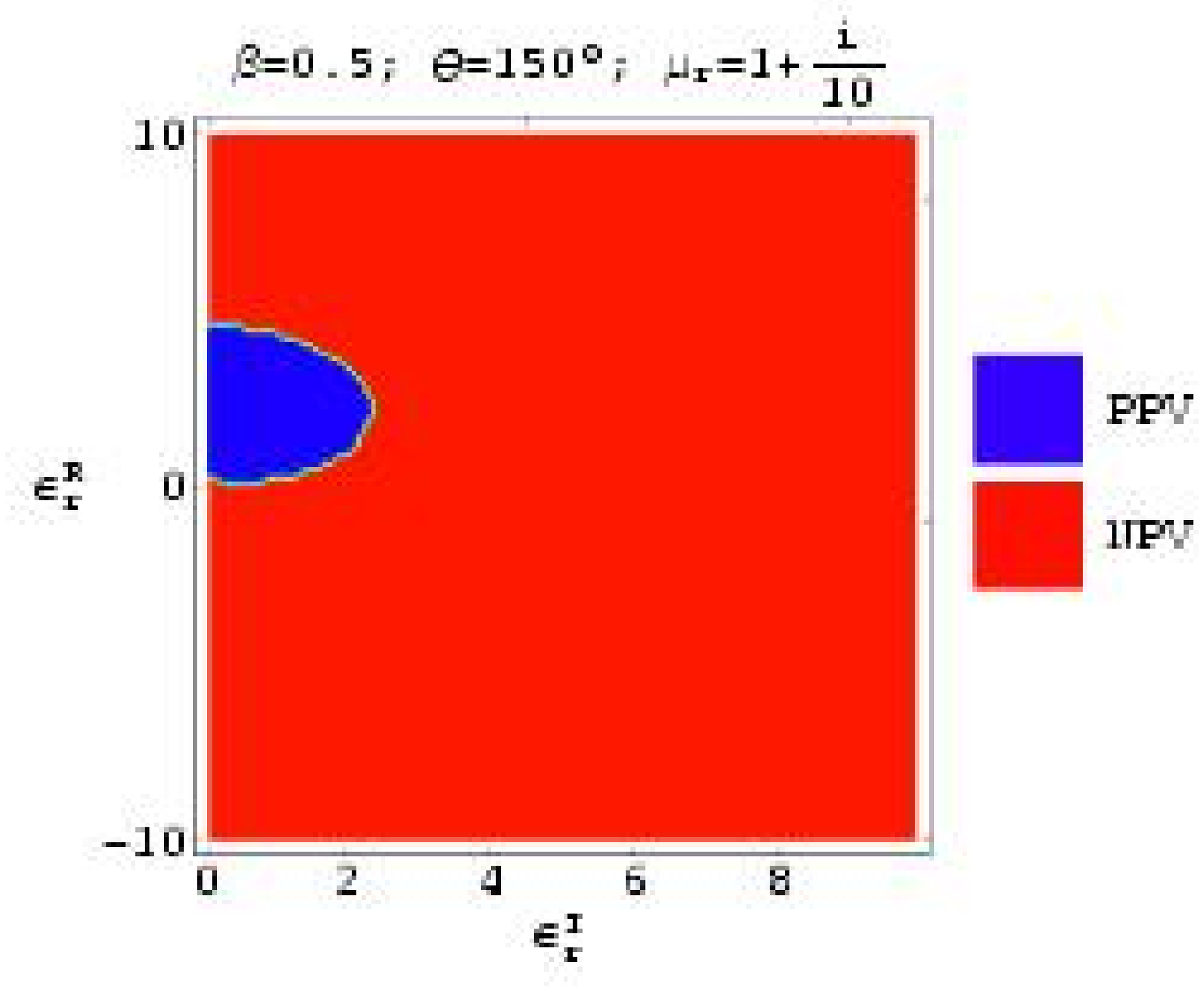,width=2.1in}
\\ \vspace{5mm}
\epsfig{file=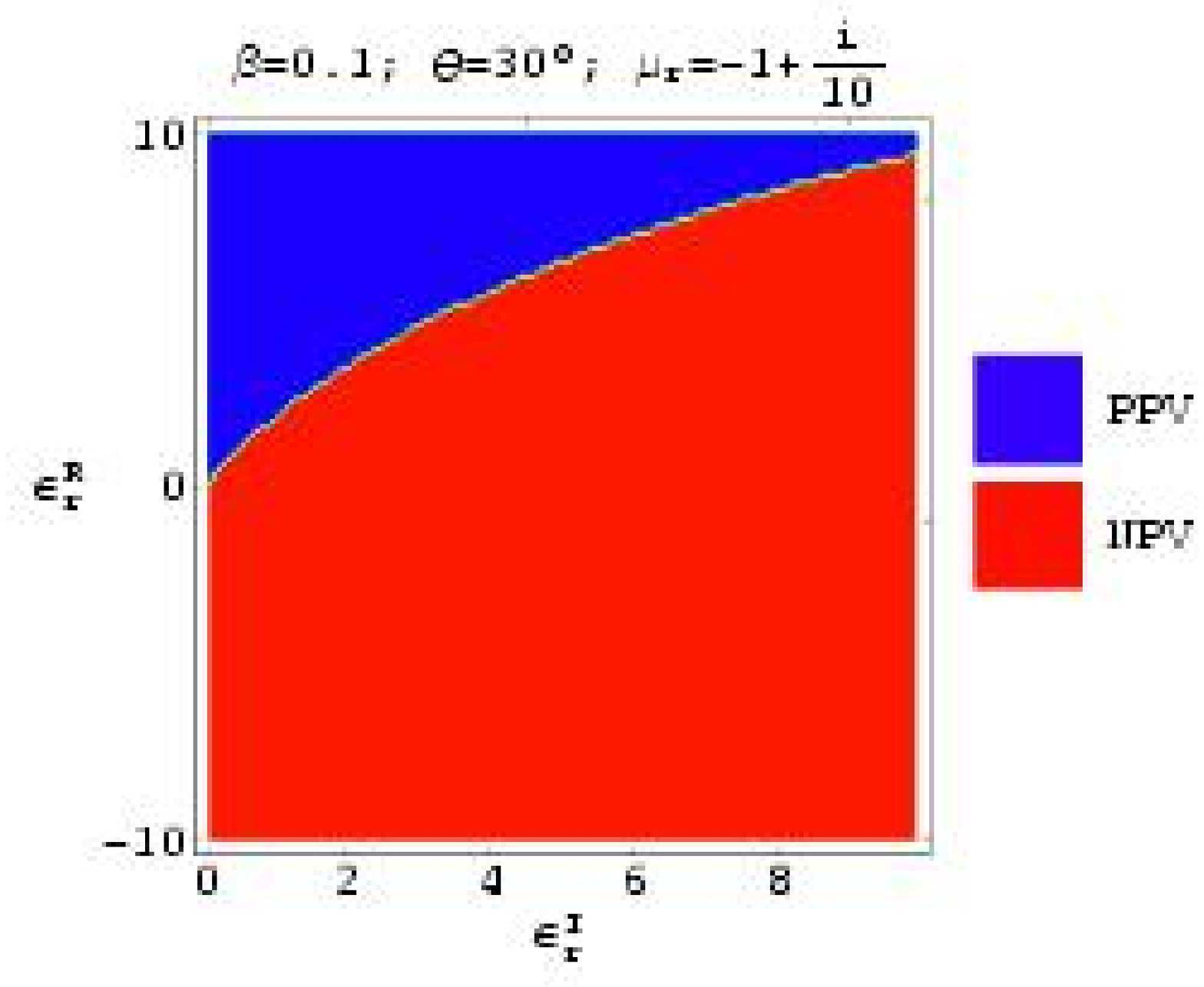,width=2.1in}
\epsfig{file=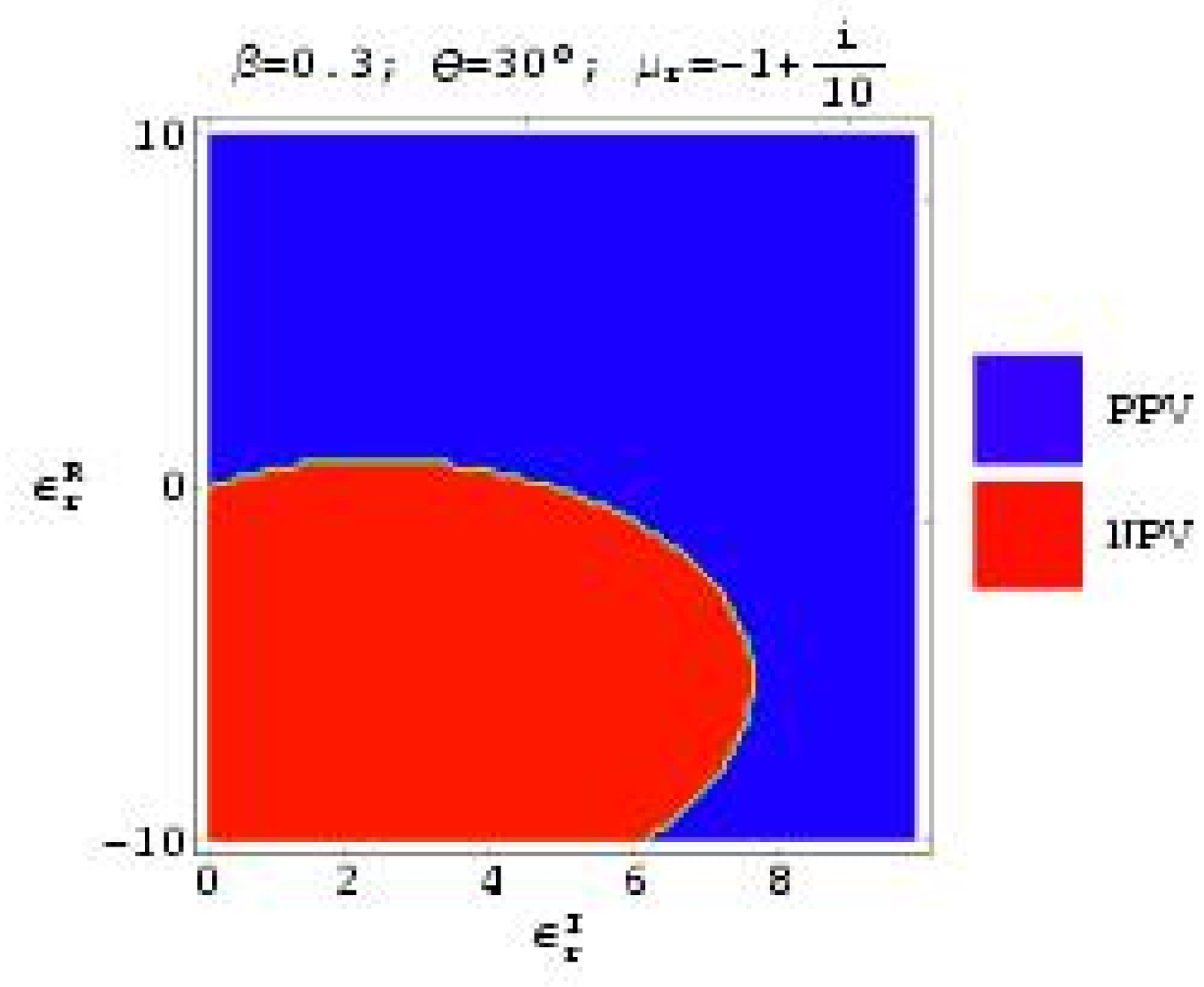,width=2.1in}
\epsfig{file=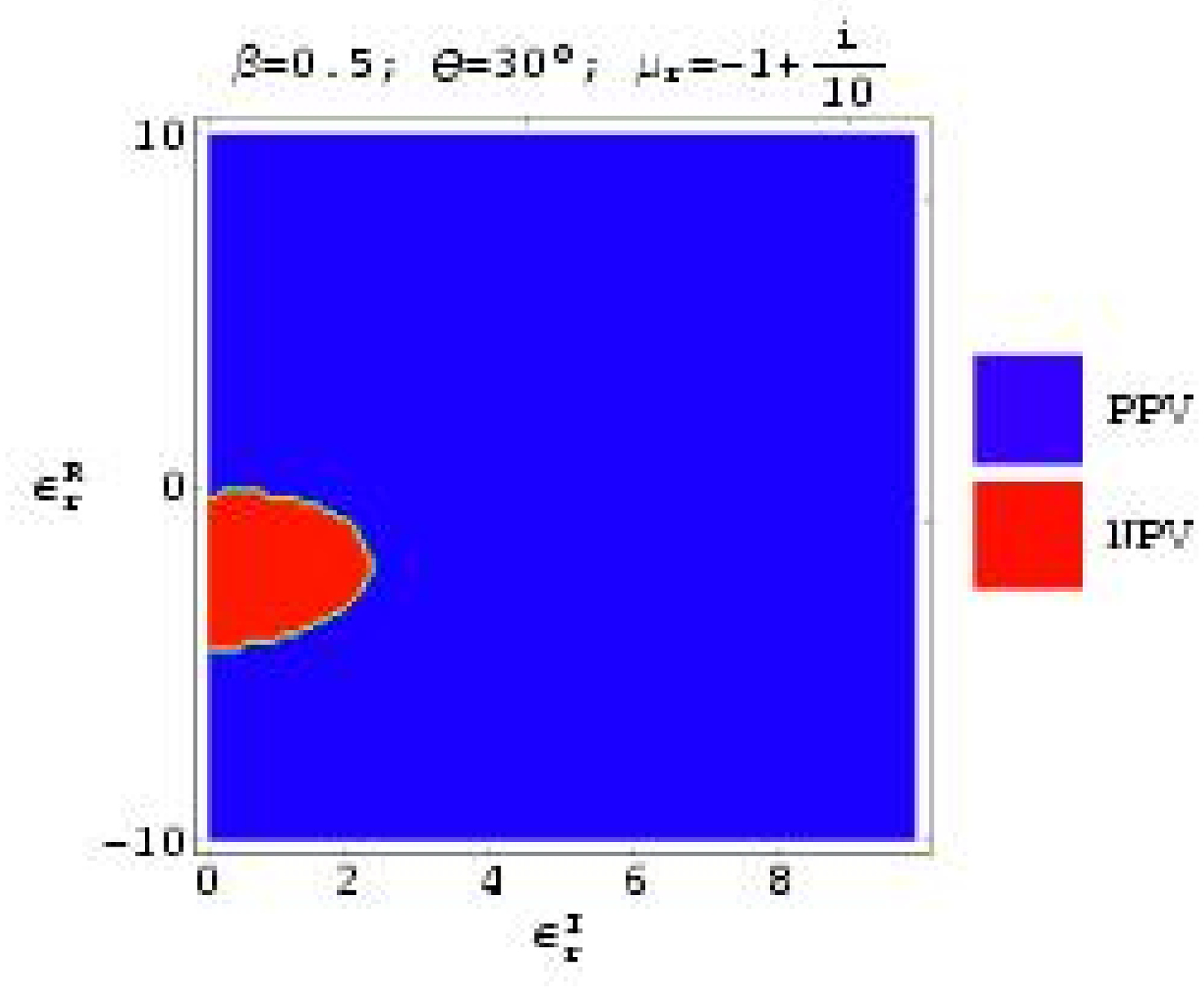,width=2.1in}
  \caption{\label{fig6}
    The distribution of positive phase velocity (PPV) and negative
  phase velocity (NPV)
in relation to $\eps^R_r \in \le -10, 10 \ri$ and $\eps^I_r \in
\le 0, 10 \ri $, for $\theta \in \lec 30^\circ, 150^\circ \ric$,
$\beta \in \lec 0.1, 0.3, 0.5 \ric$, and $\mu_r = \pm 1 + 0.1 i$.
 }
\end{figure}
\newpage

\setcounter{figure}{6}

\begin{figure}[!ht]
\centering \psfull \epsfig{file=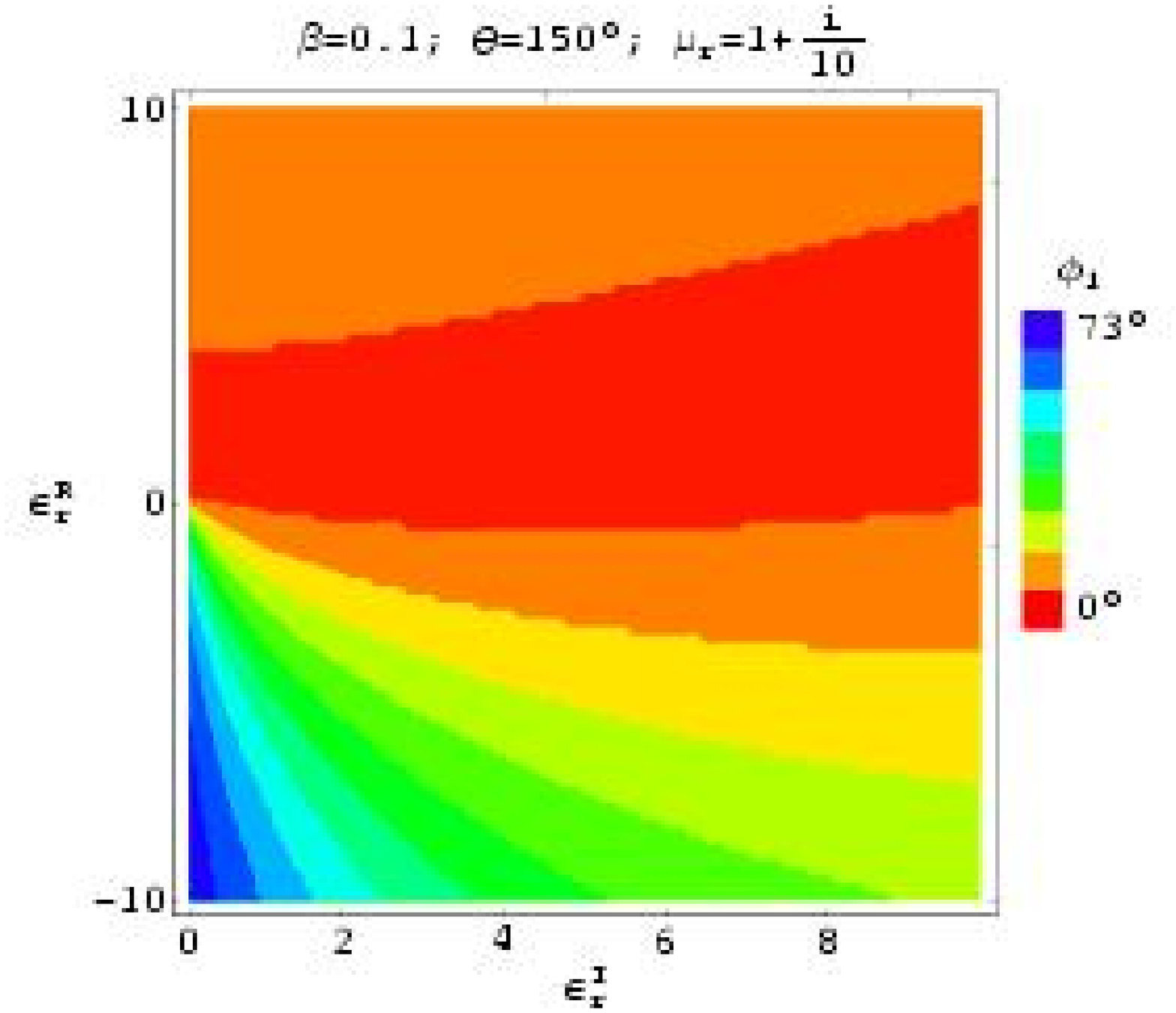,width=2.1in}
\epsfig{file=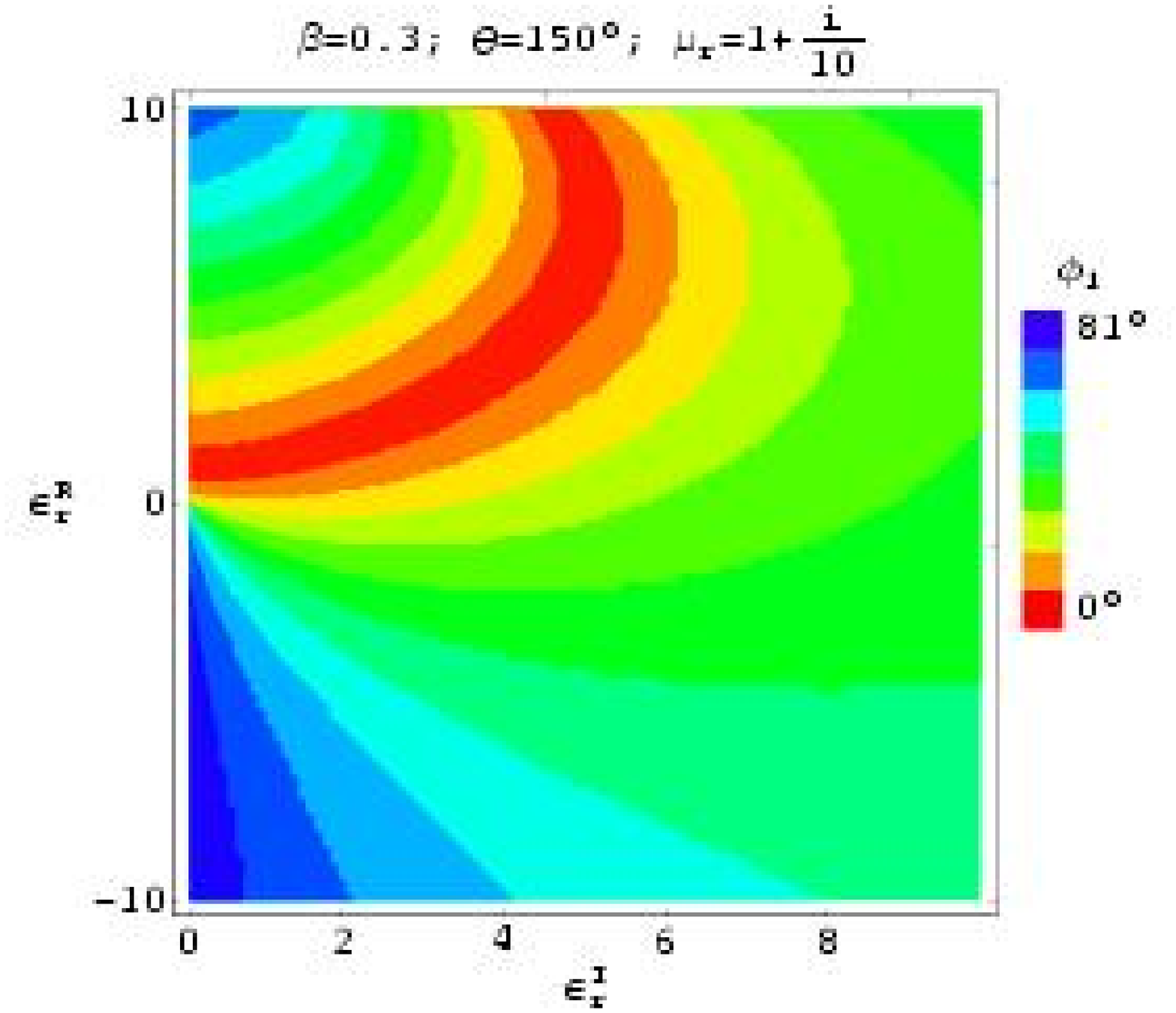,width=2.1in}
\epsfig{file=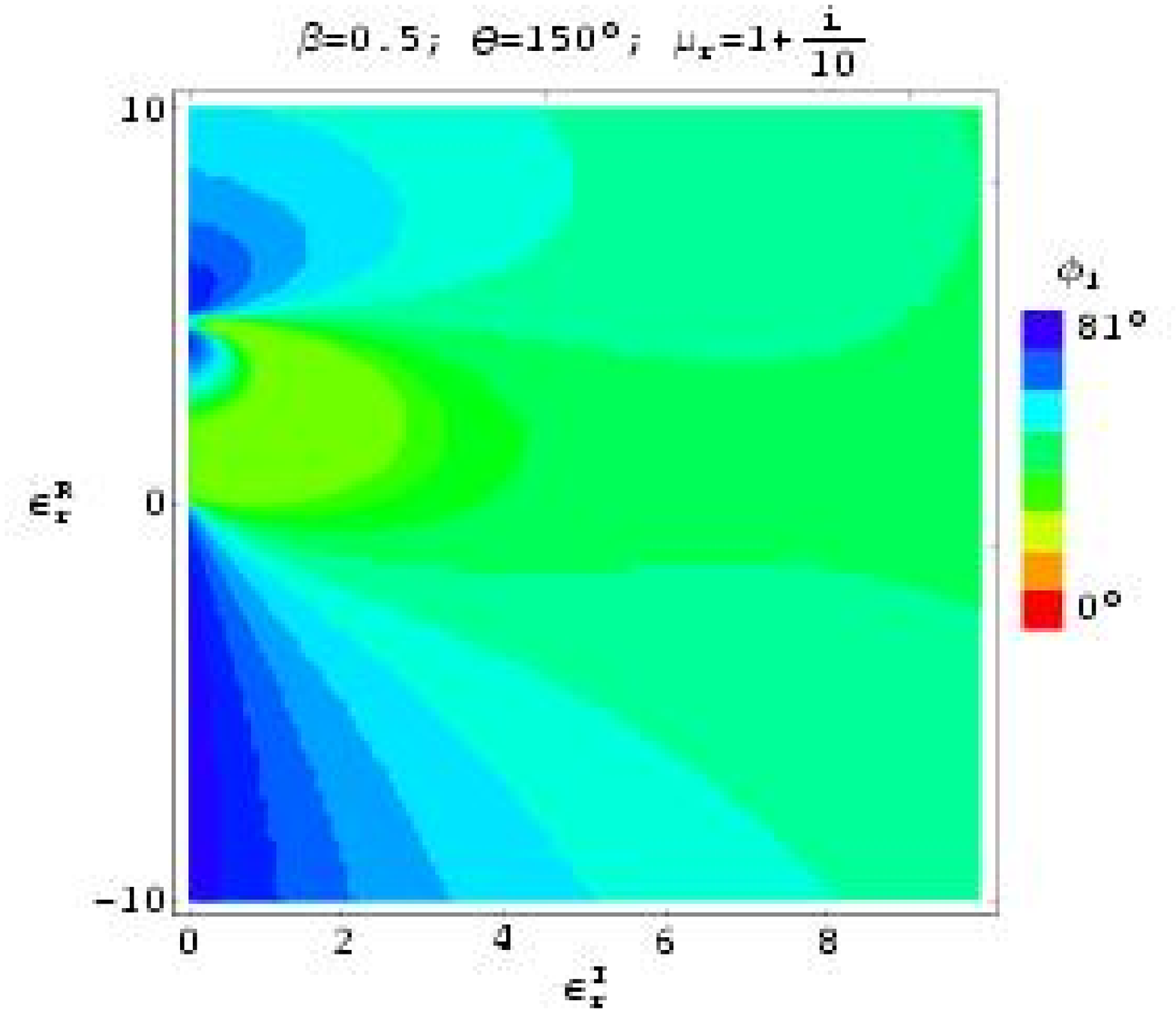,width=2.1in}
\\ \vspace{5mm}
\epsfig{file=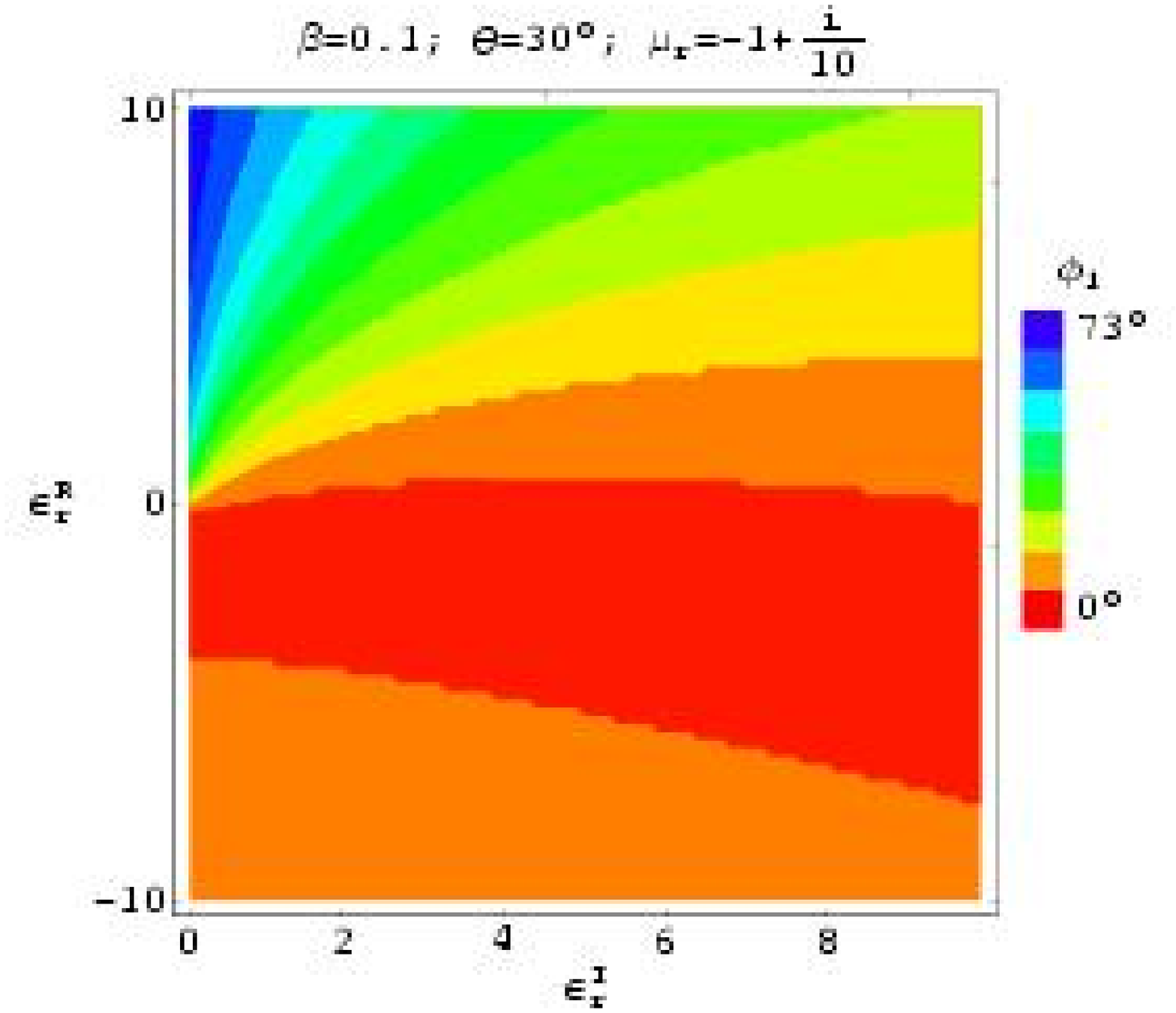,width=2.1in}
\epsfig{file=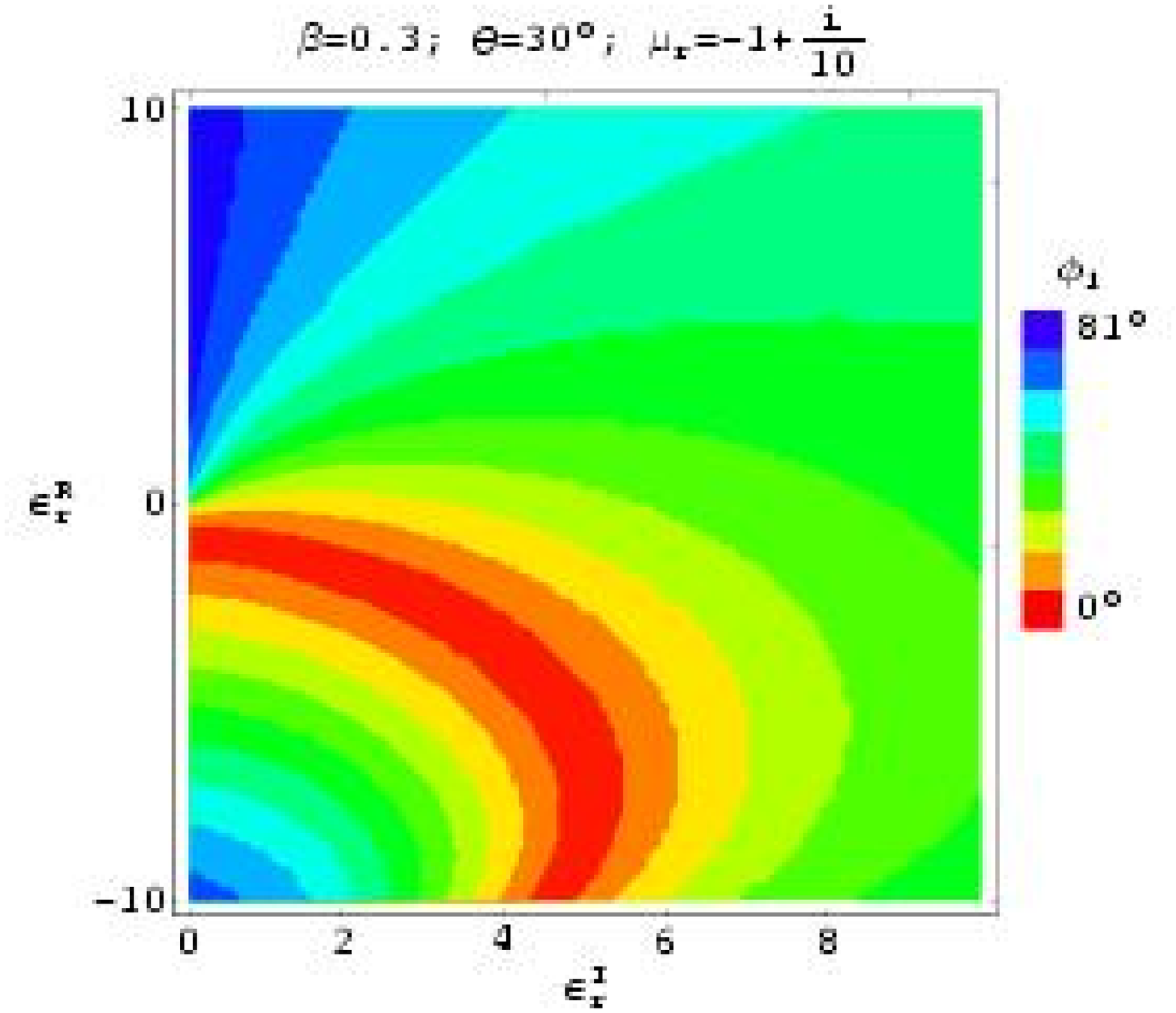,width=2.1in}
\epsfig{file=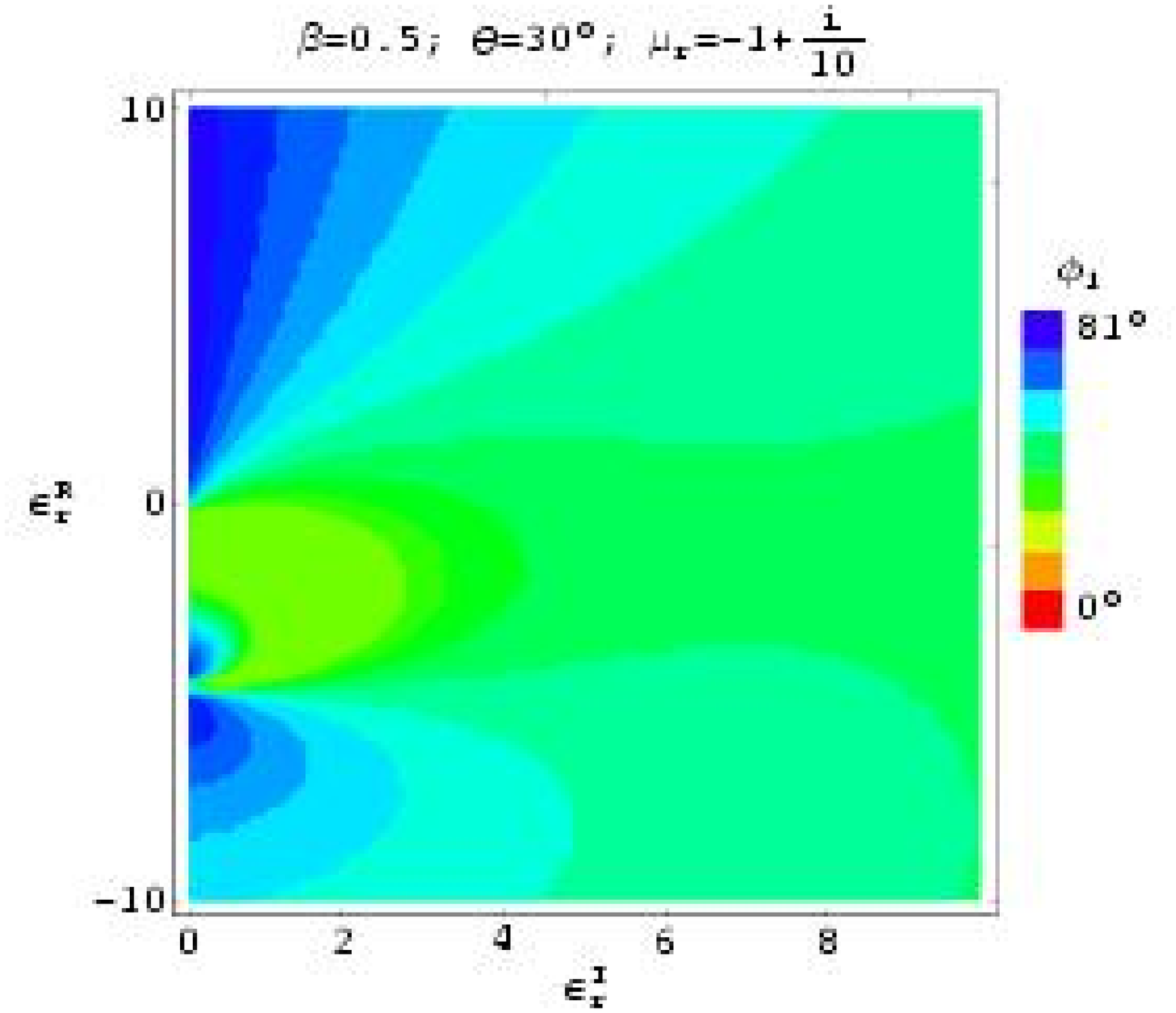,width=2.1in}
  \caption{\label{fig7a}
  The angles $\phi_1$ and $\phi_2$
    as functions of
      $\eps^R_r \in \le -10, 10 \ri$ and $\eps^I_r \in \le 0, 10
\ri $, for $\theta \in \lec 30^\circ, 150^\circ \ric$, $\beta \in
\lec 0.1, 0.3, 0.5 \ric$, and $\mu_r = \pm 1 + 0.1 i$.
 }
\end{figure}

\newpage

\setcounter{figure}{6}

\begin{figure}[!ht]
\centering \psfull \epsfig{file=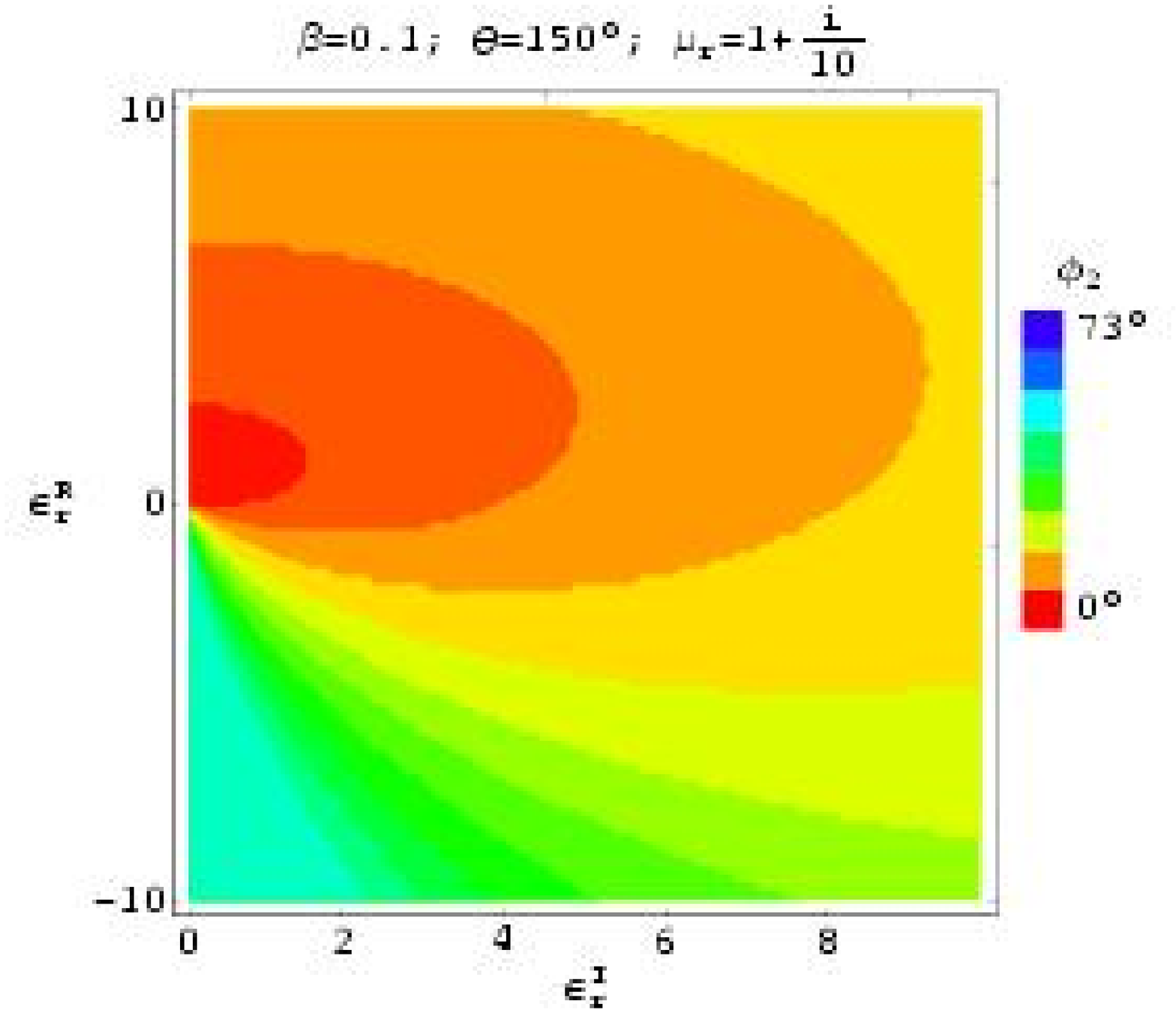,width=2.1in}
\epsfig{file=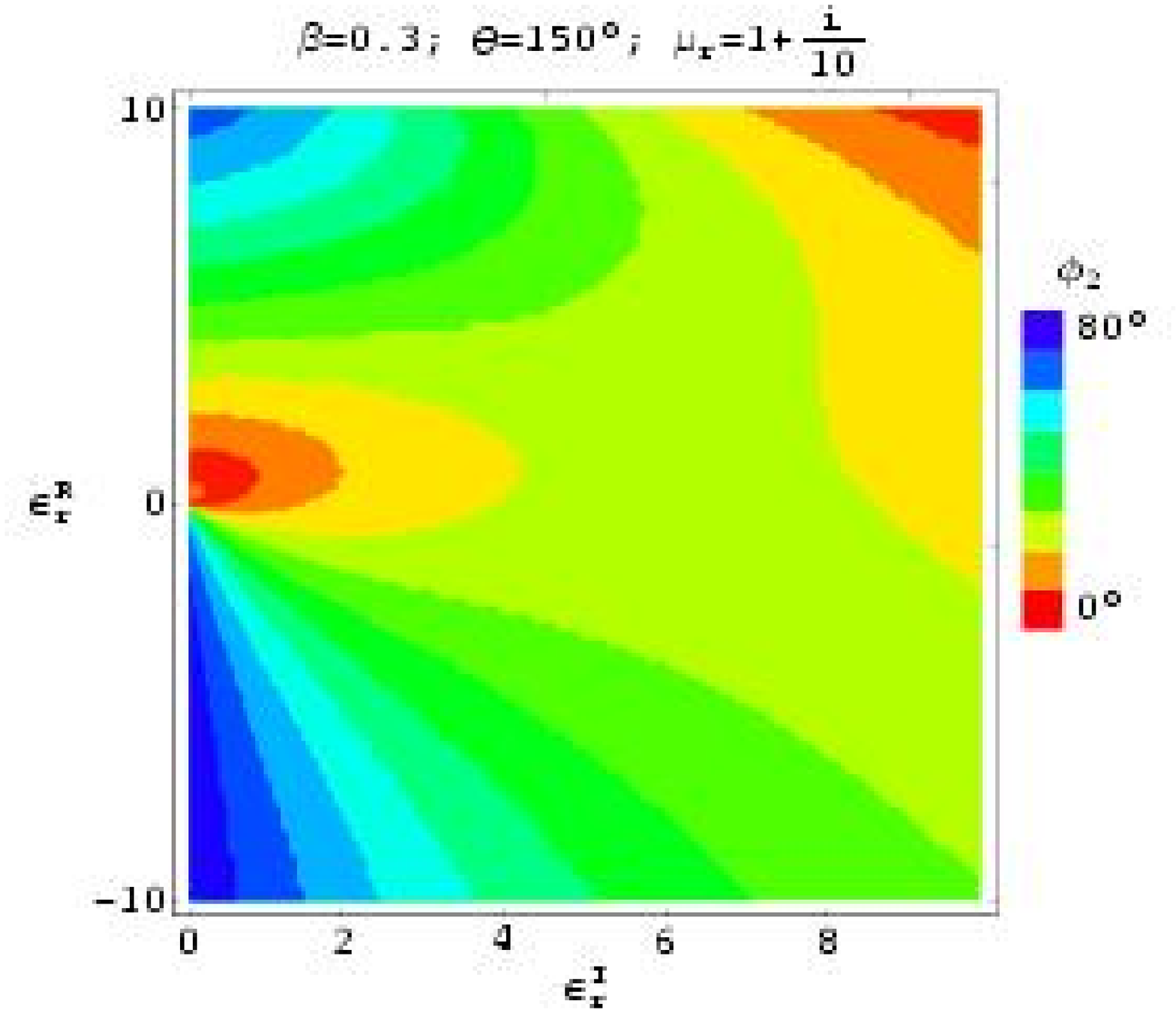,width=2.1in}
\epsfig{file=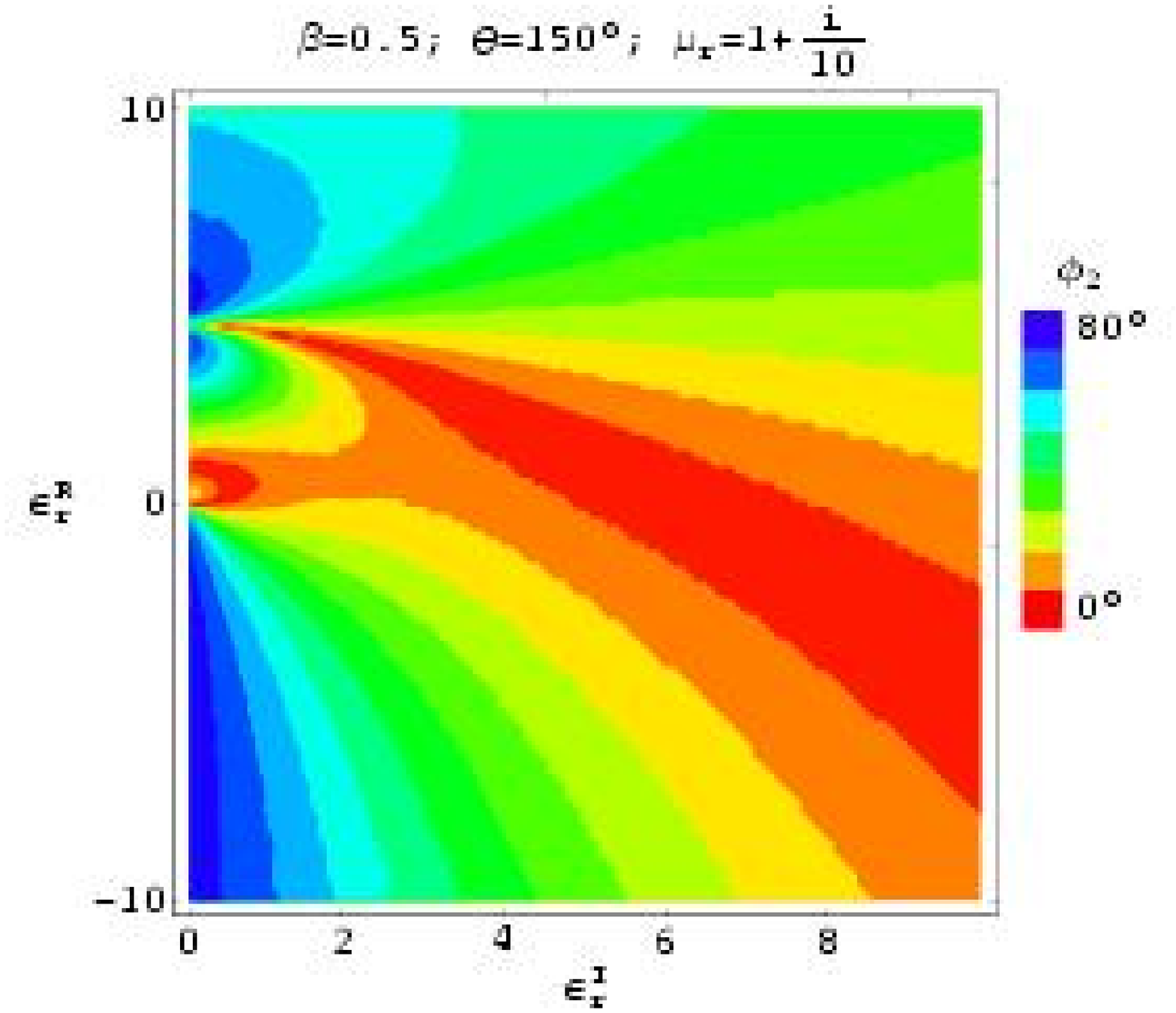,width=2.1in}
\\ \vspace{5mm}
\epsfig{file=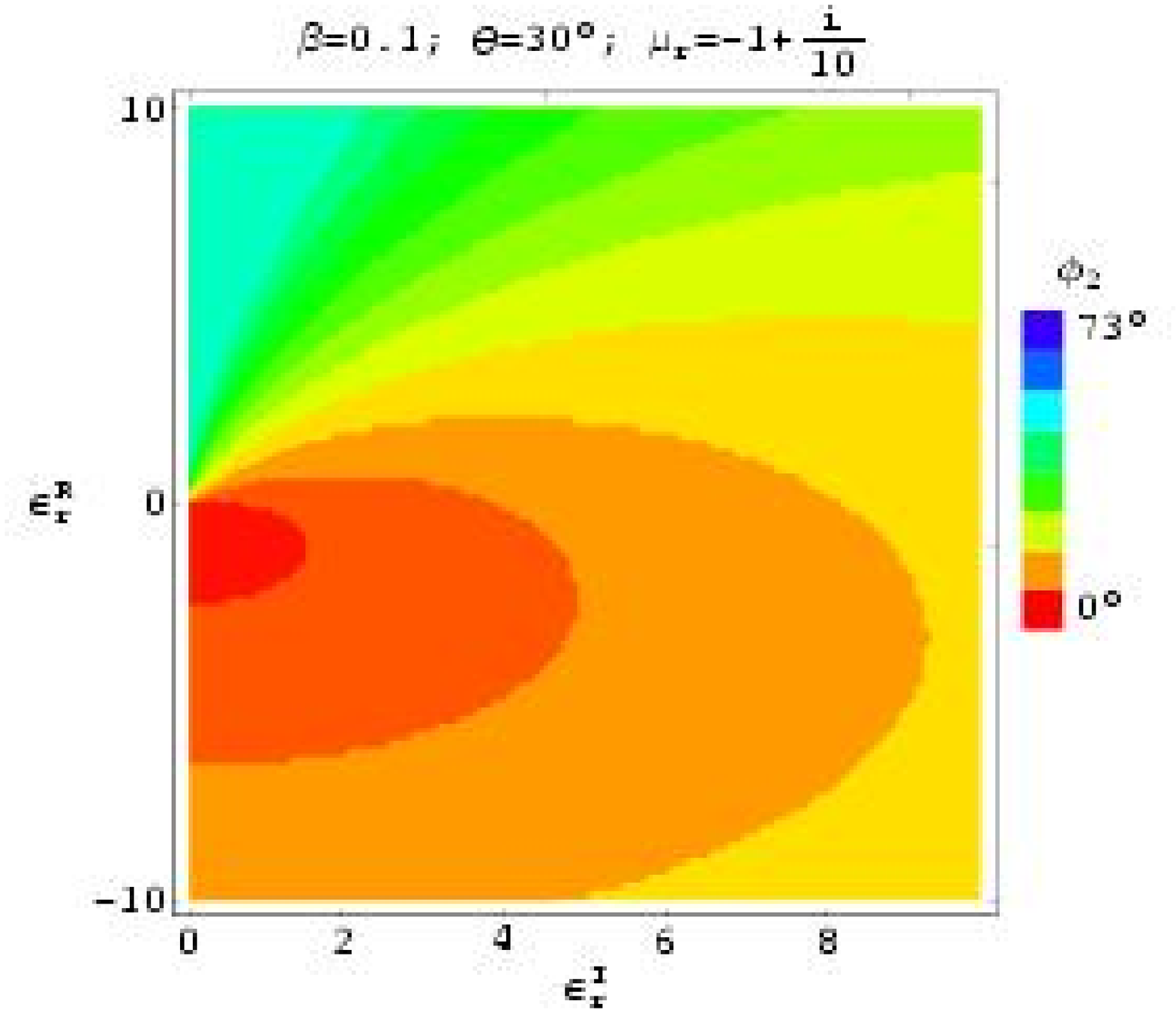,width=2.1in}
\epsfig{file=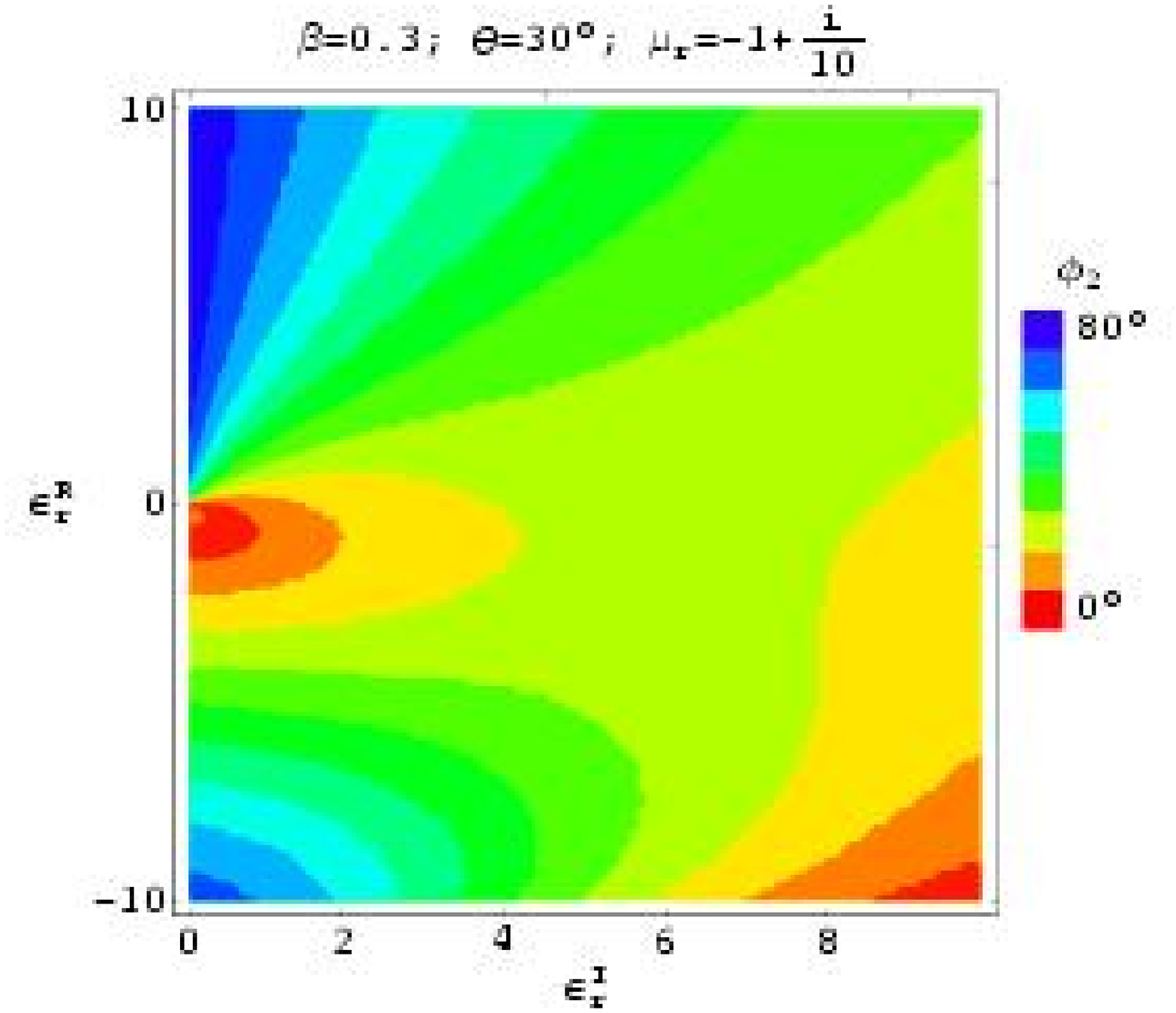,width=2.1in}
\epsfig{file=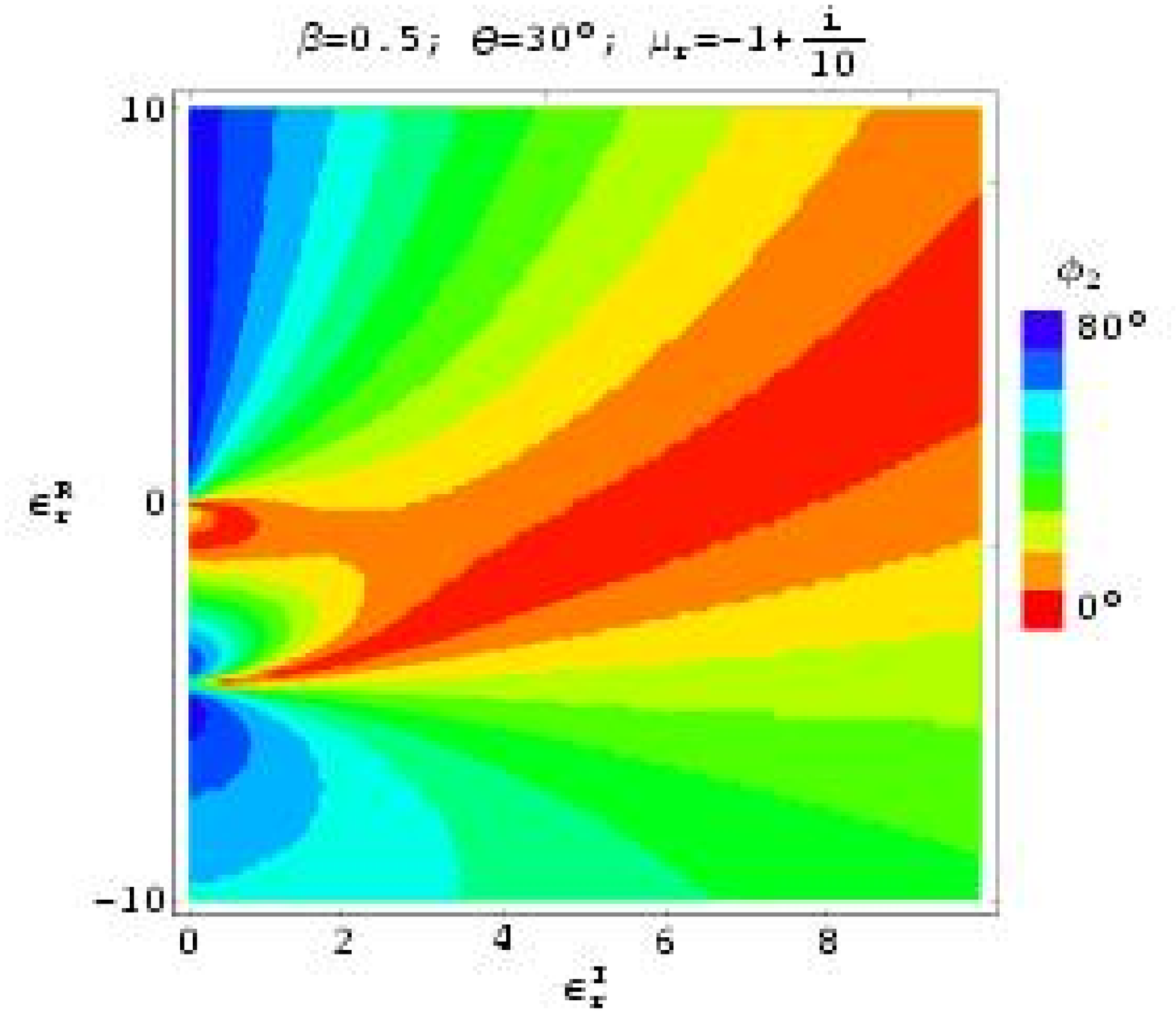,width=2.1in}
  \caption{\label{fig7b} Continued.
 }
\end{figure}

\end{document}